\documentclass[vecphys]{svmult}

\usepackage{makeidx}         
\usepackage{graphicx}        
\usepackage{multicol}        

\makeindex             

\begin{document}

\title*{Numerical solution of the Holstein polaron problem}
\author{H. Fehske\inst{1}\and
S. A. Trugman\inst{2}}
\institute{Institut f\"ur Physik, Ernst-Moritz-Arndt-Universit\"at Greifswald,
D-17487 Greifswald, Germany 
 \texttt{holger.fehske@physik.uni-greifswald.de}
\and
Theoretical Division, Los Alamos National Laboratory,
Los Alamos, New Mexico 87545, U.S.A.
\texttt{sat@lanl.gov}}
%
%
\maketitle
%

\section{Introduction }
\label{sec:in}
Noninteracting itinerant electrons in a solid 
occupy Bloch one-electron states. Phonons are collective
vibrational excitations of the crystal lattice. The basic electron-phonon
(EP) interaction process is the absorption or emission of a 
phonon by the electron with a 
simultaneous change of the electron state. From this it is clear    
that the motion of even a single electron in a deformable lattice
constitutes a complex many-body problem, in that phonons are excited
at various positions, with highly non-trivial dynamical correlations.  

The mutual interaction between the charge carrier and the lattice 
deformations may lead to the formation of a new quasiparticle,
an electron dressed by a phonon cloud. This composite entity is 
called a polaron~\cite{La33,Pe46b}. Since the induced distortion 
(polarisation) of the lattice will follow the electron when it 
is moving through the crystal, one of the most important 
ground-state properties of the polaron is an increased inertial mass. 
A polaronic quasiparticle is referred to as ``large polaron'' 
if the spatial extent of the phonon cloud is large compared to 
the lattice parameter. By contrast, if the lattice deformation 
is basically confined to a single site, the polaron is 
designated as ``small''.  Of course, depending on the strength, 
range and retardation of the electron-phonon interaction, 
the spectral properties of a polaron will also notably 
differ from those of a normal band carrier. 
Since there is only one electron in the problem, these findings are
independent of the statistics of the particle, i.e. we can think
of any fermion or boson, such as an electron-, hole-, exciton- or 
Jahn-Teller polarons (for details see 
Refs.~\cite{Fi75,Ra82,PW84}). 

The microscopic structure of polarons is very diverse. 
The possible situations are determined by the type of particle-phonon 
coupling~\cite{Ra82,SS93}. Systems characterised by optical phonons with polar 
long-range interactions are usually described by the Fr\"ohlich
Hamiltonian~\cite{Fr54,Fr74,Dev}. If the optical phonons have nonpolar 
short-range EP interactions,
Holstein's (molecular crystal) model applies~\cite{Ho59a,Ho59b}. 
For a large class of Fr\"ohlich- and Holstein-type models it has 
been proven that the ground-state energy of a polaron is an 
analytic function of the EP coupling parameter for all interaction 
strengths~\cite{PD82,Sp87a,Loe88,GL91}. 
The dimensionality of space here has no 
qualitative influence. In this sense a (formal) abrupt (nonanalytical)  
polaron transition does not exist: The standard phase 
transition concept fails to describe polaron formation.  
It is, instead, a (possibly rapid) crossover.
(We mention parenthetically that in contrast to the ground state,
the polaron first excited state may be nonanalytic in the EP coupling.)

The fundamental theoretical question in the context of polaron physics 
concerns the possibility of a local lattice instability that 
traps the charge carrier upon increasing the EP coupling~\cite{La33}. 
Such trapping is energetically favoured over wide-band Bloch states if the
binding energy of the particle exceeds the strain energy 
required to produce the trap. Since the potential itself 
depends on the carrier's state, this highly
non-linear feedback phenomenon is 
called ``self-trapping''~\cite{Fi75,Ra82,Emi95,WF98a}. 
Self-trapping does not imply a breaking of translational invariance.
In a crystal the polaron ground state is still extended allowing, 
in principle, for coherent transport although with an extremely 
narrow band. 
One way to think of this is that a hypothetical self-trapped state
can coherently tunnel with its phonon cloud to neighbouring locations, thus delocalising.
The problem of self-trapping, i.e. the crossover from 
rather mobile large polarons to quasi-immobile small polarons, 
basically could  not be addressed within the continuum approach.
Self-trapping requires a physics which is related to particle and 
phonon dynamics on the scale of the unit cell~\cite{Ra06}. 
On the experimental side, an increasing number of advanced  
materials show polaronic effects on such short length and time scales.
Self-trapped polarons can be found, e.g., in (non-stoichiometric) 
uranium dioxide, alkaline earth halides, 
II-IV- and group-IV semiconductors, 
organic molecular crystals, high-$T_c$ cuprates, charge-ordered nickelates and 
colossal magneto-resistance 
manganites~\cite{De63,TNYS90,SS93,BEMB92,SAL95,CCC93,Mi98,JHSRDE97,Eg06}.

As stated above, the generic model to capture such a situation 
is the Holstein Hamiltonian, which is most simply written
in real space~\cite{Ho59a}. Here the orbital states are identical on each site
and the particle can move from site to site exactly as in a 
tight-binding model. The phonons are coupled to
the particle at whichever site it is on. The dynamics of the lattice 
is treated purely locally with Einstein oscillators describing the 
intra-molecular oscillations.

Theoretical research on the Holstein model spans over 
five decades. As yet none of the various analytical 
treatments, based on variational approaches~\cite{Em73,To61} or on  
weak-coupling~\cite{Mi58} and strong-coupling 
adiabatic~\cite{Ho59a} and non-adiabatic~\cite{LF62,Ma95} 
perturbation expansions, are suitable for the investigation 
of the physically most interesting crossover regime where 
the self-trapping crossover of the charge carrier takes place.   
That is because precisely in this situation 
the characteristic electronic and phononic energy scales 
are not well separated and non-adiabatic effects 
become increasingly important, implying a
breakdown of the standard Migdal approximation~\cite{AM94b}. 
The Holstein polaron can be solved in infinite dimensions ($D=\infty$)
using dynamical mean-field theory~\cite{Su74,CPFF97}. 
While this method treats the local dynamics exactly, it
cannot account for the spatial correlations being of vital
importance in finite-dimensional systems.   

In principle, quasi approximation-free numerical methods like
exact diagonalisation (ED)~\cite{Ma93,AKR94,MR97,WRF96,BWF98}, 
quantum Monte Carlo (QMC)~\cite{RL83,BVL95,KP97,HEL04,Kor,HL}
and diagrammatic Monte Carlo~\cite{MN} simulations, 
the global-local (GL) method~\cite{RBL98} or the 
recently developed density-matrix
renormalisation group (DMRG) technique~\cite{JW98b,ZJW98}
can overcome all these difficulties. 
Although most of these methods give reliable results 
in a wide range of parameters, thereby closing 
the gap between the weak and strong EP coupling, 
low- and high-frequency limits, each suffers from different shortcomings. 
ED is probably the best controlled numerical method
for the calculation of ground- and excited state properties. 
In practice, however, memory limitations have restricted
brute force ED to small lattices (typically up to 20 sites). 
So results are limited to discrete momentum points. 
QMC can treat large system sizes (over 1000 sites) and 
provide accurate results for the thermodynamic properties. 
On the other hand, the calculation of spectral properties is 
less reliable mainly because of the ill-posed analytic
continuation from imaginary time. The GL method is 
basically limited to the analysis of
ground-state properties. DMRG and the recently developed dynamical
DMRG~\cite{Je02b} have proved to be extremely 
accurate for the investigation of      
1D EP systems, where they can deal with sufficiently large
system sizes (e.g., 128 sites and 40 phonons). 
The determination of spectral functions (in particular of the 
high-energy incoherent features), however, is 
computationally expensive and so far there exists no really efficient 
DMRG algorithm to tackle non-trivial problems in $D>1$.   

In this contribution we provide an exact numerical solution of the 
Holstein polaron problem by elaborate ED techniques,   
in the whole range of parameters and, at least concerning the properties 
of the ground state and low-lying excited states, in the thermodynamic limit. 
Combining Lanczos ED~\cite{CW85} with kernel polynomial~\cite{SRVK96,WWAF06} 
and cluster perturbation~\cite{SPP00,HAL03}  
expansion methods also allows the polaron's spectral and dynamical properties 
to be computed exactly. A numerical calculation is said to be exact if no 
approximations are involved aside from the restriction imposed by finite 
computational resources, the accuracy can be systematically 
improved with increasing computational effort, and actual numerical errors are 
quantifiable or completely negligible. 
In most numerical approaches to many-body problems,
the numerical error decreases as 1/log(effort), where effort means
either execution time or storage required. Thus even a large increase
in computational power will not greatly improve the accuracy.
Despite some progress by virtue of DMRG-based basis optimisation~\cite{WFWB00}
or coherent-state variational approaches~\cite{CFIP06,CFP},
ED of EP systems remained inefficient.
Recently ED methods have been developed that converge far more rapidly,
with error~$\sim 1/({\rm effort})^{\theta}$, where $\theta$ 
is a favourable power ($\theta \approx 3$ at intermediate 
coupling)~\cite{TKB04}. Thus doubling the size of the Hilbert 
space results in almost an extra significant figure in the energy. 
The algorithm~\cite{BTB99,KTB02,EBKT03} we will apply in the following 
is based on the construction of a variational Hilbert space 
on a infinite lattice and can be expanded in a systematic 
way to easily achieve greater than 10-digit accuracy
for static correlation functions and 20 digits for energies, with 
modest computational resources. The increased power makes it possible 
to solve the Holstein polaron problem at continuous wave vectors in dimensions
D=1, 2, 3, 4, \ldots .

The paper is organised as follows: In the remaining introductory part, 
Sect.~\ref{sec:mn} presents the Holstein model and outlines 
the numerical methods we will employ for its solution.  
The second, main part of this paper reviews our numerical results 
for the ground-state and spectral properties of the Holstein polaron.
The polaron's effective mass and band structure, as well as 
static electron-lattice correlations, will be analysed in Sect.~\ref{sec:gs}.
Section~\ref{sec:es} is devoted to the investigation of the 
excited states of the Holstein model. 
The dynamics of polaron formation is studied
in Sect.~\ref{sec:dy}. Characteristic results for electron and phonon 
spectral functions will be presented in Sec.~\ref{sec:sp}. 
The optical response is examined in Sec.~\ref{sec:to}. 
Here also finite-temperature properties
such as activated transport will be discussed. 
In the third part of this paper finite-density and correlation effects
will be addressed. First we investigate the possibility of 
bipolaron formation and discuss the many-polaron problem (Sect.~\ref{sec:mp}).
Second we comment on the interplay of strong electronic correlations
and EP interaction in advanced materials (Sect.~\ref{sec:sc}). 
Some open problems are listed in the concluding Sect.~\ref{sec:co}. 

\section{Model and methods}
\label{sec:mn}
\subsection{Holstein Hamiltonian}
\label{sec:hm_stuart}
With our focus on polaron formation in systems with 
short-range non-polar EP interaction only,   
we consider the Holstein molecular crystal model
on a D-dimensional hyper-cubic lattice,
\begin{equation}
H=-t\sum_{\langle i,j\rangle} c_i^{\dagger} c_j^{}
- \bar{g} \sum_i (b_i^{\dagger}  + b_i^{})  n_i^{}
+\omega_0 \sum_i  b_i^{\dagger} b_i^{}\,,
\label{hm}
\end{equation}
where $c_i^{\dagger}$ ($c_i^{}$) and $b_i^{\dagger}$ ($b_i^{}$) are,
respectively, creation (annihilation) operators for electrons and
dispersionless optical  phonons on site $i$, and 
$n_i^{}=c_i^{\dagger}c_i^{}$ is the corresponding particle number operator. 
The parameters of the model are the nearest-neighbour hopping integral $t$,
the EP coupling strength $\bar{g}$, and 
the phonon frequency $\omega_0$. The parameters  $t$,  $\omega_0$, and 
$\bar{g}$ all have units of energy, and can be used to form two independent 
dimensionless ratios. 

The first ratio is the so-called adiabaticity parameter,
\begin{equation}
\alpha =\omega_0/t\,,
\label{alpha}
\end{equation}
which determines which of the two subsystems, electrons or phonons,
is the fast or the slow one. In the adiabatic limit $\alpha\ll 1$, 
the motion of the particle is affected by quasi-static lattice
deformations (adiabatic potential surface). 
In contrast, in the anti-adiabatic 
limit $\alpha\gg 1$, the lattice deformation is presumed to
adjust instantaneously to the position of the carrier.    
The particle is referred to as a ``light'' or ``heavy'' 
polaron in the adiabatic or anti-adiabatic regimes~\cite{Ra82}.

The second ratio is the dimensionless EP coupling constant.
Here 
\begin{equation}
g =\bar{g}/\omega_0
\label{g}
\end{equation}
appears in (small polaron) strong-coupling perturbation theory.
Defining the polaron binding energy as 
$\varepsilon_p= \bar{g}^2/\omega_0=g^2\omega_0$,
the EP coupling can be parametrised alternatively as the ratio
of polaron energy for an electron confined to a single site 
and the free electron half bandwidth $2Dt$:
\begin{equation}
\lambda =\varepsilon_p/2Dt\,.
\label{lambda}
\end{equation}
 
In the limit of small particle density, a crossover from 
essentially free carriers to heavy quasiparticles    
is known to occur from early quantum Monte Carlo 
calculations~\cite{RL82}, provided that two  conditions, 
$g > 1$ and $\lambda > 1$, are fulfilled. 
So, while the first requirement is more restrictive
if $\alpha$ is large, i.e. in the anti-adiabatic case, the formation of 
a small polaron state will be determined by the second criterion in 
the adiabatic regime~\cite{CSG97,WF97}.

Perhaps it is not surprising that standard perturbative techniques 
are less able to describe the Holstein system close to the 
large- to small-polaron crossover, where $\varepsilon_p \sim 2Dt$ 
or $\varepsilon_p \sim \omega_0$.
In principle, variational approaches, that give correct results 
in the weak- and strong-coupling limits, could provide an interpolation scheme.
Most variational calculations, however, lead to a discontinuous 
transition in the wave function and the derivative of the ground-state
energy, considered as a function of the coupling parameter.
Clearly the analytical behaviour of an exact wave function may deviate
considerably from that of a variational approximation~\cite{GL91}.  
With regard to the Holstein polaron problem the nonanalytic 
behaviour found for the adapted wave function in many variational 
approaches, see, e.g.,~\cite{Feea94} and references therein,  
is an artifact of the approximations, as we will demonstrate
below for all dimensions~\cite{KTB02}. 

Nevertheless, variational calculations are an indispensable 
tool for numerical work. In the next subsection we describe 
a variational exact diagonalisation  (VED) scheme~\cite{BTB99} that  
does not suffer from the above drawback of  (ground-state)
non-analyticities at the small-polaron transition. Above all, 
in contrast to finite-lattice ED, it yields  a ground-state energy 
which is a variational bound for the exact energy in the thermodynamic
limit. As yet the VED technique is fully worked out for the single polaron
and bipolaron problem only. At finite particle densities the construction
of the variational Hilbert space becomes delicate. On this account
we will also outline some more general (robust) ED schemes,
which can be applied for the calculation of ground-state and 
spectral properties of a larger class of strongly correlated 
EP systems. 

\subsection{Numerical techniques}
\label{sec:na}
\subsubsection{Hilbert space and basis construction}
The total Hilbert space of the Holstein 
model~(\ref{hm}) can be written as the tensorial
product space of electrons and phonons, spanned by the
complete basis set 
$\left\{|b\rangle =|e\rangle\otimes |p\rangle\right\}$
with 
\begin{equation}
|e\rangle = \prod_{i=1}^N \prod_{\sigma = \uparrow,\downarrow} 
(c_{i\sigma}^{\dagger})^{n_{i\sigma,e}}  
|0\rangle_{e}\quad\mbox{and}\quad
|p\rangle = \prod_{i=1}^N \frac{1}{\sqrt{m_{i,p} !}} 
(b_i^{\dagger})^{m_{i,p}}|0\rangle_{p}\,.
\label{basis1}
\end{equation}
Here $n_{i\sigma,e} \in \{0,1\}$, i.e. with respect to the electrons  
Wannier site $i$ might be empty, singly or doubly occupied, whereas 
we have no such restriction for the phonon number, 
$m_{i,p}\; \in \{0,\ldots,\infty\}.$ Consequently, 
$e=1,\ldots,D_{e}$ and $p=1,\ldots,D_{p}$ label 
basic states of the electronic and phononic subspaces having 
dimensions $D_{e}={N \choose N_{e,\sigma}}{N\choose N_{e,-\sigma}}$
and $D_{p}=\infty$, respectively. $|0\rangle_{e/p}$ denote the 
corresponding vacua. 
This also holds including electron-electron (e.g. Hubbard-type) 
interaction terms~\cite{FWHWB04}. 
For Holstein-t-J-type models, 
acting in a projected Hilbert space without doubly occupied sites,
we have $D_{e}={N \choose N_{e,\sigma}}{N-N_{e,\sigma}\choose N_{e,-\sigma}}$ 
only~\cite{BWF98}. Since these model Hamiltonians  
commute with the total electron number operator
$\hat{N}_{e}=\sum_{\sigma } \hat{N}_{e,\sigma}$, where  
$\hat{N}_{e,\sigma}=\sum_{i=1}^Nn_{i,\sigma}$, 
and the $z$-component of the total spin $S^{z}=
\frac{1}{2}\sum_{i=1}^N(n_{i,\uparrow}-n_{i,\downarrow})$, 
the many-particle basis $\{|b\rangle\}$ can be constructed for    
fixed $N_{e}$ and $S^z$.
\paragraph{Variational approach}
Let us first describe an efficient variational 
exact diagonalisation (VED) method to solve 
the Holstein model in the single-particle subspace. 
For generalisation of this method to the case of two particles
(bipolaron) see Ref.~\cite{BKT00}. 

\begin{figure}
\centering
\includegraphics[width=0.48\textwidth]{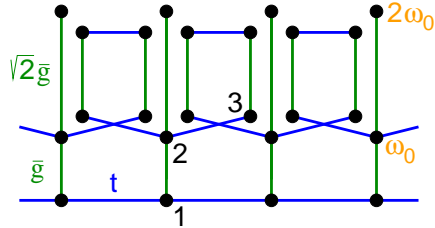}
\caption{Variational Hilbert space construction for the 1D polaron problem.
Basis states are represented by dots, off-diagonal matrix elements 
by lines. Vertical bonds create or destroy phonons with frequency
$\omega_0$. Horizontal bonds correspond to electron hops $(\propto t)$.
Accordingly, state $|1\rangle$ describes an electron at the 
origin (0) and no phonon,
state $|2\rangle$ is an electron and one phonon both at site~0,
$|3\rangle$ is an electron at the nearest-neighbour site~1,
and a phonon at site~0, and so on~\cite{BTB99}.}
\label{fig:vm}       
\end{figure}

\begin{figure}
\centering
\includegraphics[width=0.48\textwidth,angle=270]{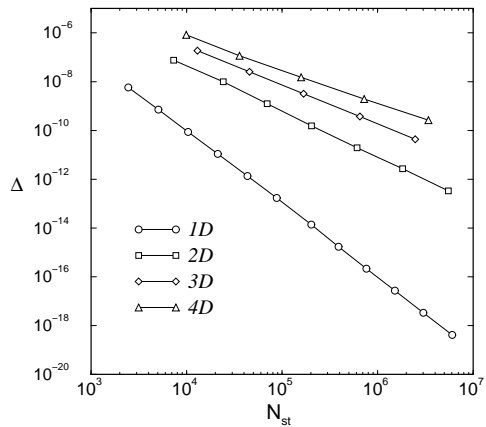}
\caption{Fractional error $\Delta$ in the ground-state energy
of a D-dimensional polaron as a function of the number of basis
states $N_{st}$ retained. Parameters are  $\lambda=0.5$, $g=1$, and $t=1$.
Figure is taken from Ref.~\cite{KTB02}.} 
\label{fig:conved}       
\end{figure}

A variational Hilbert space is constructed beginning with an initial
root state, taken to be an electron at the origin with no phonon excitations,
and acting repeatedly with the hopping ($t$) and EP coupling 
($\bar{g}$) terms of the Holstein Hamiltonian~(see Fig.~\ref{fig:vm}).
States in generation $L$ are those obtained by acting $L$ times with
these ``off-diagonal'' terms.  Only one copy of each state is retained. 
Importantly, all translations of these states on an infinite 
lattice are included. A translation moves the electron and all phonons 
$j$ sites to the right. Then, according to Bloch's theorem, 
each eigenstate can be written as $\psi=\mbox{e}^{{\rm i} kj} a_L$, 
where $a_L$ is a set of complex amplitudes related
to the states in the unit cell $j$, e.g. $L=7$ for the small variational
space shown in Fig.~\ref{fig:vm}. For each momentum $k$ the resulting 
numerical problem is then to diagonalise a Hermitian $L\times L$
matrix. The total number of states  per unit cell ($N_{st}$) 
after $L$ generations increases exponentially as $(D+1)^L$ (note that 
the bipolaron has the same exponential dependence with only a larger 
prefactor). Most notably the error in the ground-state energy $E_0$ 
decreases exponentionally, because states are added in 
a fairly efficient order. Thus in most cases 
$10^4$ -- $10^6$ basis states are sufficient to obtain an 8-16 digit
accuracy for $E_0$  (see Fig.~\ref{fig:conved}). 
The ground-state energy calculated this way is 
variational for the infinite system. 

\paragraph{Symmetrisation and phonon truncation}
Treating more complex many-particle Hamilton operators on finite lattices,
the dimension of the total Hilbert space can also be reduced.
To this end we can exploit the space group symmetries 
[translations ($G_T$) and point group operations ($G_L$)] 
and the spin rotational invariance [($G_S$); $S^z=0~$  subspace only].
Working, e.g., on finite 1D or 2D bipartite clusters 
with periodic boundary conditions (PBC),
we do not have all the symmetry properties of the underlying 1D or 
2D (square) lattices~\cite{BWF98}. Restricting ourselves to the 1D
non-equivalent irreducible representations of the group 
$G(\vec{K})=G_T\times G_L(\vec{K})\times G_S$, 
we can use the projection operator 
${\cal P}_{\vec{K},rs}=[g(\vec{K})]^{-1} 
\sum_{{\cal G} \in G(\vec{K})}\chi_{\vec{K},rs}^{(\cal G)}\;{\cal G}$
(with $[H,{\cal P}_{\vec{K},rs}]=0$, 
${\cal P}_{\vec{K},rs}^{\dagger}={\cal P}_{\vec{K},rs}$ and
${\cal P}_{\vec{K},rs}\;{\cal P}_{\vec{K}^{\prime},r^{\prime}s^{\prime}}
={\cal P}_{\vec{K},rs}\;  
\delta_{\vec{K},\vec{K}^{\prime}}\;\delta_{r,r^{\prime}}\;
\delta_{s,s^{\prime}}$) 
in order to generate a new symmetrised basis set:
$\{|b\rangle\} \stackrel{\cal P}{\to}  
\{|\tilde{b}\rangle\}$. ${\cal G}$ denotes the  $g(\vec{K})$ elements of 
the group $G(\vec{K})$ and $\chi_{\vec{K},rs}^{(\cal G)}$ 
is the (complex) character of 
${\cal G}$ in the $[\vec{K},rs]$--representation, where  
$\vec{K}$ refers to one of the $N$ allowed wave vectors in the 
first Brillouin zone,  $r$ labels the irreducible representations 
of the little group of $\vec{K}$, $G_L(\vec{K})$, and $s$
parameterises $G_S$.   
For an efficient parallel implementation of the matrix vector multiplication
(see below) it is extremely important   
that the symmetrised basis can be constructed preserving the tensor 
product structure of the Hilbert space, i.e., 
\begin{equation}
\{|\tilde{b}\rangle=
N^{[\vec{K}rs]}_{\tilde{b}}\, {\cal P}_{\vec{K},rs}\,
\left[ |\tilde{e}\rangle
\otimes |p\rangle\right]\}
\label{basis2} 
\end{equation}
with  $\tilde{e}=1,\ldots, \tilde{D}_{e}^{g(\vec{K})}$ 
$[\tilde{D}_{e}^{g(\vec{K})}\sim D_{e}/g(\vec{K})]$. The
$N^{[\vec{K}rs]}_{\tilde{b}}$ are normalisation factors. 

Since the Hilbert space associated to the phonons is infinite
even for a finite system, we use a truncation procedure~\cite{WRF96}
retaining only basis states with at most $M$ phonons:  
\begin{equation}
\{ |p\rangle\; ; \;m_p=\sum^N_{i=1} m_{i,p} \le M \}\,.
\end{equation}
Then the resulting Hilbert space has a total dimension 
$\tilde{D}=\tilde{D}_{e}^{g(\vec{K})}\times D_{p}^M$ with 
$D_{p}^M =(M+N)!/M!N!$, and a general state 
of the Holstein model is represented as   
\begin{equation}
|\psi_{\vec{K},rs}\rangle = 
\sum_{\tilde{e}=1}^{\tilde{D}_{e}^{g(\vec{K})}} 
\sum_{p=1}^{D^M_{p}} c_{\tilde{e}p}^\psi\,
|\tilde{b}\rangle\,. 
\end{equation}
The computational requirements can be further reduced
if one separates the symmetric phonon mode, 
$B_{0}=\frac{1}{\sqrt{N}}\sum_{i} b_i$, and 
calculates its contribution to $H$ analytically~\cite{SHBWF05}. 

Note that switching from a real space representation
to a momentum space description the truncation scheme takes 
into account all dynamical phonon modes, which has to be 
contrasted with the frequently used single-mode 
approach~\cite{AP98}. 
In other words, depending on the model parameters and the band filling, 
the system ``decides'' by itself how the $M$ phonons will be 
distributed among the independent Einstein oscillators related 
to the $N$ Wannier sites or, alternatively, among the different 
phonon modes in $\vec{Q}$-space. Hence with the 
same accuracy phonon dynamical effects on lattice distortions 
being quasi-localised in real space 
(such as polarons, Frenkel excitons,\ldots) or in momentum space 
(like charge-density-waves,\ldots) can be studied. 

Of course, one has carefully to check for the convergence of
the above truncation procedure by calculating the ground-state
energy as a function of the cut-off parameter $M$. 
In the numerical work below convergence is assumed to be achieved if 
$E_0$ is determined with a relative error less than $10^{-6}$\,.

\paragraph{Phonon basis optimisation}
In this section we outline an advanced phonon optimisation procedure
based on controlled density-matrix basis truncation~\cite{WFWB00}. 
The method provides a natural way to dress the particles with phonons
which allows the use of a very small optimal basis without significant 
loss of accuracy.

Starting with an arbitrary normalised quantum state, 
\begin{equation}
  |\psi\rangle = \sum_{e=0}^{D_e-1} \sum_{p=0}^{D_p-1} 
c^\psi_{ep} [|e\rangle\otimes |p\rangle]\,,
\end{equation}
expressed in terms of the basis of
the direct product space, we wish to reduce the dimension $D_p$ 
of the phonon space $H_p$ by introducing a new basis, 
\begin{equation}
  |\tilde p\rangle = \sum_{p=0}^{D_p-1} 
  \alpha_{\tilde{p} p}|p\rangle\,,
\end{equation}
with $\tilde{p}=0,\ldots ,(D_{\tilde{p}}-1)$ and $D_{\tilde{p}}<D_p$.
We call $\{|\tilde{p}\rangle\}$ an optimised basis, if the projection 
of $|\psi\rangle$ on the corresponding subspace 
$\tilde{H} = H_{e}\otimes H_{\tilde{p}} \subset H$ is as 
close as possible to the original state. Therefore we minimise
\begin{equation}
 \| |\psi\rangle - |\tilde\psi\rangle \|^2 = 1 - {\rm Tr}\;({\bf \alpha \rho \alpha^\dagger})  
\end{equation}
with respect to the 
$\alpha_{\tilde{p} p}$ under the condition 
$\langle \tilde{p}'|\tilde{p}\rangle = 
\delta_{\tilde{p}' \tilde{p}}$, where  
\begin{equation}
  |\tilde\psi\rangle = \sum_{e=0}^{D_e-1} 
  \sum_{\tilde{p}=0}^{D_{\tilde{p}}-1} \sum_{p,p'=0}^{D_p-1} 
  \alpha_{\tilde{p}p} \alpha_{\tilde{p} p'}^{*} c^\psi_{p'e} 
  [|e\rangle\otimes|p\rangle]
\end{equation}
is the projected state. 
${\bf \rho} = \sum_{e=0}^{D_e-1} (c^\psi_{ep})^{*}c^{\psi}_{ep}$ is
called the density matrix of the state $|\psi\rangle$. 
Clearly the states $\{|\tilde{p}\rangle\}$ are optimal if they are
elements of the eigenspace of ${\bf \rho}$ corresponding to its
$D_{\tilde{p}}$ largest eigenvalues $w_{\tilde{p}}$.
If we interpret  $w_{\tilde{p}}\sim \exp (-a\tilde{p})$ as the 
probability of the system to 
occupy the corresponding optimised state $|\tilde{p}\rangle$, 
we immediately find that the probability for the complete phonon basis state 
$\otimes_{i = 0}^{N-1} |\tilde{p}_i\rangle$ is proportional to
$\exp (-a\sum_{i = 0}^{N-1} \tilde{p}_i)$. This is reminiscent of an 
energy cut-off, and we therefore propose the following
choice of a mixed phonon basis $\{|\mu_i\rangle\}$ 
at each site,
\begin{eqnarray}
  \forall\ i:\ \{|\mu_i\rangle\} & = & \textrm{ON}(\{|\mu\rangle\})\\ 
    |\mu\rangle & = & \left\{
    {\hspace*{-0.3cm}\textrm{optimal state }\;\;\;|\tilde{p}\rangle,\ 0\le \mu < M_{\rm opt}\atop
      \textrm{bare state }\qquad |p \rangle,\ M_{\rm opt}\le \mu < M}
  \right.\,,
\end{eqnarray}
and for the complete phonon basis 
$\left\{\otimes_{\Sigma_i \mu_i < M} |\mu_i\rangle\right\}$,
yielding $D_{\rm ph} = {N+M-1\choose N}$.

After a first initialisation the optimised states 
are improved iteratively through the following steps:
(1) calculate the requested eigenstate $|\psi\rangle$ of the 
  Hamiltonian $H$ in terms of the actual basis,
(2) replace $\{|\tilde{p}\rangle\}$ with the most important (i.e.
largest eigenvalues $w_{\tilde{p}}$) eigenstates of the density matrix 
$\rho$; (3) change the additional states $\{|p\rangle\}$ in the set 
$\{|\mu\rangle\}$; (4) orthonormalise 
the set $\{|\mu\rangle\}$, and return to step (1).

\begin{figure}[t]
\centering
\includegraphics[width=0.48\textwidth]{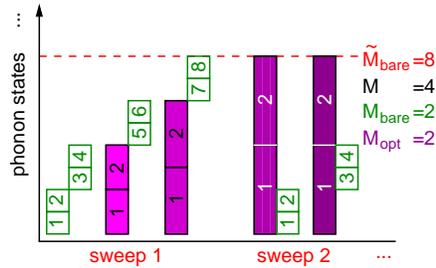}
\caption{Sweep technique in constructing optimised phonon states.}
\label{fig:sweep} 
\end{figure}

A simple way to proceed in step (3) is to sweep the bare states 
$\{|p\rangle\}$ through a sufficiently large part of the infinite 
dimensional phonon Hilbert space. One can think of the algorithm as 
``feeding'' the optimised states with bare phonons, thus allowing the 
optimised states to become increasingly perfect linear combinations of 
bare phonon states (see Fig.~\ref{fig:sweep}).

\subsubsection{Solution of the eigenvalue problem}
In all the above cases the numerical problem that remains is to find 
the eigenstates of a (sparse) Hermitian matrix.
Here iterative (Krylov) subspace methods like Lanczos~\cite{CW85} 
and variants of Davidson~\cite{Da75} diagonalisation techniques 
are frequently applied. These algorithms 
contain basically three steps:
(1) project the problem matrix ${\rm A}$ $ \in {\rm R}^n$ onto a 
subspace $\bar{\rm A}^k \in {\rm V}^k$ $(k\ll n)$;
(2) solve the eigenvalue problem in ${\rm V}^k$ 
using standard routines; 
(3) extend the subspace ${\rm V}^k \to {\rm V}^{k+1}$ by 
a vector $\vec{t}\perp {\rm V}^k $ and go back to (2).
This way we obtain a sequence of approximate inverses of the
original matrix ${\rm A}$. A  powerful and widely used technique
is the Lanczos algorithm which recursively generates a set of orthogonal
states (Lanczos vectors):
\begin{equation}
|\varphi_{l+1}\rangle=
{\rm H}^{\tilde{D}}|\varphi_{l}\rangle
-a_l|\varphi_{l}\rangle
-b_l^2|\varphi_{l-1}\rangle\,,
\label{lr1}
\end{equation} 
where
\begin{equation}
a_l=\frac{\langle\varphi_{l}|{\rm H}^{\tilde{D}}
|\varphi_{l}\rangle}{\langle\varphi_{l}|
\varphi_{l}\rangle}\,,\quad b_l^2=\frac{\langle\varphi_{l}|\varphi_{l}\rangle}{\langle\varphi_{l-1}|
\varphi_{l-1}\rangle}\,,\quad b_0^2=0\,,
\label{lr2}
\end{equation} 
and $|\varphi_{-1}\rangle=0$. Obviously, the 
representation matrix  $[{\rm T}^L]_{l,l^{\prime}}=
\langle \varphi_{l}|{\rm H}^{\tilde{D}}|
\varphi_{l^{\prime}}\rangle$ of ${\rm H}^{\tilde{D}}$ is tridiagonal 
in the $L$-dimensional Hilbert space spanned by the
$\{|\varphi_{l}\rangle\}_{l=0,\ldots,L-1}$, where $L\ll\tilde{D}$. 
The eigenvalue spectrum of ${\rm T}^L$ can be easily
determined using standard routines from libraries 
such as EISPACK (see http://www.netlib.org). Note that the convergence 
of the Lanczos algorithm  is excellent at the
edges of the spectrum (the ground state for example is obtained with
high precision after at most $\sim 100$ Lanczos iterations) 
but rapidly worsens inside the spectrum. 

In general the computational
requirements of these eigenvalue algorithms are determined by
matrix-vector multiplications (MVM), which have to be implemented in 
a parallel, fast and memory saving way on modern supercomputers. 
The MVM step can be be done in parallel by using a 
parallel library such as PETSc
(see http://www-unix.mcs.anl.gov/petsc/petsc-as/).

Our matrices are extremely sparse 
because the number of non-zero entries per row of our Hamilton matrix 
scales linearly with the number of electrons.
Therefore a standard implementation of the MVM step uses a sparse storage
format for the matrix, holding the non-zero elements only. 
The typical storage requirement per non-zero entry
is 12-16 Byte, i.e. for a matrix dimension of
$\tilde{D}=10^9$ about one TByte main memory is required 
to store only the matrix elements of the EP Hamiltonian.
To extend our EP studies to even larger matrix sizes we 
no longer store the non-zero matrix elements but generate them
in each MVM step. Of course, at that point standard 
libraries are no longer useful and a parallel code 
tailored to each  specific class of Hamiltonians must be developed. 
Clearly the parallelisation approach follows the 
inherent natural parallelism of the Hilbert space.
Assuming that the electronic dimension ($\tilde{D}_{e}$) is a multiple 
of the number of processors used we can easily 
distribute the electronic basis states among the processors.
As a consequence of this choice only the electronic hopping term 
generates inter-processor communication in the MVM
while all other (diagonal electronic) contributions can be 
computed locally on each processor.
Using  supercomputers with hundreds of processors and one TByte 
of main memory, such as the IBM p690 cluster, at the moment, one is 
able to run simulations up to a matrix dimension of 
$3 \times 10^{10}$.     
\subsubsection{Determination of dynamical correlation functions}
The numerical calculation of dynamical spectral functions, 
\begin{eqnarray}
A^{\cal O}(\omega)&=&-\lim_{\eta\to 0^+}\frac{1}{\pi} \mbox{Im} \left[
\langle\psi_{0}
|{\rm O}^{\dagger}\frac{1}{\omega - {\rm H} +E_0 +i\eta}{\rm O}^{} 
|\psi_{0}\rangle\right]\nonumber\\&=&
\sum_{n=0}^{\tilde{D}-1}|\langle\psi_{n}|{\rm O}|
\psi_{0}\rangle |^{2}\delta [\omega - (E_{n} - E_{0})]\,,
\label{specfu}
\end{eqnarray}
where ${\rm O}$ is the matrix representation of a certain 
operator ${\cal O}$ [e.g., the creation operator
$c_{\vec{K}}^\dagger$ of an electron with wavevector $\vec{K}$ if one wants 
to calculate the single-particle spectral function; or the current operator
$\hat{\jmath} = - \mbox{i} e t\sum_{i}(c_{i}^{\dagger} 
  c_{i+1}^{} - c_{i+1}^{\dagger} 
  c_{i}^{})$ if one is interested in the optical conductivity],  
involves the resolvent of the Hamilton 
matrix ${\rm H}$. Once we have obtained the 
eigenvalues and eigenvectors of $H$ we can plug them into
Eq.~(\ref{specfu}) and directly obtain the corresponding  
dynamical correlation or Green functions.
For the typical EP problems under investigation    
we deal with Hilbert spaces having total dimensions $\tilde{D}$ 
of $10^6$ - $10^{11}$. Finding all eigenvectors and eigenstates 
of such  huge Hamilton matrices is impossible, 
because the CPU time required for
exact diagonalisation of ${\rm H}$ scales as 
$\tilde{D}^3$ and memory as $\tilde{D}^2$.  
So in practice this ``naive'' approach is applicable  for small
Hilbert spaces only, where the complete 
diagonalisation of the Hamilton matrix is feasible.
Fortunately, there exist very accurate and well-conditioned 
linear scaling algorithms for a direct approximate calculation of 
$A^{\cal O}(\omega)$. 
\paragraph{Kernel polynomial method (KPM)}
The idea of the KPM (for a review see~\cite{WWAF06}), is 
to expand  $A^{\cal O}(\omega)$ in a finite series 
of $L+1$ Chebyshev polynomials $T_m(x)= \cos [m
\arccos(x)]$. Since the Chebyshev polynomials are defined 
on the real interval $[-1,1]$, first a simple
linear transformation to the Hamiltonian and all energy scales
has to be applied:
${\rm X}=({\rm H}-b)/a$, $x=(\omega -b)/a$, 
$a=(E_{max}-E_{min})/2(1-\epsilon)$, and $b=(E_{max}+E_{min})/2$  
(the small constant $\epsilon$ is introduced in order to avoid
convergence problems at the endpoints of the interval 
 -- a typical choice is $\epsilon \sim 0.01$ which has only
1\% impact on the energy resolution~\cite{SR97}). 
Then the expansion is
\begin{equation}
A^{\cal O}(x)=\frac{1}{\pi
\sqrt{1-x^{2}}}\left(\mu_{0}^{\cal O}+
2\sum_{m=1}^{L}\mu_{m}^{\cal O}T_{m}(x)\right)\,,
\label{akpm}
\end{equation}
with moments  
\begin{equation}
\mu_m^{\cal O}=\int_{-1}^{1}dx\,T_{m}(x)A^{\cal O}(x)=\langle
\psi_{0}| {\rm O}^{\dagger}T_{m}({\rm X}){\rm O}^{}|\psi_{0}\rangle\,.
\label{mkpm}
\end{equation}
Eq.~(\ref{akpm}) converges to the correct function for $L\to\infty$.
The moments 
\begin{equation}
\mu_{2m}^{\cal O}=2\langle\phi_m|\phi_m\rangle -\mu_0^{\cal O}\quad\mbox{and}\quad
\mu_{2m+1}^{\cal O}=2\langle\phi_{m+1}|\phi_m\rangle -\mu_1^{\cal O}
\label{moments2}
\end{equation}
can be efficiently obtained by repeated parallelised MVM, 
where $|\phi_{m+1}\rangle
=2{\rm X}|\phi_m\rangle -|\phi_{m-1}\rangle$
but now $|\phi_1\rangle={\rm X}|{\rm O}\phi_0\rangle$ 
with $|\phi_{0}\rangle\equiv|\psi_0\rangle$ 
determined by Lanczos ED.

As is well known from Fourier expansion,  
the series~(\ref{akpm}) with $L$ finite 
suffers from rapid oscillations (Gibbs phenomenon)
leading to a poor approximation to $A^{\cal O}(\omega)$.
To improve the approximation 
the moments $\mu_n$ are modified $\mu_n \to g_n \mu_n$, 
where the damping factors $g_n$ 
are chosen to give the ``best'' approximation for a given $L$.
In more abstract terms this truncation of the infinite series to order $L$
together with the corresponding modification 
of the coefficients is equivalent to a convolution of the 
spectral function with a smooth approximation kernel 
$K_L(x,y)$. 
The appropriate (optimal) choice of this kernel, that is of $g_n$, 
e.g. to guarantee positivity of $A^{\cal O}(\omega)$, 
lies at the heart of KPM. 
We mainly use the Jackson kernel 
which results in a uniform approximation whose resolution 
increases as $1/L$, but for the determination of the 
single-particle Green functions below we use a Lorentz kernel 
which mimics a finite imaginary part $\eta$ 
in Eq.~(\ref{specfu})~\cite{WWAF06}.

In view of the uniform convergence of the expansion,  
the accuracy of the spectral functions can be made 
as good as required by just increasing $L$.

\paragraph{Cluster perturbation theory (CPT)}
The spectrum of a finite system of $N$ sites which we obtain through
KPM differs in many respects from that in the thermodynamic limit
$N\to\infty$, especially in that it is obtained for a 
finite number of momenta $K=\pi\, m/N$ only.
While we cannot easily increase $N$ without going beyond
computationally accessible Hilbert spaces
we can try to extrapolate from a finite to the infinite system. 

For this purpose we first calculate the Green function $G^c_{ij}(\omega)$
for all sites $i,j=1,\dots,N_c$ of a $N_c$ -- size cluster with open
boundary conditions, and then recover the infinite lattice
by pasting identical copies of this cluster at their edges~\cite{SPP00}.
The ``glue'' is the hopping $V$
 between these clusters, where  $V_{kl}=t$ for $|k-l|=1$ and
$k,l \equiv 0,1 ({\rm mod} N)$, 
which is dealt with in first order perturbation theory.
Then the Green function $G_{ij}(\omega)$ of the infinite lattice
is given through a Dyson equation 
$G_{ij}(\omega) = G^c_{ij}(\omega) + \sum_{kl} G^c_{ik}(\omega) V_{kl} 
  G_{lj}(\omega)$, where indices of $G^c(\omega)$ are counted modulo $N_c$. 
The Dyson equation is solved by Fourier transformation over momenta
$K= k N_c$ corresponding to translations by $N_c$ sites
\begin{equation}
  G_{ij}(K,\omega) 
= \left[ \frac{G^c(\omega)}{1-V(K)G^c(\omega)} \right]_{ij}\,,
\end{equation}
from which one finally obtains
\begin{equation}
G(k,\omega) = \frac{1}{N_c} \sum_{i,j=1}^{N_c} G^c_{ij}(N_c k,\omega) 
\exp(- \mathrm{i} k (i-j) )\;.
\end{equation}

In this way we obtain a Green function $G(k,\omega)$ with continuous
momenta $k$ from the cluster Green function $G^c$. 
Two approximations are made, 
one by using first order perturbation theory in $V=t$,
the second on assuming translational symmetry in $G_{ij}(\omega)$
which is only approximatively satisfied.

\section{Ground state results}  
\label{sec:gs}
The VED method can compute polaron properties in 
dimensions 1 through 4 and higher.
The energies for 1D to 4D polarons at $\vec k = 0$ 
for intermediate to weak coupling
on linear, square, cubic, and hypercubic lattices are listed in the 
Tab.~\ref{t:e0}.  
The number of significant digits is
determined by comparing the energy as the size of the Hilbert 
space is increased.
The accuracy is high compared to that of other numerical methods, 
even when limited by the
single-processor desktop workstations of several years ago, 
on which the results were obtained~\cite{KTB02}.
Correlation functions are generally less accurate than energies. 
\begin{table}[t]
\centering
\caption {Polaron ground state energies at
$\vec k=0$ in 1D - 4D for $\lambda=0.5$ and $g=1.0$
In the following $t=1$ unless specifically noted.} 
\begin{tabular}{ccccc} 
\hline\noalign{\smallskip}
   &  1D & 2D   & 3D  & 4D \\
\noalign{\smallskip}\hline\noalign{\smallskip}  
\quad $E_0/t\quad$  & -2.46968472393287071561$\;$  & -4.814735778337$\;$  
&  -7.1623948409$\;$ &
 -9.513174069$\;$\\ 
\noalign{\smallskip}\hline
\label{t:e0}
\end{tabular}
\end{table}
\begin{figure}[h]
\centering
\includegraphics[width=0.6\textwidth, angle=-90]{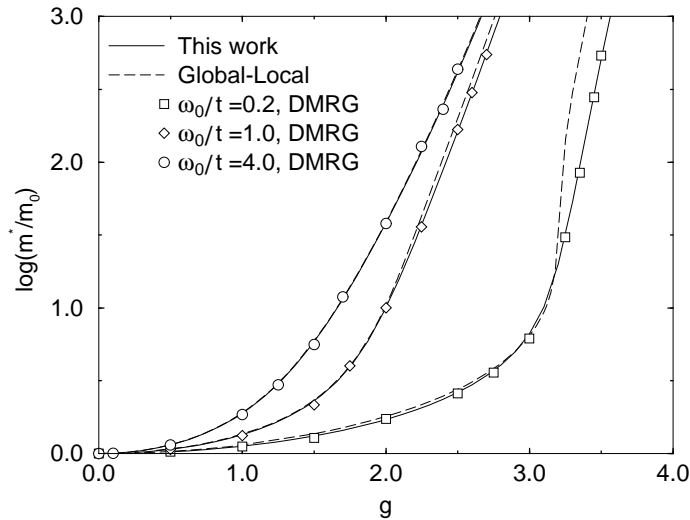}
\caption{Logarithm of the polaron effective mass in 1D as a function
of $g$.  VED results (full lines) were obtained operating  
repeatedly $L=20~$ times with the off-diagonal pieces 
of the Holstein Hamiltonian (cf. Sec.~\ref{sec:na}). For comparison 
GL (global-local) results (dashed lines) are included~\cite{RBL98}.
Open symbols, indicating the value of $\omega _0/t$,
are DMRG results~\cite{JW98b}.}
\label{fig:mass0}
\end{figure}

Figure~\ref{fig:mass0} shows the effective mass
computed by VED~\cite{BTB99} in comparison with GL and DMRG methods. 
$m^*$ is obtained from 
\begin{equation}
  {m_0 \over m^*} = \left. {1 \over {2t}} 
  {\partial ^ 2 E_k \over {\partial k ^ 2}}\right|_{k=0},
\label{eq:mass}
\end{equation}
where $m_0 = 1/(2t)$ is the effective mass of a free electron. 
The second derivative is evaluated by small finite differences in the
neighbourhood of ${\vec k} = 0 $.  Note that although the calculated
energy $E_{\vec k}$ is a variational bound for the exact energy, there is no
such control on the mass, which may be either above or below the exact
answer, and is expected to be more difficult to obtain accurately.
Nevertheless, in the intermediate coupling regime where VED at
$L=20$ gives an energy accuracy of 12 decimal places, one can
calculate the effective mass accurately (6-8 decimal places)
by letting $\Delta k \to 0$.

In Fig.~\ref{fig:mass0} the parameters span 
different physical regimes including
weak and strong coupling, and low  and high 
phonon frequency.  We find good agreement between VED and GL away from
strong coupling and excellent agreement in all regimes 
with DMRG results.  DMRG
calculations are not based on finite-$k$ calculations due to a
lack of periodic boundary conditions, so they
extrapolate the effective mass from the ground state data using chains
of different sizes, which leads to larger error bars and demands more
computational effort. Notice that their discrete data is slightly
scattered around the VED curves.  Nevertheless, both methods agree well. 
We have compared the VED results for effective mass 
obtained on different systems
from $L=16$ with $N_{st}= 178617$ states to $L=20$ with $N_{st}=2975104$
states and obtained convergence of results to at least 4 decimal
places in all parameter regimes presented in Fig.~\ref{fig:mass0}. 
Our error is therefore well below the linewidth. 
Even though there is no phase transition in the ground state of the model,
the polaron becomes extremely heavy in the strong coupling
regime.  The crossover to a regime of large polaron mass is more rapid in
adiabatic regime, i.e. at small $\omega_0/t$.

\begin{figure}
\vspace*{.5cm}
\centering
\includegraphics[width=0.65\textwidth]{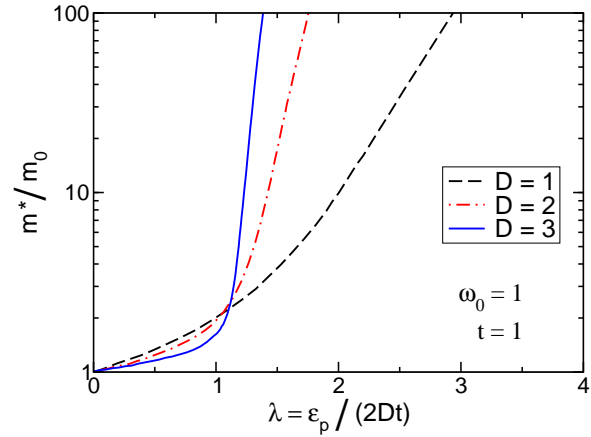}
\caption{Effective mass of the Holstein 
polaron in dimensions 1, 2, and 3.}
\label{fig:mass_D123}
\end{figure}

The polaron effective mass in higher dimensions is shown 
as a function of the EP coupling in Fig.~\ref{fig:mass_D123}.
The mass increases exponentially for large $\lambda$.   
The crossover to larger effective mass is more rapid, 
though still continuous, in higher dimensions.

\begin{figure}[b]
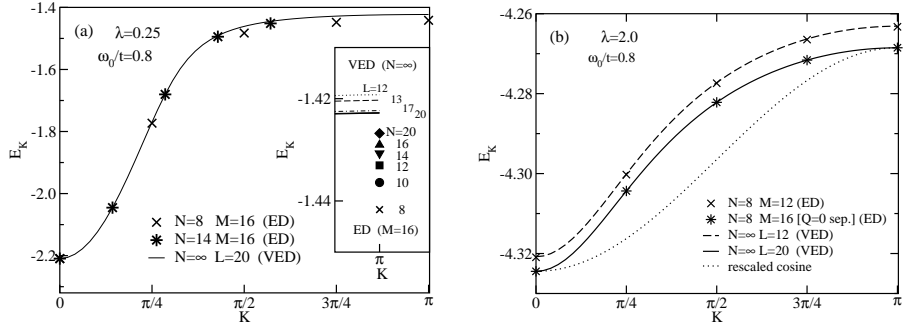

\vspace*{0.5cm}
\centering
\includegraphics[width=0.48\textwidth]{bandis_wc_vl.eps}\hspace*{.5cm}
\includegraphics[width=0.48\textwidth]{bandis_sc_vl.eps}\\[0.2cm]
\caption{Band structure of the 1D Holstein model in the weak (a) and
strong (b) EP coupling regimes. $M$ and $L$ denote the number of phonons
and the depth of the basis in ED and VED calculations, respectively. 
Within ED  the ($Q=0$) centre of mass motion has been separated if indicated. 
The inset shows the finite-size dependence being significant 
for weak EP coupling only. In the strong EP coupling case ED results 
basically agree with those obtained by VED.} 
\label{fig:bandis}       
\end{figure}

Of course it is of special interest to understand  
the evolution of the polaronic band structure 
as the EP coupling strength increases.  
Figure~\ref{fig:bandis} presents the results 
for the wavevector dependence of the ground-state energy  
$E_{k}$ in 1D at weak (a) and strong (b) EP couplings. 
As might be expected, for $\lambda=0.25$ the coherent bandwidth,  
$\Delta E= \sup_{k} E_{k}-\inf_{k} E_{k}$,
is approximately given by the phonon energy 
($\Delta E=0.782\sim \omega_0 =0.8$).  
If the EP interaction is enhanced a band collapse appears. 
Note, however, that even in the relatively strong EP coupling regime 
displayed Fig.~\ref{fig:bandis}~(b) the standard Lang-Firsov 
formula, $ \Delta E_{LF}= 4Dt\exp[-g^2]$ 
(obtained by performing the 
Lang-Firsov canonical transformation~\cite{LF62} and taking 
the expectation value of the kinetic energy over the transformed 
phonon vacuum), does not give a satisfactory estimate
of the bandwidth. So we found $\Delta E_{LF}=0.0269$ which has to be 
contrasted  with the exact result $\Delta E=0.0579$. 
Besides the band narrowing effect, there are several 
other features worth mentioning for polaronic band states 
in the crossover region. Most notably, throughout the whole Brillouin zone
the band structure differs significantly from that of a 
rescaled tight-binding (cosine) band containing only nearest-neighbour 
hopping~\cite{WF97}. Obviously the residual 
polaron-phonon interaction generates longer-ranged 
hopping processes~\cite{Fi75,WF97}. Concomitantly, the 
mass enhancement due to the EP interaction is weakened at the band minimum. 
It is important to realize that these effects are most pronounced 
at intermediate EP couplings and phonon frequencies. In this parameter 
region even  higher-order strong-coupling perturbation theory, with its
internal states containing some excited phonons,  seems to be 
intractable because the convergence of the series 
expansion is  poor~\cite{St96}. 
Of course the dispersion is barely changed from a rescaled tight-binding 
band in the very extreme small polaron limit ($\lambda,\;g^2 \gg 1$).

Further information about the quasiparticle may be obtained by
computing the quasiparticle residue, the overlap 
squared between a bare electron and a polaron,
\begin{equation}
Z_{\vec k}^{} = \vert\langle\psi_{\vec k}^{}\vert c_{\vec k}^\dagger
\vert0\rangle\vert^2,
\label{zk_fac}
\end{equation}
where $\vert0\rangle$ is the state with no electron and
no phonon excitations,
and $\vert \psi_{\vec k}\rangle$ is the polaron wavefunction
at momentum $\vec k$. 
$Z_{\vec k}$ can be measured by angle-resolved photoemission,
and gives the bare electronic contribution
of the polaronic state. The phonon contribution 
to the quasiparticle is described  by the 
$k-$dependent mean phonon number
\begin{equation}
N^{ph}_{\vec k}=\sum_i  \vert\langle\psi_{\vec k}^{}\vert b_i^\dagger b_i^{}\vert\psi_{\vec k}^{}
\rangle\vert^2.
\label{Nk}
\end{equation}
\begin{figure}[t]
\vspace*{0.5cm}
\centering
\includegraphics[width=0.6\textwidth]{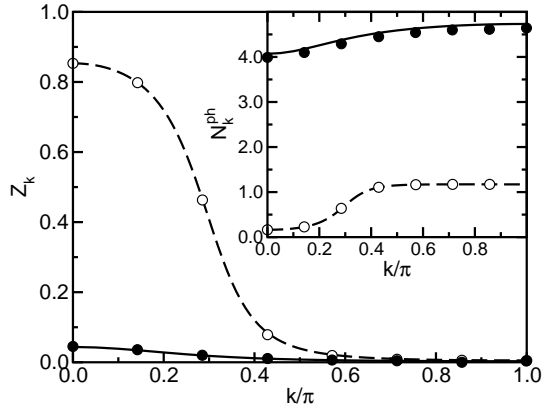}\\[0.2cm]
\caption {The quasiparticle weight $Z_{\vec k}$ and in the inset 
the total number of phonons $N^{ph}_{\vec k}$ as a function of the wavevector 
${\vec k}$ for the 1D Holstein model with $(\bar{g} / t )^2 =0.4$ [dashed line (VED);
open circles (ED)], $3.2$ [solid line (VED);
filled circles (ED)]. 
$\alpha =\omega_0/t= 0.8.$
Data are taken from Refs.~\cite{BTB99,WF97}.}
\label{fig:zk}
\end{figure}

Figure~\ref{fig:zk} shows the spectral weight $Z_{\vec k}$ 
and the mean phonon number $N_{\vec k}^{ph}$ as a function of $\vec k$.
The two sets of parameters  correspond to the large and small polaron 
regime respectively.
The DMRG cannot straightforwardly compute this quantity.
There is excellent agreement between the VED and ED methods 
in the weak coupling case.
In the strong coupling regime there
is an approximately $1\%$ disagreement in $N^{ph}_{\vec k}$ due to a lack of 
phonon degrees of freedom in the variational space of the
ED calculation. The results in the weak coupling case show a smooth crossover
from predominantly electronic character of the wavefunction 
for small $k$ (large $Z_{\vec k}$ and small $ N^{ph}_{\vec k} \approx 0$)
to predominantly phonon character around ${\vec k} =\pi$ characterised by
$Z_{\vec k}\approx 0$ and $N^{ph}_{\vec k} \approx 1$. 
In the strong coupling regime there is less $\vec{k}$-dependence. 
The $Z_{\vec k}$ is already close to zero at small $k$, indicating strong
EP interactions that lead to a large polaron mass.  
Concomitantly an appropriately defined polaron quasiparticle
residue $\tilde{Z}_{0}$ tends to one~\cite{FLW97,LHF06}.   
So we arrive at the conclusion that a well-defined electronic 
(polaronic) quasiparticle exists for $k=0$ at weak (strong) EP coupling.

\begin{figure}
\centering
\includegraphics[width=0.6\textwidth, angle=-90]{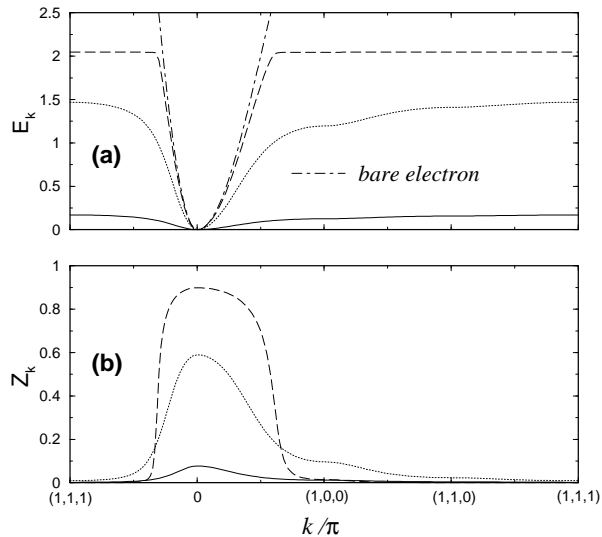}
\caption{Ground-state energy $E_{\vec k}$ (shifted by $E_0$) 
of the 3D polaron in panel (a) and
quasiparticle weight $Z_{\vec k}$ in panel (b) 
for three different EP coupling constants, 
$\bar g =4.5$ (solid line), $\bar g =3.5$ (dotted line),
and $\bar g =2.0$ (dashed line), for $\omega_0/t= 2$.   
The dot-dashed line in (a) is the dispersion of a bare electron. 
The corresponding ground state energies $E_0/t$ are 
$-10.608348,\; -8.0642850$, and $-6.588526818$ respectively. }
\label{fig:band}
\end{figure}

VED is one of the few methods
that can also calculate the polaron band dispersion in 3D systems  
(QMC is another, but is limited to the condition 
that the polaron bandwidth is much smaller than
the phonon frequency, which corresponds to the strong-coupling regime.)
Figure~\ref{fig:band}(a) shows the band dispersion for the 
3D polaron along symmetry directions in the Brillouin zone 
at various EP coupling constants $\bar g=g\omega_0$.  
Starting with weak coupling
$\bar g = 2.0$ (dashed line), the polaron band is close to the bare
electron band at the lower band edge. The deviation between 
them increases as ${\vec k}$ increases.  When $E_{\vec k}-E_0$ 
approaches $\omega_0$, we observe a band flattening
effect, similar to the 1D and 2D cases, accompanied by a sharp drop of
quasiparticle weight $Z_{\vec k}$ (see Fig.~\ref{fig:band}(b)). 
The large $\vec k$ lowest energy state can be
considered roughly as ``a $\vec k=0$ polaron ground state'' plus ``an itinerant
(or weakly-bound) phonon with momentum ${\vec k}$''. It is the 
phonon that carries the momentum so as to make $Z_{\vec k}$ 
essentially vanish and give an approximate bandwidth
$\omega _0$. Due to the large spatial extent of the EP correlations in
the flattened band, our results are less accurate in this 
regime. In the case of intermediate coupling $\bar g=3.5$ ($\lambda=1.0208$), 
the polaron bandwidth is narrower than the phonon frequency.  The upper
part of the band has much less dispersion than the lower part but with
a substantial $Z_{\vec k}$. This indicates a distinct mechanism for the
crossover as a function of ${\vec k}$.  In the case of $\bar g =4.5$, 
the strong EP interaction leads to a significant suppression 
of $Z_{\vec k}$ at all ${\vec k}$.
$Z_{\vec k =0}$ approaches the strong-coupling result 
$Z=\exp(-g^2)$ for $\lambda\,,g^2 \gg
1$. Note that the inverse effective mass $m^*/m_0$ and $Z$ differ if the
self-energy is strongly ${\vec k}$-dependent. This discrepancy has its
maximum in the intermediate-coupling regime for 1D systems, but vanishes in
the limit $\lambda\to \infty$.
In the Holstein model with on-site electron-phonon interactions, $Z$
and the inverse effective mass are closely related.  However,
the two can be made arbitrarily different by 
increasing the range of the EP interaction~\cite{KTB02}.

\begin{figure}
\centering
\vspace*{0.5cm}
\includegraphics[width=0.6\textwidth]{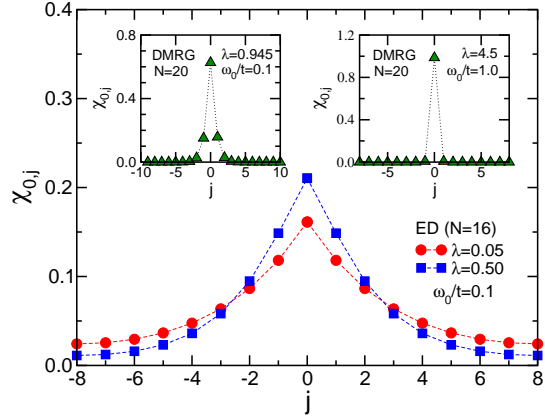}
\caption{Renormalised EP correlation function $\chi_{0,j}= \langle
  n_0(b^\dagger_{0+j} + b^{}_{0+j})\rangle/2g\langle n_0\rangle $
 as a function of the electron-phonon separation~$j$
in the $\vec k=0$ ground state of the 1D Holstein model.
Results are taken from~\cite{WF98a,FAHW06}.}
\label{fig:chi}
\end{figure}

The correlation function
between the electron position and the phonon displacement is
\begin{equation}
\chi_{i,j}^{}  
= \langle\psi_{\vec k}^{}\vert c_i^\dagger c_i^{} (b_j^{} + b^\dagger_j)
\vert\psi_{\vec k}^{}\rangle.
\label{chi_eq}
\end{equation}
This correlation function can be considered as a measure of the polaron
size. It should not be confused with the ``polaron radius'' in the
extreme adiabatic limit, which refers to the spatial extent of a hypothetical
symmetry-breaking localised state.
The ground-state EP correlation function is plotted for the 
1D Holstein polaron in Fig.~\ref{fig:chi}.
For parameters close to the (adiabatic) weak EP coupling regime (main panel), 
the amplitude of $\chi_{0,j}$ is small and the spatial extent 
of the electron-induced lattice deformation
is spread over the whole (finite) lattice.
The DMRG results shown in the left inset indicate a substantial 
reduction of the polaron's size near the crossover region. Finally
a small polaron is formed at large couplings (right inset); now 
the position of the electron and the phonon displacement is strongly
correlated. 

\begin{figure}
\vspace*{0.5cm}
\centering
\includegraphics[width=0.8\textwidth]{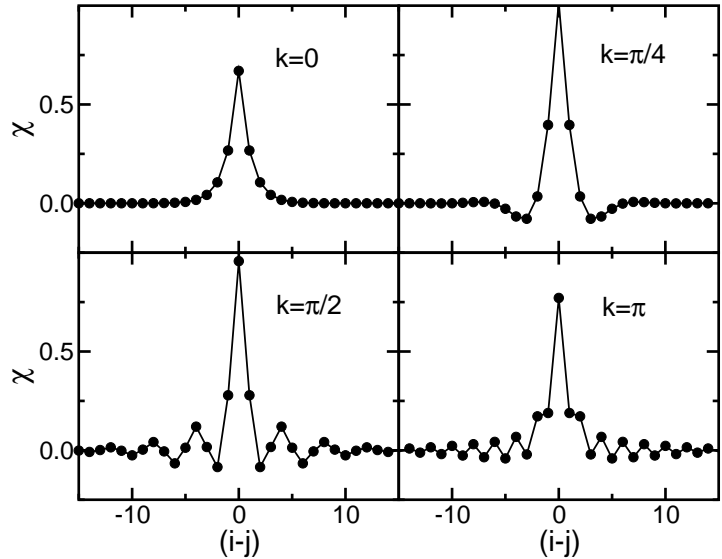}
\caption{Lattice deformation $\chi$ for 1D as a function  of $(i-j)$ for
$\omega_0 =0.8$, $t=1$, $\bar g ^2=0.4$ and $L=18$, 
for four different values of
momentum $k$.  The variational Hilbert space for
$L=18$ allows nonzero correlations up to a maximum distance
$\vert i-j\vert_{max}=17$.  Only distances up to 15 are plotted.
} 
\label{fig:chi_k} 
\end{figure}

How does the electron-phonon correlation function change
as the polaron acquires a nonzero momentum $\vec k$?
The answer is shown for 1D in Fig.~\ref{fig:chi_k}~\cite{BTB99}.
The parameters correspond to weak coupling.
At $k=0$, where the group velocity is zero,
the deformation is limited to only a few
lattice sites around the electron.  The correlation is always positive and
exponentially decaying.  At finite but small $k=\pi/4$, the
deformation around the electron increases in amplitude 
and rings (oscillates in sign) as the polaron acquires
a finite group velocity.
At $k=\pi/2$ the ringing is strongly
enhanced.  Note also that the spatial extent of the deformation
increases in comparison to $k=0$. The range of the deformation is
maximum at $k=\pi$, where it extends over the entire region
shown in the figure.  In keeping with the larger extent of the lattice
deformation near $k=\pi$, the ground-state energy $E_\pi$ converges
more slowly with the size of the Hilbert space. 
We have also computed the $k-$dependent $\chi$ for 
the strong-coupling case $\omega_0=0.8,
\bar g ^2=3.2$ (not shown).  We find only  weak  $k-$dependence, which is a
consequence of the crossover to the small polaron regime.  The
lattice deformation is localised predominantly on the electron site.

Over four decades ago, a simple and intuitive 
variational approach to the 1D polaron
problem was proposed by Toyozawa~\cite{To61}. 
This method has been successfully applied to
various fields and revisited in a number of guises over
the years. It is generally believed to provide a 
qualitatively correct description of the polaron
ground state, aside from predicting a spurious 
discontinuous change in the mass at intermediate coupling.
The Toyozawa  variational wavefunction consists
of a product of coherent states (displaced oscillators) around the
instantaneous electron position.  The phonons create a symmetrical cloud
around the electron.  Numerical studies of the 1D electron-phonon correlation
function (two-point function) are in semi-quantitative agreement with
the Toyozawa variational wavefunction.   The numerically exact
three-point function, however, disagrees wildly.  
Denoting the instantaneous electron
position as 0, the Toyozwa variational wavefunction requires 
that the probability to find
phonon excitations, for example, on both sites 3 and 4, 
should be identical
to finding them on sites (-3) and 4.  Numerically, 
however, the latter is many
orders of magnitude less probable~\cite{KTB02}.  
This suggests a physical picture in which
the $k=0$ polaron is viewed not as 
an electron surrounded by a symmetrical cloud of
phonons, but is instead a coherent superposition 
of two ``comets,'' one with a tail
extending to the right, and the other to the left.

\begin{figure}[b]
\centering
\includegraphics[width=0.6\textwidth, angle=-90]{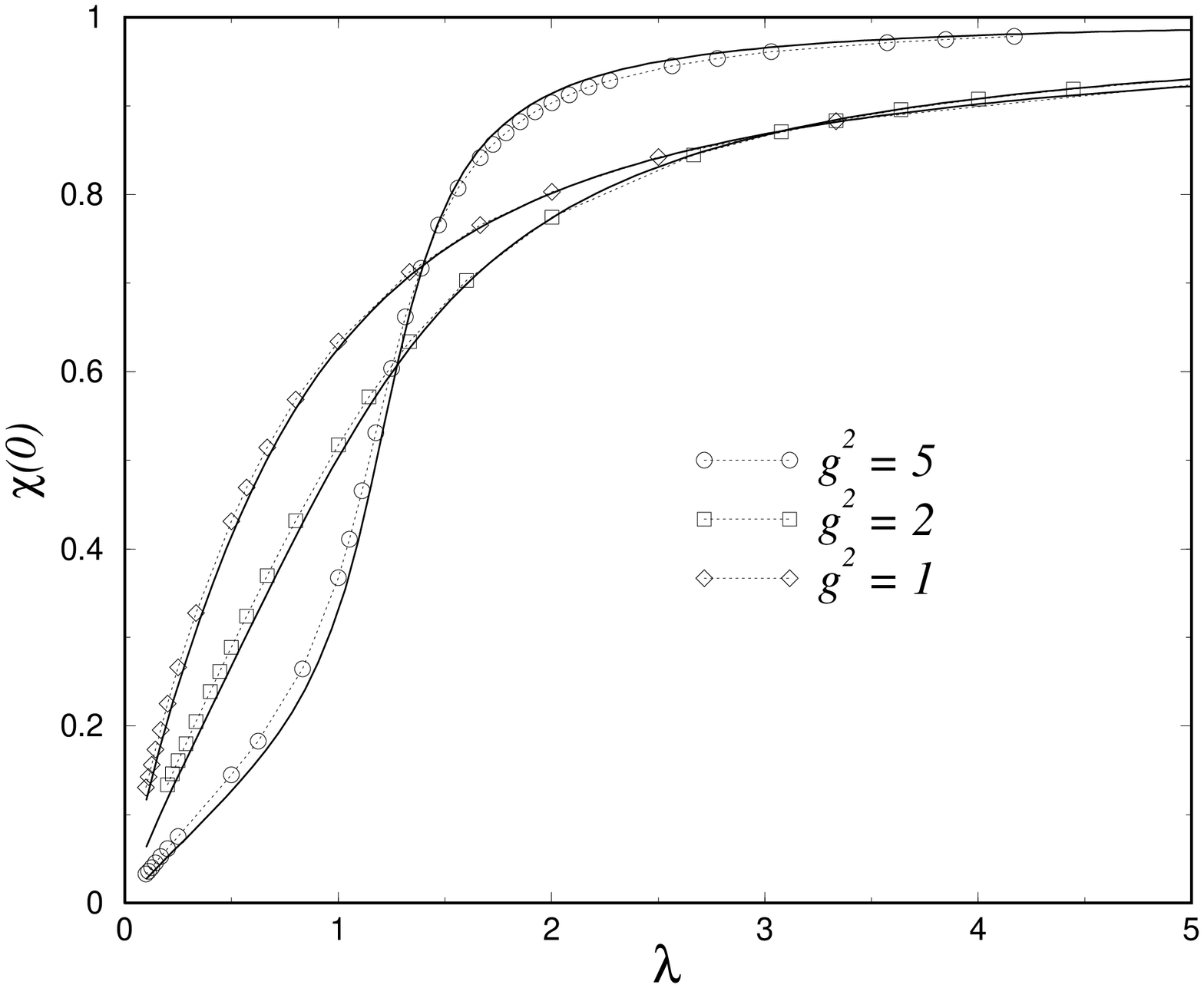}
\caption{On-site correlation $\chi(0)$ for the 3D polaron. VED results
(solid lines)~\cite{KTB02} are compared to DMFT (dotted lines with 
symbols)~\cite{CPFF97}. $\chi(0)$ is normalised to 1 when 
$\lambda$ is infinite (i.e. $t\to 0$) according to 
$\mathop {\lim}\limits_{t\to 0} \chi(0) = 2g$. Parameters 
are $\omega _0 = t = 1$.}
\label{fig:chi0}
\end{figure}
Studying the properties of Holstein polarons, 
DMFT is exact in infinite dimensions.  An
interpolation to 3D lattices is made possible by using a semielliptical free
density of states $N(E)$ to mimic the low-energy features. Here DMFT is 
accurate in the strong-coupling regime, where 
the surrounding phonons are predominately on
the electron site. This is also the regime where strong-coupling
perturbation theory works well. However, DMFT fails to 
compute quantities such as the polaron mass correctly in the weak-coupling
regime. The reason is that in DMFT, the lattice problem is mapped onto a
self-consistent local impurity model, which preserves the 
interplay of the electron and the phonons only at the local level. 
The spatial extent of the EP correlations increases as the EP 
coupling decreases, which explains the significant discrepancy in the
weak-coupling regime. Therefore only the on-site 
EP correlation has been studied by DMFT, and the results are 
compared with VED~\cite{KTB02} in Fig.~\ref{fig:chi0}. There is
good qualitative agreement.  The curves show a rapid change
in slope only for large $g$, where DMFT is less accurate. 
It is worth noting that DMFT neglects the $\vec k$
dependence of self-energy, i.e., the inverse effective mass is 
always equal to the quasiparticle spectral weight.  
Clearly the difference between $m_0/m^*$ and the 
spectral weight $Z_{\vec k}$ is not negligible in 
the intermediate- to weak-coupling regime.

\section{ Excited states}
\label{sec:es}
\begin{figure}[t]
\centering
\includegraphics[width=0.6\textwidth, angle=90]{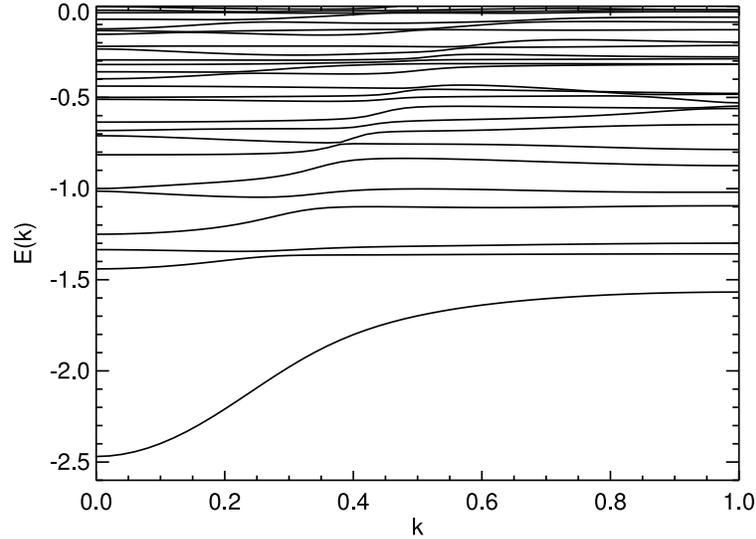}
\caption{The ground and excited state 1D polaron energy eigenvalues
(those energies below 0) are plotted 
as a function of $k$ (in units of $\pi$)
for $\bar g = \omega_0 = t =1$, $L = 9$, $N_{st}=1185$.
}
\label{fig:spaghetti}
\end{figure}

In this section we turn our attention from the ground state to
the excited states of the Holstein model.
Figure~\ref{fig:spaghetti}  plots the energy eigenvalues
for a  small variational space
containing a maximum of 9 phonon excitations.
The lowest curve is the polaron ground state at momentum $k$.
Excited states are the polaron with additional bound or unbound (or both)
phonon excitations.
A ripple can be discerned near the bare electron dispersion.
The figure superficially resembles a ``band structure,''
which however encodes ground and excited state information for
the many-body (many phonon) polaron problem.  
The ac conductivity of the polaron,
for example, appears as an ``interband'' transition in
this mapping.

\begin{figure}
\vspace*{0.3cm}
\centering
\includegraphics[width=0.6\textwidth]{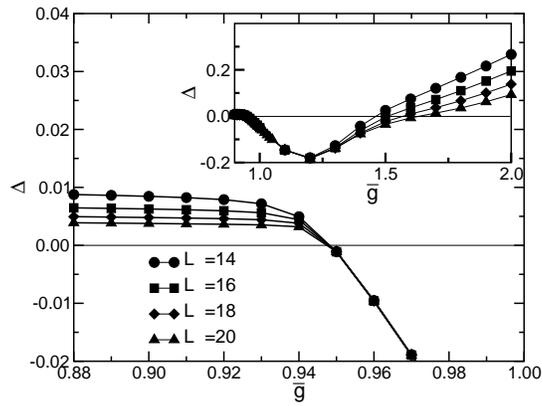}\\[0.4cm]
\caption{First excited state binding energy 
$\Delta = E_1 - E_0 - \omega _0$ as a function of $\bar g~$~\cite{BTB99}. Results
are for $\omega _0 =0.5$ and various Hilbert space sizes $L$.
Inset:  Binding energy over a wider range of $\bar g$.
} 
\label{fig:de}
\end{figure}

What is the nature of the first excited state?
We focus on the question of whether an extra phonon excitation 
forms a bound state with the polaron,
or instead remain as two widely separated entities. 
Using numerical and analytical approaches we show evidence that there
is a sharp phase transition (not a crossover) between these two types of states,
with a bound excited state appearing only as the EP coupling is increased.
Although the ground state energy $E_0$ is an analytic
function of the parameters in the Hamiltonian,
there are points at which the energy $E_1$
of the first excited state is nonanalytic.
In previous work, Gogolin has found bound states of the
polaron and additional phonon(s), but he does not
obtain a phase transition between bound and unbound
states because his approximations are limited to
strong coupling $g  \gg 1$~\cite{Go82}.
A phase transition between a bound and unbound first
excited state has been calculated for 3D 
using a dynamical CPA (coherent potential approximation)~\cite{Su74} 
and DMFT~\cite{CPFF97}.

We compute the energy difference $\Delta E = E_1
- E_0$, where $E_1$ and $E_0$ are the first excited state and the
ground state energies at $k=0$ (the two lowest
bands in Fig.~\ref{fig:spaghetti}).
In the case where the first excited
state of a polaron can be described as a polaron ground state
and an unbound extra phonon excitation, this energy
difference should in the thermodynamic limit equal the phonon
frequency, $\Delta E = \omega _0$.  In Fig.~\ref{fig:de} we
plot the binding energy $\Delta = \Delta E -\omega _0$ for $\omega _0=0.5$
as a function of the EP coupling $\bar g$ for various 
sizes of the variational space.
We see two distinct regimes. Below $\bar{g} _c\sim 0.95$, $\Delta$ 
varies with the system size but remains positive $( \Delta >0 )$.  
Physically, for $\bar g < \bar{g}_c$, the additional phonon excitation
would prefer to be infinitely separated from the polaron,
but is confined to a distance no greater than $L -1$
by the variational Hilbert space.
As the system size increases, $\Delta $ slowly approaches zero from above
as the ``particle in a box'' confinement energy decreases.
In the other regime, $\bar g > \bar{g}_c$, the data has clearly converged
and $\Delta <0$.  This is the regime where the extra phonon excitation
is absorbed by the polaron forming an excited polaron bound state.  
Since the excited polaron forms an exponentially decaying 
bound state, the method
already converges at $L=14$.  In the inset of Fig.~\ref{fig:de} we
show the binding energy $\Delta$ in a larger interval of EP 
coupling $\bar{g}$.  Although the results cease to converge at larger
$\bar g$, we notice that the binding energy $\Delta$ reaches a minimum as a
function of $\bar g$.  As one can demonstrate within the strong coupling
approximation, the true binding energy approaches zero exponentially from
below with increasing $\bar g$.  

\begin{figure}[t]
\vspace*{0.5cm}
\centering
\includegraphics[width=0.6\textwidth]{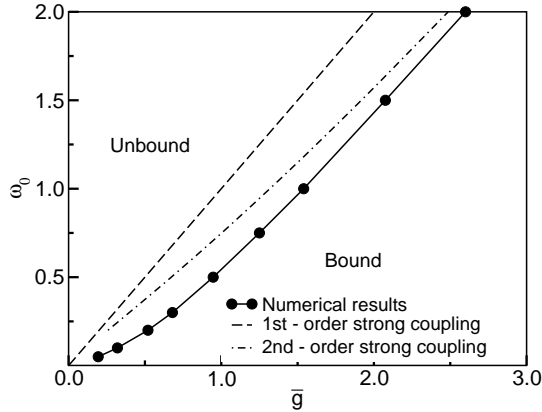}\\[0.3cm]
\caption{The phase diagram for the bound to unbound
transition of the
first excited state, obtained using 
the condition $\Delta (\omega _0 ,\bar g)=0$. 
The corresponding phase diagram for the ground state would be
blank:  there is no phase transition
in the ground state, only a crossover.
} 
\label{fig:phd}
\end{figure}

Figure~\ref{fig:phd} shows the phase diagram 
for $k=0$ separating the two regimes.  The phase boundary, 
given by $\Delta=0$, was obtained 
numerically, and compared to strong coupling perturbation theory in $t$ 
to first and second order. 
The phase transition where $\Delta$ becomes negative at
sufficiently large $\bar g$ is also seen in ED calculations.

\begin{figure}[t]
\vspace*{0.5cm}
\centering
\includegraphics[width=0.8\textwidth]{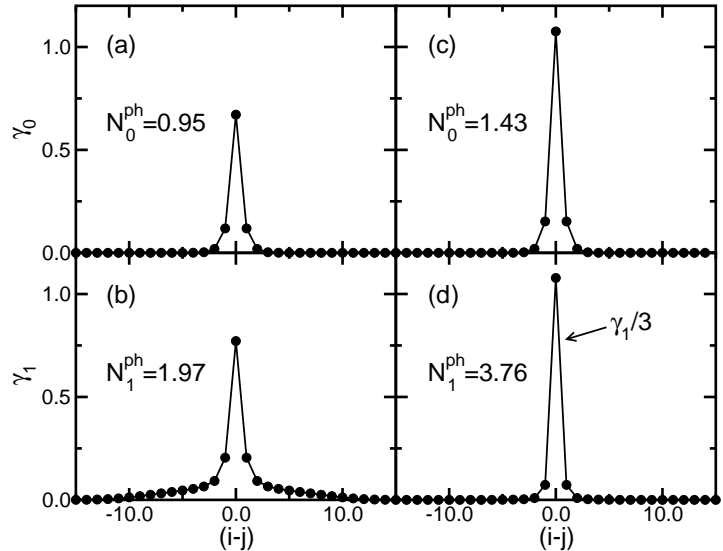}
\caption{The phonon number $\gamma$ as a function of the distance
from the electron position for the ground state (a) and the
first excited state (b), both computed at $\bar g=0.9$; and the same in
(c) and (d) for $\bar g = 1.0$. All data are computed at 
phonon frequency $\omega _0=0.5$ and $L=18$.  
Note that (d) is a plot of $\gamma_1 / 3$.
In (a,b), $\gamma_1-\gamma_0$ drops to zero around
$\vert i-j\vert=15$. This is a finite-size effect. Computing the same
quantity with larger $L$ below the phase transition would lead
to a larger extent of the correlation function indicating 
that the extra phonon excitation is not bound to the polaron.  
}
\label{fig:gamma}
\end{figure}

The distribution of the number of  phonons in the vicinity
of the electron is defined as
\begin{equation}
\gamma(i-j)  = \langle\psi_k^{}\vert c_i^\dagger c_i^{} b^\dagger_jb_j^{}
\vert\psi_k^{}\rangle.
\label{gamma}
\end{equation}
In Fig.~\ref{fig:gamma} we compute
this distribution for the
ground state $\gamma_0$ and the first excited state $\gamma_1$
slightly below $(\bar g =0.9)$, and above $(\bar g=1.0)$  
the transition for
$\omega_0 =0.5$.

The central peak of the correlation function $\gamma_1$
below the transition point is
comparable in magnitude to $\gamma_0$  (Fig.~\ref{fig:gamma}~(a,b)). 
The main difference between the two curves is the long range
decay of $\gamma_1$ as a
function of distance from the electron, onto which the central peak is
superimposed. 
The extra phonon that is represented by this long-range
tail extends throughout the whole system and is not bound to
the polaron.  
The existence of an unbound, free phonon is confirmed
by computing the difference of total phonon number $N^{ph}_{0,1}=\sum_l
\gamma_{0,1}(l)$. This difference should equal one below the
transition point. Our numerical values give $N^{ph}_1-N^{ph}_0\sim
1.02$.  We attribute the deviation from the exact result to
the finite relative separation allowed.
Correlation functions above the transition point (Fig.~\ref{fig:gamma}~(c,d)) 
are physically different.  First, phonon correlations in $\gamma_1$ decay
exponentially, which also explains why the
convergence in this region is excellent.  Second, the size of the
central peak in $\gamma_1$ is 3 times higher than $\gamma_0$.
(Note that to match scales in Fig.~\ref{fig:gamma}~(d) 
we divided $\gamma_1$ by 3).
The difference in total phonon number gives
$N^{ph}_1-N^{ph}_0\sim 2.33$.  We are thus facing a totally different
physical picture: The excited state is composed of a polaron which
contains several extra phonon excitations (in comparison to the 
ground-state polaron) and the binding energy of the excited
polaron is $\Delta <0$. The extra phonon excitations are located
almost entirely  on the electron site. 
The value of $\gamma_1-\gamma_0$ at $j=0$ is 2.16,
which almost exhausts the phonon sum.

\begin{figure}[t]
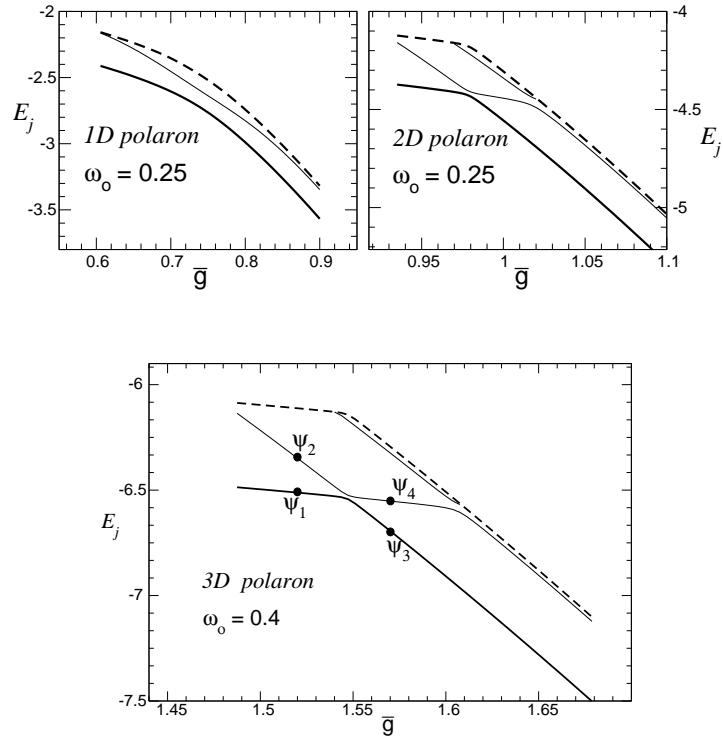

\vspace*{0.5cm}
\centering
\includegraphics[width=0.8\textwidth]{Level12Dn_new.eps} \vskip 0.9cm
\centering
\includegraphics[width=0.6\textwidth]{Level3Dnx_new.eps} \vskip 0.4cm
\caption{ Eigenvalues of low-lying states as functions of coupling constant in
1D through 3D. Hopping $t=1$ in all panels.
In the adiabatic regime in higher dimensions, the ground
state (thick solid lines) shows a fairly abrupt change in slope.
In the 3D panel,
$\psi_1$ and $\psi_4$ are a lightly-dressed electron state;  $\psi_2$ and
$\psi_3$ are a heavy polaron state. 
The dashed lines are the beginning of the
lowest continua.}
\label{fig:levels_3D}
\end{figure}

Next we discuss the role of dimensionality
in the excited states.
The effect of dimensionality on static properties has been
studied previously \cite{EH76,JE83,KM93,KTB02}.
The eigenvalues of the low-lying $k=0$ states
are shown as functions of $\bar g$ in
Fig.~\ref{fig:levels_3D}.
The energy spectra in D$>$1 are
qualitatively different than in 1D. The 1D polaron ground state becomes
heavy gradually as $\bar g$ increases. However, in D$\geq$2, the ground state
crosses over to a heavy polaron state by a narrow avoided level crossing, 
which is consistent with the existence of a potential barrier \cite{EH76}.
In the lower panel of
Fig.~\ref{fig:levels_3D}, $\psi_1$ and $\psi_4$ are
nearly free electron states;
$\psi_2$ and $\psi_3$ are heavy polaron states. The inner
product $|\langle \psi_1 | \psi_4 \rangle|$ is equal to 0.99.
Just right of the crossing region the effective
mass 
(approximately equal to the inverse of the spectral weight) 
of the first
excited state can be smaller than the ground state by 2 or 3 orders of magnitude,
while their energies can differ by much less than $\omega_0$.
The narrow avoided crossing description works less
well for larger $\omega_0$.

\section{ Dynamics of polaron formation} 
\label{sec:dy}
How does a bare electron time evolve to become a polaron quasiparticle?
The bare electron can be injected by inverse photoemission or tunnelling, or a hole
by photoemission, or an exciton (electron-hole bound state) by fast optics.

One approach is to construct a variational many-body Hilbert space including
multiple phonon excitations, and to numerically integrate the many-body
Schr\"odinger equation, 
\begin{equation}\label{schrodinger}
i { {d \psi} \over {d t} } = H \psi 
\end{equation}
in this space~\cite{KG96}. The full many-body wave function is obtained.
This method includes the full quantum dynamics
of the electrons and phonons.  
Note that alternative
treatments, such as the semiclassical approximation that treats the phonons
classically, fail for this problem, particularly in the limit
of a wide initial electron wavepacket.

\begin{figure}
\vspace*{0.9cm}
\centering
\includegraphics[width=0.9\textwidth]{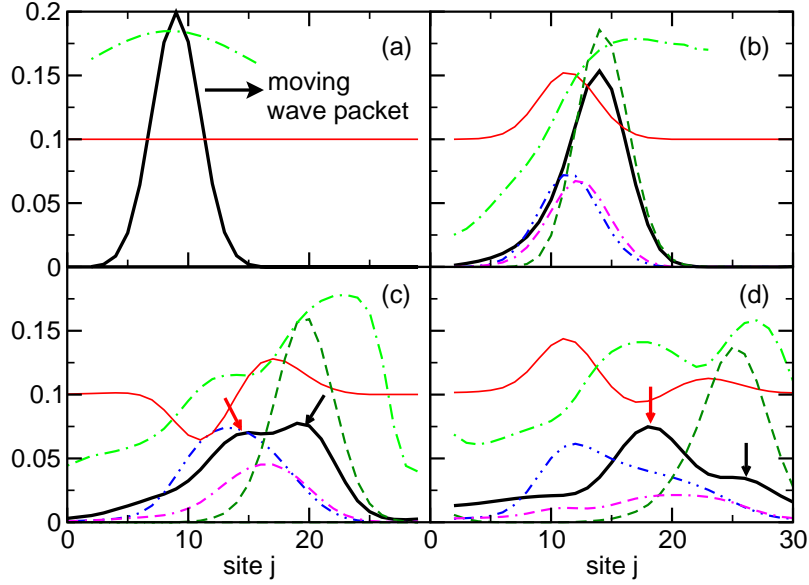}
\vspace*{0.9cm}
 \caption{ Snapshots of the polaron-formation process, for
$t=\omega_0=1$, and $\bar g=0.4$. The calculation is performed on a 30-site
periodic lattice. Time is measured in units of the phonon period. 
Shown are the electron density $\langle c_j^\dag c_j^{} \rangle$
(thick solid black line), phonon density $\langle b_j^\dag b_j^{} 
\rangle$ (blue double-dot dashed line), lattice displacement 
$\langle \hat{x}_j \rangle \equiv \langle b_j^{} + b_j^\dag \rangle$
(thin solid red line), 
velocity in units of lattice constant per phonon period
(green dot dashed line), and EP
correlation function $\langle c_j^\dag c_j^{} b_j^\dag b_j^{} \rangle $
(magenta dot double-dashed line). The green dashed line gives the 
free-electron wave packet for reference. For clarity, the origins of the thin solid red
and green dashed curves are offset by 0.1 and their values are rescaled by a factor of
0.2 and $0.05/(2\pi)$ respectively.  The blue double-dot dashed curve has 
been rescaled by a factor of 0.5.
\label{fig:snapshot}} 
\end{figure}

Figure~\ref{fig:snapshot} shows snapshots of polaron formation at weak
coupling.
An initial bare electron wave packet is launched
to the right as
shown in panel (a). This initial condition is relevant to the recent
experiments~\cite{Geea98,SSKYK01,DVBS00,TNSSTK98,AT02},
and to electron injection from a time-resolved STM 
(scanning tunneling microscope) tip~\cite{DRT00}.  
Although polarons injected optically or by STM 
can have a range of initial momenta, it would be more 
realistic to take $k = 0$ for an optically created exciton.
In panel (b) the electron is not yet dressed and thus is moving roughly as fast
as the free electron (green dashed line). In addition, there exists a
back-scattered current (which later evolves into a left-moving polaron) on the
left side of the wave packet (green dot-dashed and thick black curves).
In panel (c) after an elapsed time of one phonon
period, the electron density consists of two peaks. The peak on the right
(black arrow) is essentially a bare electron. The peak on the left is a polaron
wave packet moving more slowly.   As
time goes on, the bare electron peak decays and the polaron peak grows. Some
phonons are left behind (blue double-dot dashed line), mainly near the injection point. These
phonons are of known phase with displacement
shown in thin solid red. Some phonon excitations travel with the
polaron (magenta dot double-dashed line).
Finally, a coherent polaron wave packet is observed when
the polaron separates from the uncorrelated phonon excitations.
The velocity operator is defined as 
\begin{equation}\label{velocity}
\hat{v}_j \equiv {2  \hat{\jmath}_{j,j+1} \over e \langle c_j^\dag c_j^{}
+ c_{j+1}^\dag c_{j+1}^{}  \rangle } ,
\end{equation} 
where $j$ is
the site index and $\hat{\jmath}$ is the current operator.
$\langle  \hat{v}_j   \rangle$ is shown as a green dot-dashed line.

There are regimes where the polaron formation time is
a calculable constant of
order unity times a phonon period $T_0$, as
seen in some experiments and in Fig.~\ref{fig:snapshot},
but there are other regimes in which the phonon period is not the relevant
timescale. The limit of hopping $t \rightarrow 0$ is 
instructive~\cite{KAK02,BLW86}.
After a time
$T_0 / 4$, the expectation of the lattice displacement 
$\langle \hat{x}_j \rangle$ on the
electron site has the same value as a static polaron.  It is tempting (but we
would argue incorrect) to identify this as the polaron formation time.  At
later times, $\langle \hat{x}_j \rangle$ overshoots by a factor of two, and after a
time $T_0$,  $\langle \hat{x}_j \rangle$ 
and all other correlations are what they
were at time zero when the bare electron was injected. The system oscillates
forever.  In general an electron emits phonons enroute to becoming a polaron,
and we propose that the polaron formation time be defined as the time required
for the polaron to physically separate from the radiated phonons. The
polaron formation time for hopping $t \rightarrow 0$ is thus infinite, because
the electron is forever stuck on the same site as the radiated phonons.
\begin{figure}[t]
\centering
\includegraphics[width=0.7\textwidth]{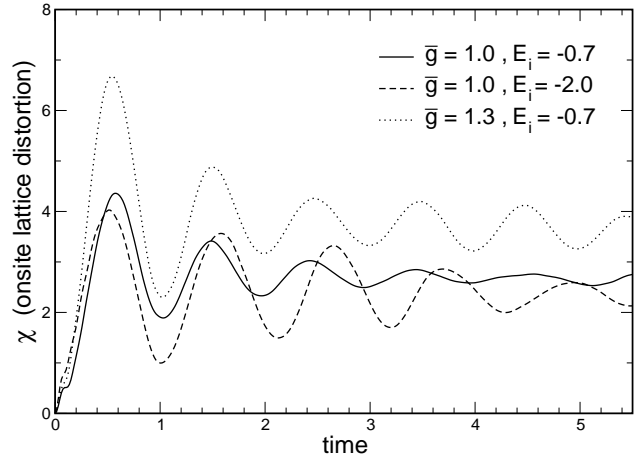}
\vspace*{0.9cm}
\caption{The on-site electron-phonon correlation function
$\chi = \langle c_j^\dag c_j^{} (b_j^{} + b^\dag_j) \rangle$
as a function of time measured in phonon periods.  For all curves,
$\omega_0=0.5$ and hopping $t=1$.
The solid line is for a bare electron injected with
nonzero initial momentum at energy $E_i=-0.7$,
where the bottom of the bare band is at energy -2.
The phonon displacement is larger and more weakly damped
for larger electron-phonon coupling $\bar g$, dotted line.
In contrast to a bare electron, an exciton
(bound particle-hole pair)
is generally created with
an initial momentum zero, corresponding
to $E_i=-2$, dashed line.}
\label{fig:Sugita}  
\end{figure}

An electron injected at several times the phonon energy $\omega _0$
above the bottom of the band is another instructive example. The electron
radiates successive phonons to reduce its kinetic energy to near the bottom of
the band, and then forms a polaron. For very weak EP coupling,
the rate for radiating the first phonon can be computed by Fermi's golden rule,
$ \tau_{FGR}^{-1} = \bar g ^2 / [ \hbar~ t \sin(k_f) ], $
where $k_f$ is the electron momentum after emitting a phonon.  The
phonon emission time can be arbitrarily long for 
small $\bar g$. For strong coupling, the rate approaches $ \tau_{SC}^{-1}
= \bar g / \hbar $ because the polaron spectral function smoothly 
spans numerous narrow bands and its standard deviation is equal to $\bar g$.

\begin{figure}[t]
\centering
\includegraphics[width=0.7\textwidth]{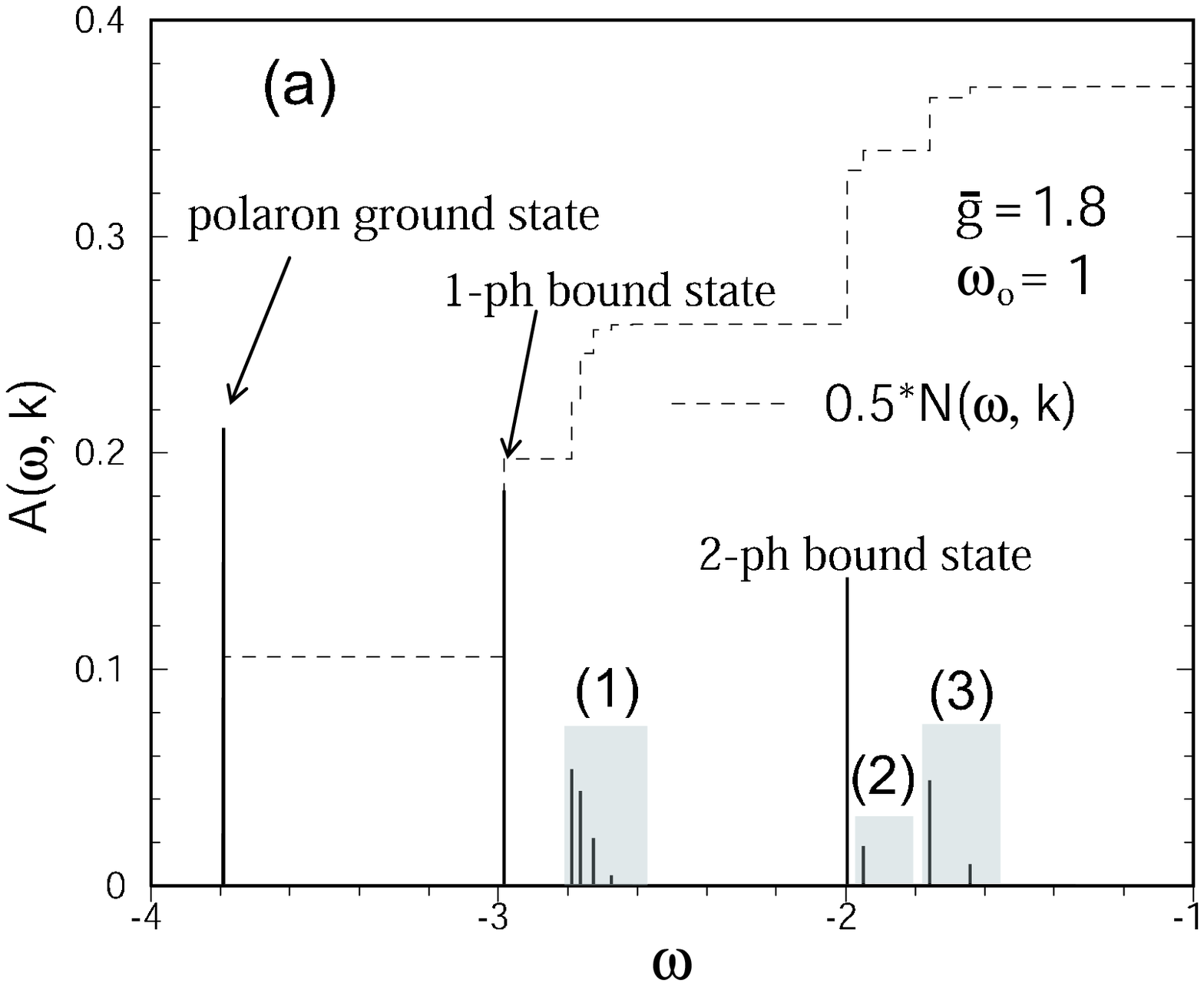}
\centering
\includegraphics[width=0.7\textwidth]{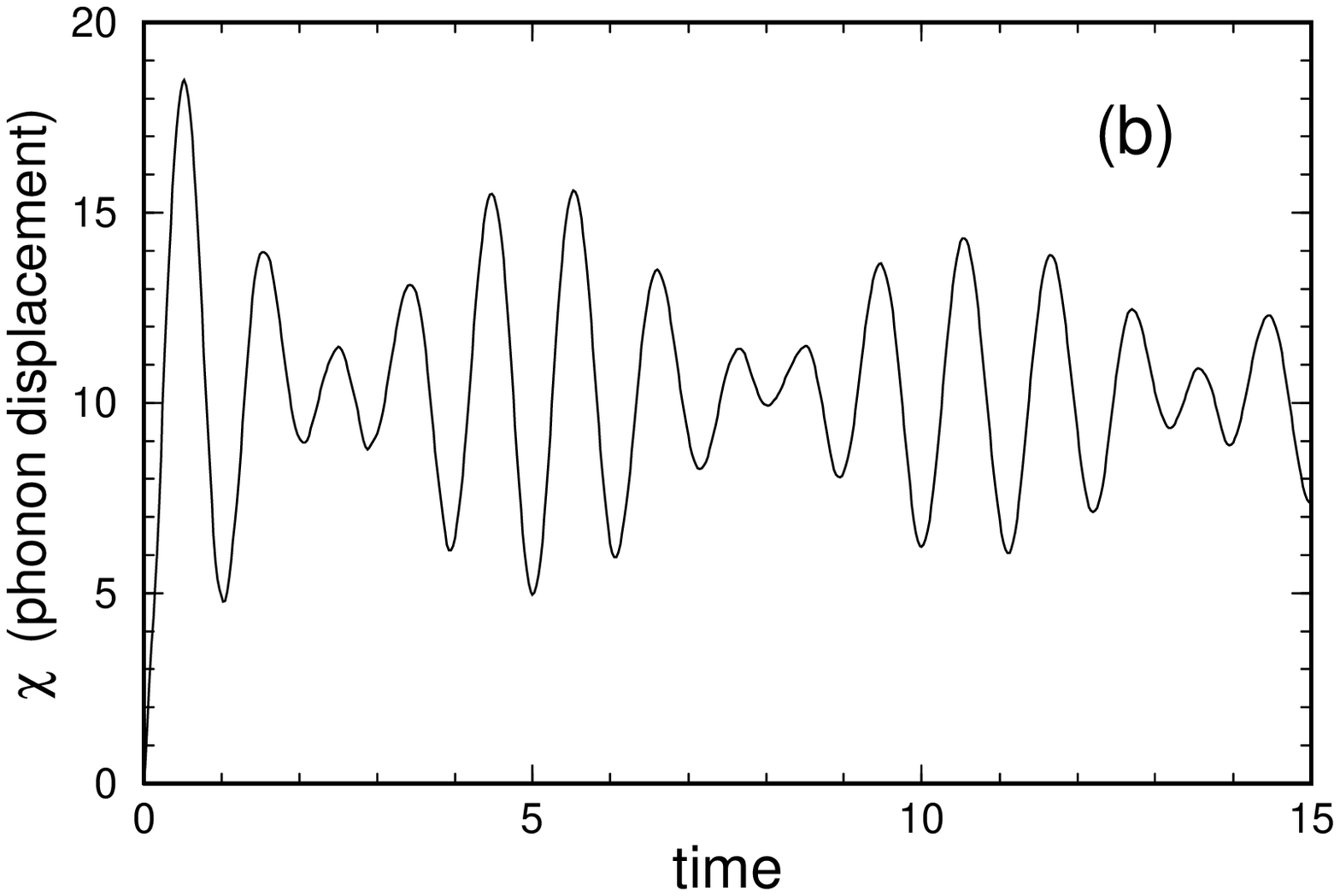}
 \caption{Panel (a): Spectral function at strong coupling. There are
 three quasiparticle excited states split off from the continua.
 Shaded areas (1) and (2) correspond to continuum
 states.   
 (b): Quantum beat formed by multiple
 excited states and continua. } 
 \label{fig:spc2}
 \end{figure}

Decaying oscillations in polaron formation (actually
the formally equivalent problem of an exciton coupled to phonons~\cite{Ra82})
have been observed in a pump-probe experiment that measures
reflectivity after a bare exciton is created~\cite{SSKYK01}. The observed oscillatory
reflectivity was interpreted as the lattice motion in the phonon-dressed (or
``self-trapped'') exciton level.  Assuming the modulation in the exciton level goes as
$\Delta E = -\lambda c_j^\dag c_j^{} \hat{x}_j^{}$, where
$\hat{x}_j^{} = \langle b_j^{} + b^\dag_j \rangle$ is the 
lattice displacement, the
model Hamiltonian applies directly to the experiment. We calculate the
corresponding EP correlation function
in Fig.~\ref{fig:Sugita}.
In this regime, the polaron formation time (damping time)
increases as the electron-phonon coupling $\bar g$ increases,
and also as the initial electron (exciton) energy
approaches the band bottom.
We find satisfactory agreement when
compared to Fig.~2b of Ref.~\cite{SSKYK01}. Both show a damped oscillation
with a delayed phase.
(Numerical calculations in Figs.~\ref{fig:Sugita}-\ref{fig:spc2}
are performed on an extended system, not a
finite cluster.)

Figure~\ref{fig:spc2} shows the spectral function at strong coupling. 
Three delta functions are visible, corresponding
to polaron ground and excited states, along with three continua
containing unbound phonons.
There is additional structure at higher energy (not
shown). 
The probability to remain in the initial bare particle state
$ P(\tau) \equiv | \langle \psi (\tau ) |c_k^\dag|0 \rangle | ^2 $
for this spectrum is complicated, and
includes oscillating terms that do not decay to zero at zero temperature from
the polaron ground and excited states beating against each other.
The branching ratios into the various channels are 
calculated in Ref.~\cite{Ku03}.

\section{Spectral signatures of Holstein polarons} 
\label{sec:sp}
As already stressed in the introduction the crossover from
quasi-free electrons or large polarons to small polarons
becomes manifest in the spectral properties above all. 
Here of particular interest is whether an ``electronic'' 
or ``polaronic''  (quasi-particle) excitation exists in the spectrum. 
This question has been partially addressed by calculating 
the wavefunction renormalisation factor [(electronic) quasi-particle
weight] $Z_k$ in Sec.~\ref{sec:gs} (see Fig.~\ref{fig:zk}).
More detailed information can be obtained from the one-particle
spectral function  $A(k,\omega)$.  This quantity of great importance
can be probed by direct (inverse) photoemission, where a bare electron 
with momentum $k$ and energy $\omega$ is removed (added) 
from (to) the many-particle system. 
The intensities (transition amplitudes) of these processes 
are determined by the imaginary part of the retarded one-particle
Green's function, i.e. by 
\begin{equation}
  A(k,\omega) 
  = 
  -\frac{1}{\pi}
  {\rm Im}\,G(k,\omega) 
  = A^+(k,\omega)+A^-(k,\omega)
  \,,
\label{aimg}
\end{equation}
with
\begin{eqnarray}\label{aspekt}
A^{\pm}(k,\omega)&=&
-\frac{1}{\pi}{\rm Im}\!\!\lim_{\eta\to 0^+}
\langle \psi_0 | c_k^\mp \frac{1}{\omega 
+{\rm i}\eta\mp H} c_k^\pm |\psi_0\rangle\nonumber\\[0.1cm]
&=& \sum_{m} 
|\langle \psi_m^{\pm}|c_{k}^{\pm} 
|\psi_0^{}\rangle|^2
\delta [\,\omega\mp(E_m^{\pm}-E_0^{})]\,,
\end{eqnarray}
where $c_{k}^{+}=c_{k}^{\dagger}$ and $c_{k}^{-}=c_{k}^{}$
($T=0$; 1D spinless case). These functions test both excitation 
energies ($E_m^{\pm}-E_0^{}$) and overlap ($\propto Z_k$) of the 
$N_e$-particle ground state $| \psi_0\rangle$ 
with the exact eigenstates $|\psi_m^{\pm}\rangle$ of
a $(N_e \pm 1)$--particle system.  
The electron spectral function of the single-particle Holstein model
corresponds to $N_{e} = 0$, i.e., $A^-(k,\omega)\equiv 0$. 
$A(k,\omega)$ can be determined, e.g., by a combination of KPM and CPT 
(cf. Sec.~\ref{sec:na}). 

\begin{figure}
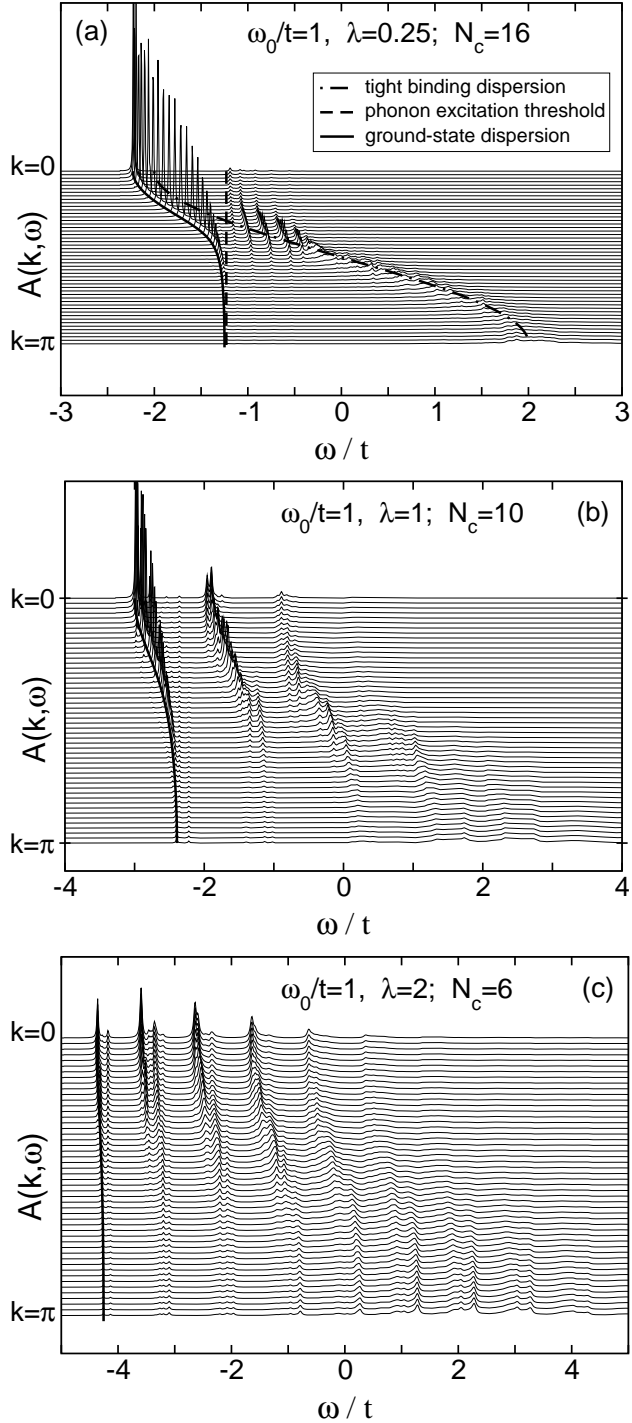

\centering
\includegraphics[width=0.7\textwidth]{sf_cpt_wc.eps}\\[0.2cm]
\includegraphics[width=0.7\textwidth]{sf_cpt_ic.eps}\\[0.2cm]
\includegraphics[width=0.7\textwidth]{sf_cpt_sc.eps}\\
\caption{Spectral function of the 1D Holstein polaron 
  in the weak (a), intermediate (b), and strong (c) EP coupling regimes. 
  CPT data based on finite-cluster ED with $N_c$ sites, and $M=7$
  ($\lambda=0.25$),  $M=15$ ($\lambda=1$), 
   $M=25$ ($\lambda=2$) phonon quanta. Note that the non-monotonic heights
of the lowest energy peaks in~(a) are an artifact of the 
CPT calculation, where some of the wavevectors fit the $N_c=16$ cluster 
size, and some don't. Also the dispersionless absorption feature 
in (c), just above the small-polaron peak, is due to a finite-size effect,
but not the double-peak structures of the higher excitation bands.
This has been proved by determining 
the spectral function in the $k=0$--sector 
by VED.}
\label{f:cpt_elecspe}
\end{figure}

Figure~\ref{f:cpt_elecspe}~(a) shows that at weak EP coupling,
the electronic spectrum is nearly unaffected for energies below the phonon
emission threshold. Hence, for the case considered with $\omega_0$ lying
inside the bare electron band $\varepsilon_k=-2t\cos k$, 
the signal corresponding to the renormalised dispersion $E_k$  
nearly coincides with the tight-binding cosine band  
(shifted $\propto \varepsilon_p$) up to some $k_X$, 
where the phonon intersects the bare electron band. 
At $k_X$ electron and phonon states  ``hybridise'' and repel each
other, forming an avoided-crossing like gap. 
For $k>k_X$  the lowest absorption signature in each $k$ sector
follows the dispersionless phonon mode, leading to the  
flattening effect~\cite{WF97}. Accordingly the (electronic) 
spectral weight of these peaks is very low. The high-energy incoherent 
part of the spectrum is broadened $\propto \varepsilon_p$, 
with the $k$-dependent maximum again following the bare cosine dispersion.

Reaching the intermediate EP coupling  (polaron crossover) regime 
a coherent band separates from the rest of the spectrum 
[$k_X\to\pi$; see panel (b)].  At the same time its spectral weight
becomes smaller and will be transferred to the incoherent part, where
several sub-bands emerge.  

The inverse photoemission spectrum in the strong-coupling case is shown
in Fig.~\ref{f:cpt_elecspe}~(c). First, we observe all
signatures of the famous polaronic band-collapse. The 
coherent quasi-particle absorption band becomes 
extremely narrow. Its bandwidth approaches 
the strong-coupling result $4Dt\exp(-g^2)$ for $\lambda\,,g^2 \gg 1$. 
If we had calculated the polaronic instead of the
electronic spectral function, nearly all spectral 
weight would reside in the coherent part
of the spectrum, i.e. in the small-polaron band. 
This has been demonstrated quite 
recently~\cite{LHF06}. In our case the incoherent part of the 
spectrum carries most of the spectral weight.
It basically consists of a sequence of sub-bands 
separated in energy by $\omega_0$, which correspond  
to excitations of an electron and one or more phonons. 

Let us emphasise that for all couplings the lowest signature in
$A(k,\omega)$ almost perfectly coincides with the coherent polaron
band-structure (solid lines) obtained by VED (see  
Sec.~\ref{sec:gs}), which underlines the high precision of the 
CPT-KPM approach used here.

Of course, the phonon modes are unaffected by one electron in the solid,
i.e. the phonon self-energy is zero. Actually this is true in the 
thermodynamic limit only. In a finite-cluster calculation there will be 
an influence of order ${\cal O}(1/N)$ and the phonon spectra 
provide additional useful information concerning the polaron dynamics. 
For this purpose, we calculate the $T=0$
phonon spectral function 
\begin{equation}
B(q,\omega)=-\frac{1}{\pi}{\rm Im} D^R(q,\omega) 
\label{ph_specf}
\end{equation}
which is related to the phonon Green's function 
\begin{equation}
\label{ph_green}
  D^R(q,\omega) 
  = 
  \lim_{\eta\to 0^+}\ 
  \langle \psi_0 | \hat{x}_{q}
  \frac{1}{\omega + {\rm i}\eta - H}
  \hat{x}_{-q} |\psi_0\rangle\,,
\end{equation}
where $\hat{x}_{q}= N^{-1/2} \sum_j \hat{x}_j e^{-{\rm i} qj}$
and $\hat{x}_j=(b^\dagger_j+b^{}_j)/\sqrt{2\omega_0}$. 

For the Holstein model~(\ref{hm}), $B(q,\omega)$ is
symmetric in $q$. The bare propagator 
$D_0(q,\omega) = 2\omega_0/(\omega^2-\omega_0^2)$ is dispersionless.   
Then, adapting the CPT-KPM approach to the calculation of the phonon
spectral function, the cluster expansion is identical to 
replacing the full real-space Green's function $D_{ij}$ by $D^c_{ij}$. 

\begin{figure}
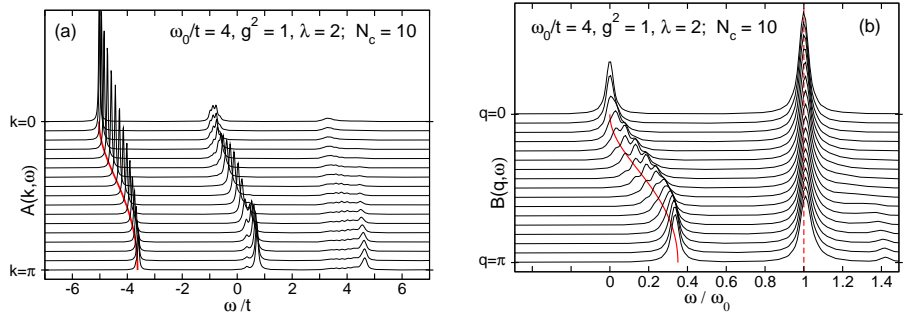

\centering
\includegraphics[width=0.48\textwidth]{sf_cpt_aa.eps}\hspace*{.5cm}
\includegraphics[width=0.48\textwidth]{bsf_w4.0_Ep4.0.eps}\\[0.2cm]
\caption{Electron (a) and phonon (b) spectral functions in the 
anti-adiabatic intermediate EP coupling regime. Solid (dashed) lines
give $E_k$ ($\omega_0$) determined by VED. Note that abscissae are scaled 
differently.}
\label{f:epspec}       
\end{figure}

Figure~\ref{f:epspec} compares electron~(a) and phonon~(b)
spectra in the high phonon-frequency limit, where the small polaron 
crossover is determined by $g^2$. Obviously the phonon spectrum 
is also renormalised by the EP interaction 
due to the finite ``particle density'' $N_e/N_c=1/10$. 
So we can detect a clear signature of the polaron band having a width 
$W\simeq 1.5t$ (cf. Fig.~\ref{f:epspec}~(a)). 
The dispersionless excitation at $\omega/\omega_0=1$ 
is obtained by adding one phonon with momentum $q$ to the $k=0$ 
ground state.  Above this pronounced peak, we find an 
``image'' of the lowest polaron band -- shifted by $\omega_0$ -- with
extremely small spectral weight, hardly resolved in Fig.~\ref{f:epspec}~(b).

\section{Transport and optical response}
\label{sec:to}
The investigation of transport properties has been playing a central
role in condensed matter physics for a long time.
Optical measurements, for example, proved the importance of 
EP interaction in various novel materials
such as the 
cuprates, nickelates or 
manganites and, in particular, corroborated
polaronic scenarios for modelling their electronic transport properties
at least at high temperatures~\cite{Emi93,AM95,WMG98}.
Actually, the optical absorption of small polarons is distinguished
from that of large (or quasi-free) polarons by the shape and the 
temperature dependence of the absorption bands which arise from exciting
the self-trapped carrier from or within the potential well 
that binds it~\cite{Emi95}.  Furthermore, as was the case with 
the properties of the ground state, the optical spectra of light and 
heavy electrons, small and large polarons differ 
significantly as well~\cite{Ra82}. 
In the most simple weak-coupling and anti-adiabatic strong EP 
coupling limits, the absorption associated with photoionization 
of Holstein polarons is well understood and the optical 
conductivity can be analysed analytically (\cite{Emi93,Mah00,RH67,Lo88,Feea94};
for a detailed discussion of small polaron transport phenomena 
we refer to Refs.~\cite{Fi95,Fir}).   
The intermediate coupling and frequency regime, however,  
is as yet practically inaccessible for a rigorous analysis 
(here the case of infinite spatial dimensions, where dynamical mean-field 
theory yields reliable results, is an exception~\cite{FC03,FC06}). 
So far unbiased numerical studies of the optical absorption in the 
Holstein model were limited to very small 2 to 
10-site 1D and 2D clusters~\cite{AKR94,CSG97,FLW97,WF98a}.   
In the following we will exploit the 
VED and KPM schemes~\cite{EBKT03,SWWAF05},
in order to calculate the optical conductivity numerically in 
the whole parameter range on fairly large systems.
\subsection{Optical conductivity at zero-temperature } 
\label{sec:sigma0}
Applying standard linear-response theory, the real part of the conductivity
takes the form
\begin{equation}
\label{re_sigma} 
{\rm Re} \sigma(\omega)={\cal D} \delta(\omega)+ \sigma^{reg}(\omega)\,,
\end{equation}
where ${\cal D}$ denotes 
the so-called Drude weight at $\omega=0$ and $\sigma^{reg}$
is the regular part (finite-frequency response) for $\omega>0$
which can be written in spectral representation at $T=0$ as~\cite{Mah00}  
\begin{equation}\label{sigma_reg}
   \sigma^{reg}(\omega) = \frac{\pi}{\omega N} \sum_{E_m>E_0} 
  |\langle \psi_m|\hat{\jmath}|\psi_0\rangle |^2\ 
\delta[\omega - (E_m - E_0)]
\end{equation}
with  the (paramagnetic) current operator 
$\hat{\jmath} = - \mbox{i} e t\sum_{i}(c_{i}^{\dagger} c_{i+1}^{} -
c_{i+1}^{\dagger} c_{i}^{})$.
   
Introducing the $\omega$-integrated spectral weight,
\begin{equation}
S^{reg} (\omega)=\int_{0^+}^\omega\,d\omega' \sigma^{reg}(\omega')\,,
\label{S_reg}
\end{equation} 
we arrive at the f-sum rule
\begin{equation}
-E_{kin}/2= {\cal  D} + S^{reg}/\pi\qquad {\rm (1D\;case)}\,,
\label{sumrule}
\end{equation} 
where $E_{kin}=-t\sum_i(c_{i}^{\dagger} c_{i+1}^{} +
c_{i+1}^{\dagger} c_{i}^{})$ is the kinetic energy and 
$S^{reg}=S^{reg} (\infty)$. 
Since for the Holstein model the Drude weight can be calculated
independently from Kohn's formula or the effective mass,  
\begin{equation}
{\cal  D}=\frac{1}{2N}
\frac{\partial^2E_0(\Phi)}{\partial \Phi^2}\Big|_{\Phi=0}=\frac{1}{2N}
\frac{\partial^2E_k}{\partial k^2}\Big|_{k=0}=\frac{1}{2m^*}\,, 
\label{drudemass}
\end{equation} 
the f-sum rule may be used to test the numerics. In Eq.~(\ref{drudemass}),
the first equality relates ${\cal D}$ to the second derivative of the
(non-degenerate) ground-state energy with respect to a field-induced 
phase $\Phi$ coupled to the hopping.

\begin{figure}
\centering
\includegraphics[width=0.7\textwidth]{sigma_w02.eps}\\[0.2cm]
\includegraphics[width=0.7\textwidth]{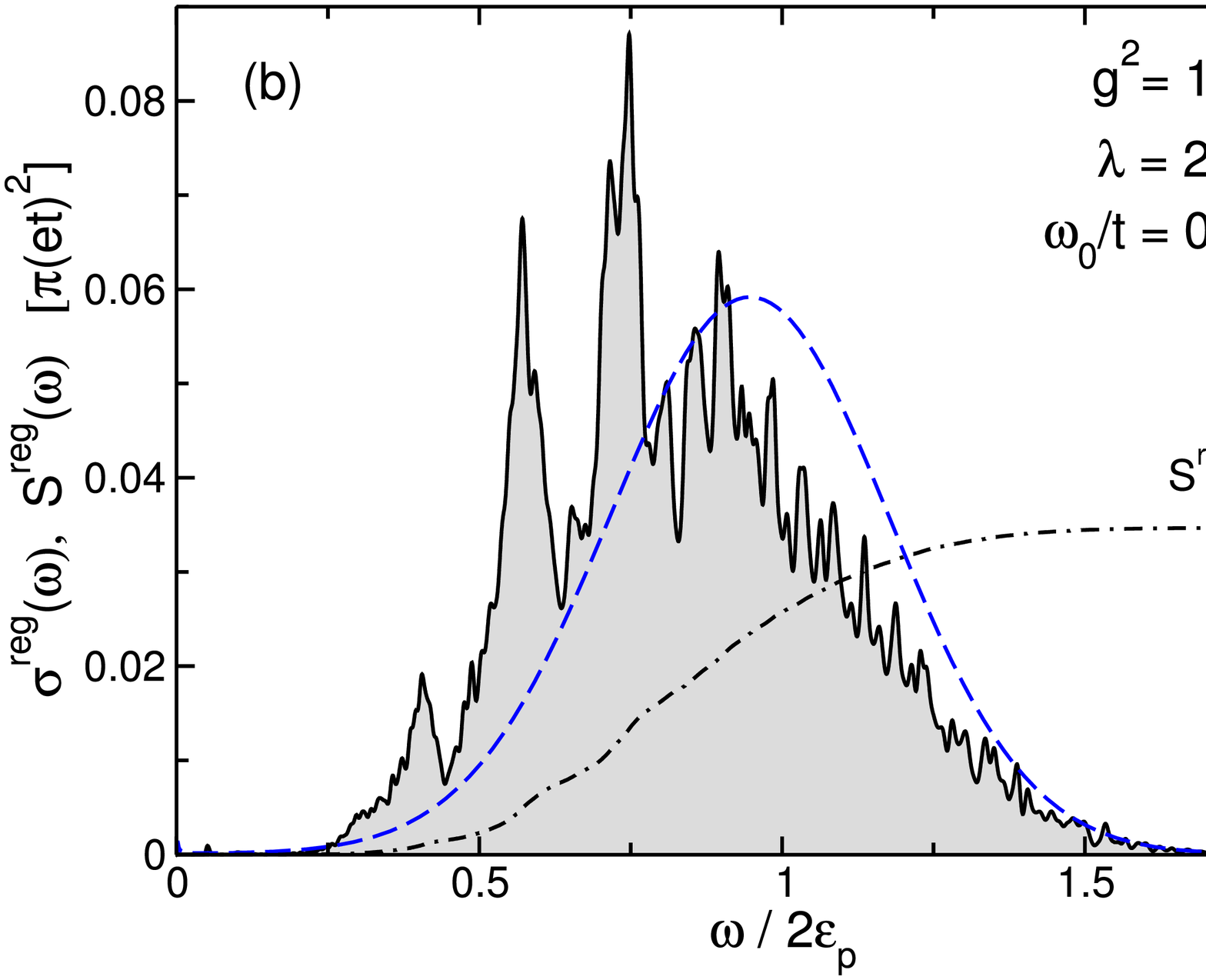}\\[0.2cm]
\includegraphics[width=0.7\textwidth]{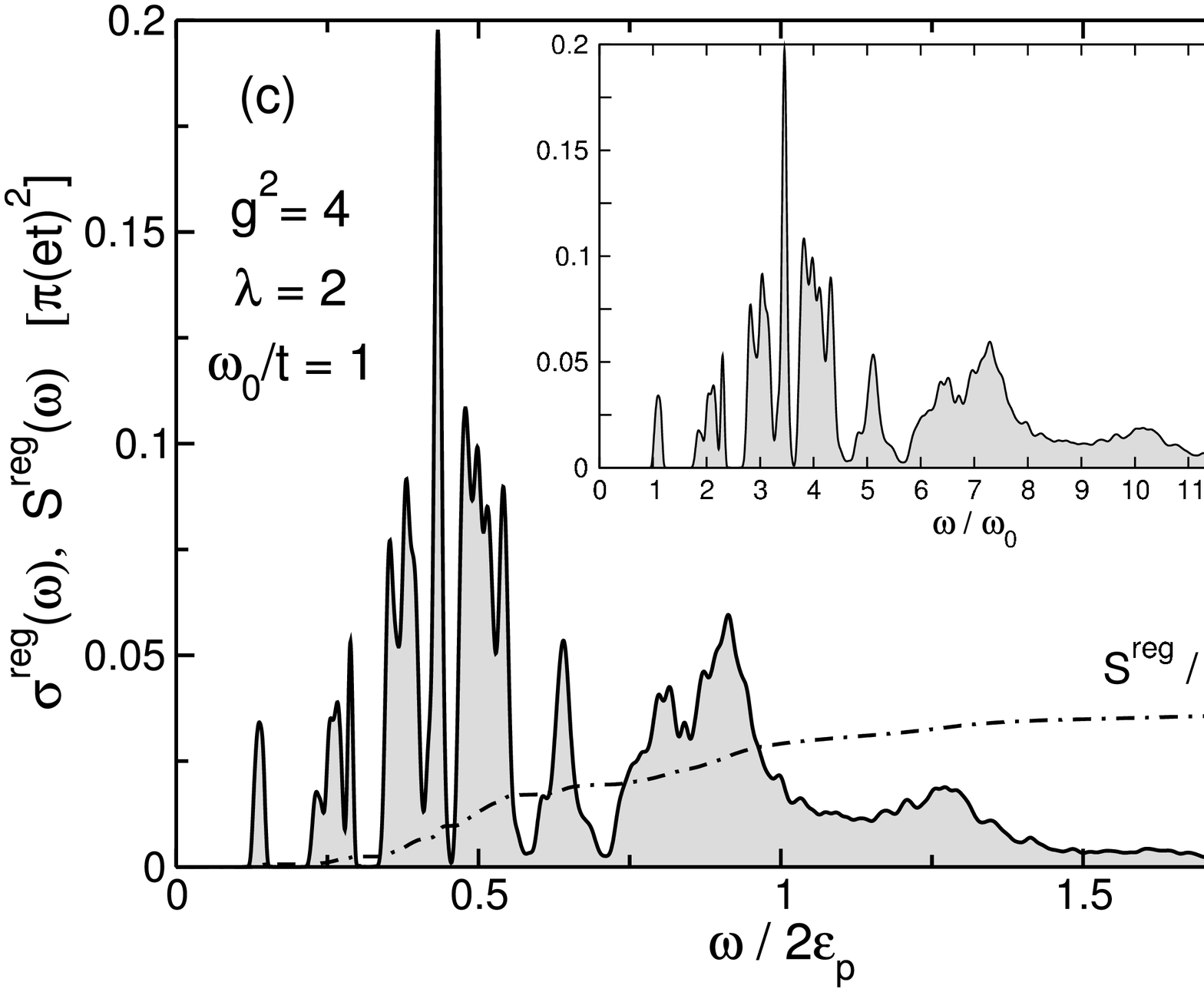}
\caption{Optical conductivity 
of the 1D Holstein polaron at $T=0$.
Results for $\sigma^{reg}(\omega)$ and $S^{reg}(\omega)$ 
are obtained by VED. In order to reduce finite-$M$ effects, 
data calculated for $M=15$ to 20 were averaged. 
The dashed line in (b) displays the analytical small polaron result,
$\sigma^{reg} (\omega) = 
  \frac{\sigma_0}{\sqrt{\varepsilon_p\omega_0}} \frac{1}{\omega} 
  \exp \left[- \frac{(\omega-2\varepsilon_p )^2}{4 
\varepsilon_p \omega_0}\right]$ (cf. Ref.~\cite{Emi93}), 
where $\sigma_0$ was determined to give the same integrated spectral
weight as $\sigma^{reg}(\omega>0)$.}
\label{f:sig}
\end{figure}

We first present $\sigma^{reg}(\omega)$ and its integral 
$S^{reg}(\omega)$ for the 1D Holstein model 
in Fig.~\ref{f:sig}. The upper panel (a) 
gives the results for intermediate-to-strong EP coupling, i.e. near the
polaron crossover, in the adiabatic (light electron) regime.
Of course, the optical conductivity threshold is $\omega=\omega_0$
for the infinite system we deal with using VED. In this respect standard  
ED, defined on finite lattices, suffers from pronounced
finite-size effects due to the discreteness in $k$-space.
Knowing that at about $\lambda\simeq 1$ a coherent polaron band
with bandwidth much smaller than $\omega_0$ splits off, 
the first (few) isolated peak(s) at the lower bound of the spectrum 
can be attributed to one- (few-) ($q=0-$) phonon emission processes
(cf. also Fig.~\ref{f:cpt_elecspe}~(b)). Of course, these transitions 
have little spectral intensity. At higher energies 
transitions to the incoherent part of the spectrum 
take place by ``emitting'' phonons with finite momentum 
(to reach the total momentum $k=0$ ground-state sector).
The main signature of $\sigma^{reg}(\omega)$ is that the spectrum
is strongly asymmetric, which is characteristic for rather
large polarons. The weaker decay at the high-energy side
meets the experimental findings for many polaronic materials like 
${\rm TiO_2}$~\cite{KMF69} even better than standard small-polaron theory.

For $\lambda=2$ and $\omega_0=0.4$, i.e., at larger EP coupling,
but not yet in the small-polaron limit, we find a more 
pronounced and symmetric maximum in the low-temperature optical 
response (see Fig.~\ref{f:sig}~(b)).
The maximum is located below the corresponding one 
for small polarons at $T=0$, 
which on its part lies somewhat below  $2\varepsilon_p=2t 
\lambda = 2g^2\omega_0$ (being the maximum of the Poisson distribution) 
because of the $1/\omega$ factor contained in the conductivity.
In this case the polaron band structure is more strongly renormalised, 
but, more importantly, the phonon distribution function in the ground state
becomes considerably broadened. 
Since the current operator connects only different-parity states having 
substantial overlap as far as the phononic part is concerned, 
in the optical response multi-phonon emissions/absorptions (i.e., 
non-diagonal transitions~\cite{Mah00}) become  
increasingly important.  Again deviations from the 
analytical small-polaron result (dashed 
line in  Fig.~\ref{f:sig}~(b)) might be important for relating
theory to experiment. 

The optical response obtained if the phonon frequency becomes 
comparable to the electron transfer amplitude is illustrated in 
Fig.~\ref{f:sig}~(c). Now the lowest one-phonon absorption 
(threshold) signal is well separated. In contrast to the light 
electron case Fig.~\ref{f:sig}~(a), the different absorption bands appearing
for a  heavier electron can be classified according to the 
number of phonons involved in the optical transition
(see inset).  
Increasing $\omega_0$ at fixed $g^2$ this becomes even more manifest 
(at the same time a Poisson distribution of the different sub-bands
evolves). The sub-bands are broadened by transitions to different 
``electronic'' levels.
For our parameters, a scattering continuum appears above
$\omega > 5$ to $6 ~ \omega_0$. 
Note that the ``fragmentation'' of the spectrum 
appearing at smaller energy transfer is not caused by finite-size effects.

\begin{figure}
\centering
\vspace*{1cm}
\includegraphics[width=0.9\textwidth]{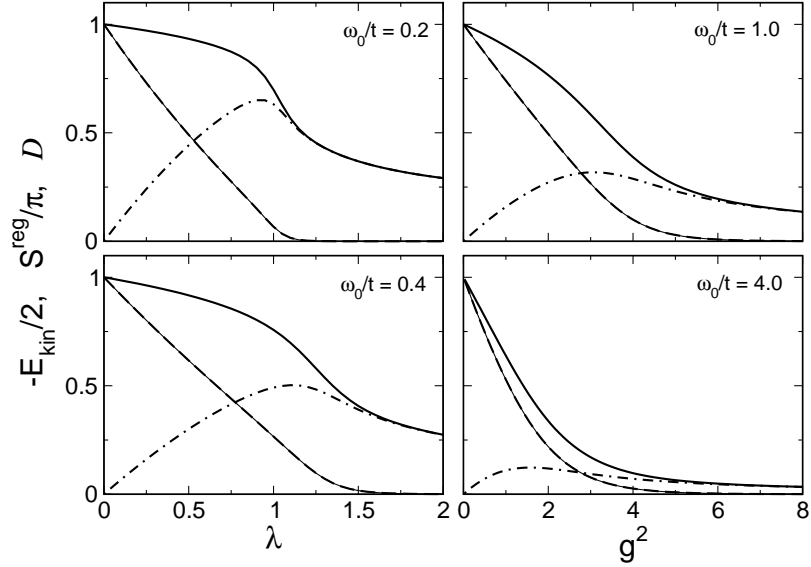}\\[0.2cm]
\caption{Renormalised kinetic energy ($E_{kin}$; solid line) 
and contribution of $\sigma^{reg}$ to the f-sum rule 
($S^{reg}$; dot-dashed line) as a function of the EP 
couplings $\lambda$ and $g$ in the adiabatic (left panels) 
and non-to-antiadiabatic (right panels) regimes, respectively. 
The Drude weights were obtained from the f-sum rule [(Eq.~(\ref{sumrule}); thin
solid line)] and effective mass [(Eq.~(\ref{drudemass}); dashed line)].}
\label{f:sumrule}       
\end{figure}

Turning to the sum rules presented in Fig~\ref{f:sumrule}, 
we notice a monotonic decrease of the total sum rule
$S^{tot}/\pi=-E_{kin}/2$, which indicates a suppression of
the electronic kinetic energy with increasing EP coupling.
In agreement with previous numerical results~\cite{RL82,WF98a,HEL04},  
the kinetic energy clearly shows the crossover from a large polaron, 
characterised by a $E_{kin}$ that is only weakly reduced from 
its non-interacting value [$E_{kin}(\lambda =0)=-2t$], to a less 
mobile small polaron in the strong EP-coupling limit,
where the strong-coupling perturbation theory result,
\begin{equation}
E_{kin}^{SCPT}=
-\frac{4t}{\omega_0} \Big\langle\frac{1}{s}\Big\rangle_{\kappa=2g^2}^{}
-\mbox{e}^{-g^2}\left[ 2 
+ \frac{4t}{\omega_0}\Big\langle\frac{1}{s}
\Big\rangle_{\kappa=g^2}^{}\right]
\end{equation}
($\langle \ldots\rangle_\kappa$ denotes the average with respect
to the Poisson distribution with parameter $\kappa$),
gives a sufficiently accurate description 
in both the adiabatic and antiadiabatic regimes.

For light electrons (adiabatic regime $\omega_0/t=0.2,\;0.4$; left panels), 
we found a rather narrow transition region. 
The drop of $S^{tot}$ in the crossover region $\lambda\simeq 1$
is driven by the sharp fall of the Drude weight, 
which is a measure of the coherent transport properties of a polaron. 
By contrast the optical absorption due to inelastic scattering processes,  
described by the regular (dissipative) part of the optical conductivity, 
becomes strongly enhanced around $\lambda\simeq 1$~\cite{FLW97} 
(cf. the behaviour of $S^{reg}$). 

The large to small polaron crossover is considerably broadened
for heavy electrons (non-to-antiadiabatic case $\omega_0/t=1,\;4$; 
right panels). Here $E_{kin}$ decreases more gradually and
 $S^{reg}$ exhibits a less pronounced maximum at about $g^2=1$.

As quoted above, we can calculate the Drude weight 
independently from the effective mass of the Holstein
polaron. Using this data, it is worth mentioning that 
the f-sum rule~(\ref{sumrule}) is satisfied numerically 
to at least six digits in the whole parameter regime~\cite{EBKT03}.  
\subsection{Thermally activated transport}
\label{sec:sigmaT}
If the polaron effects are assumed to be dominant the coherent 
bandwidth is extremely small. Then the physical picture is that the 
particle is trapped at a certain lattice site and that 
hopping occurs infrequently from site to site. There are two kinds
of transfer processes~\cite{Ho59b}. All phonon numbers might 
remain the same during the hop (diagonal transition) or,
alternatively, the number of phonons is changed (non-diagonal transition).
In the latter case each hop 
may be approximated as a statistically independent event
and the particle loses its phase coherence by this phonon emission 
or absorption (inelastic scattering). Diagonal and non-diagonal 
transitions show a different temperature dependence. While
the rate of diagonal (band-type) transitions decreases with 
increasing temperature, small-polaron theory predicts that the
non-diagonal (incoherent hopping) rate is thermally activated
and may become the main transport process at higher temperatures
(cf., e.g., Ref.~\cite{Mah00}). Deviations from standard 
small-polaron theory are expected to occur in the intermediate 
coupling regime. By means of ED and KPM techniques we are able to study 
the optical response of Holstein polarons precisely in this regime,
at least for small lattices. 
\paragraph{ac conductivity}
Our starting point is the Kubo formula for the electrical conductivity
at finite temperatures~\cite{Mah00},
\begin{equation}
  \sigma^{reg}(\omega;T) = \frac{\pi}{\omega N} \frac{1}{Z} 
  \sum_{m,n>0}^{\infty} \left[{\rm e}^{-\beta E_n} - {\rm e}^{-\beta E_m}
\right] 
  \,|\langle \psi_n|\hat{\jmath}|\psi_m\rangle|^2 \,
  \delta(\omega - \omega_{mn})\,,
  \label{si_1}
\end{equation}
where $Z = \sum_{n}^{\infty} \E^{-\beta E_n}$ is the partition function
and $\beta=T^{-1}$ denotes the inverse temperature ($k_B=1$).  
Since in practice the contribution of highly excited
phonon states is negligible at the temperatures of relevance, 
the system is well approximated by a truncated phonon space with at most
$M(\lambda,g,\omega_0;T)$ phonons~\cite{BWF98}. Then
$|\psi_n\rangle$ and $|\psi_m\rangle$ are the eigenstates of $H$ within our
truncated Hilbert space. $E_n$ and $E_m$ are the corresponding
eigenvalues with $\omega_{mn} = E_m - E_n$.

In order to evaluate temperature-dependent response functions
like~(\ref{si_1}), recently a generalised ``two-dimensional'' 
KPM scheme has been proposed~\cite{WWAF06,SWWAF05}, which, in our case, 
can be set up using a current operator density
\begin{equation}
 j(x,y) = \sum_{m,n} |\langle \psi_n|\hat{\jmath}|\psi_m\rangle|^2 
  \ \delta(x-E_n)\ \delta(y-E_m)\,.
\label{j_xy}
\end{equation}
For the regular part of the conductivity we obtain 
\begin{equation}
    \sigma^{reg}(\omega) = 
    \frac{2\pi}{\omega N} \frac{1}{Z}
\int\limits_{0^+}^{\infty}
    j(y+\omega,y)
    \big[{\rm e}^{-\beta y} - {\rm e}^{-\beta (y+\omega)}\big]\, dy\,,
\label{si_2}
\end{equation}
where the partition function $Z=2\int_{0^+}^{\infty} \rho(E)\exp 
(-\beta E)$ is easily obtained by integrating over the density of states
$\rho(E) = \sum_{n=0}^{D-1} \delta (E-E_n)$, which can be expanded in
parallel to $j(x,y)$ (here $D$ is the dimension of
the Hilbert space).  One advantage of this
approach is that the current operator density that enters the conductivity 
is the same for all temperatures, i.e., it needs to be expanded only
once.

\begin{figure}
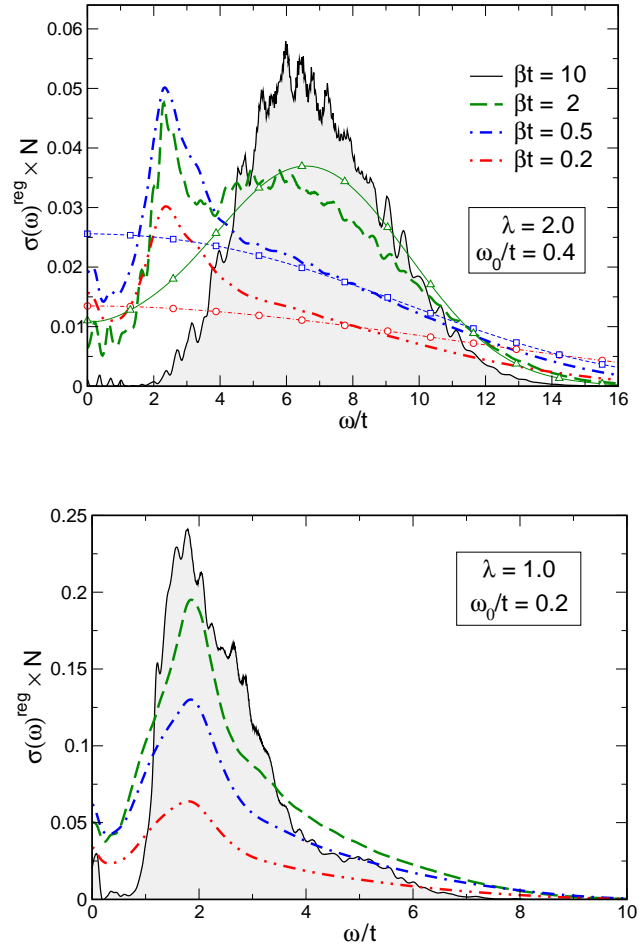

\centering\vspace*{1cm}
\hspace*{0.48cm}\includegraphics[width=0.75\textwidth]{sigma_w04_T.eps}\\[1cm]
\includegraphics[width=0.7\textwidth]{sigma_w02_T.eps}\\[.5cm]
\caption{Optical absorption by Holstein polarons at finite temperatures
in the adiabatic strong (upper panel) and intermediate (lower panel)  
EP coupling regime. Results are obtained by ED for a $N=6$ site lattice
with $M=45$ phonons. In the upper panel, 
thin lines with symbols give the analytical results  
    for the small polaron transport~\cite{BB85,Mah00}
    at temperatures $\beta t=2$ (triangles), $0.5$ (squares), $0.2$ (circles). 
    The deviations observed for high excitation energies
    at very large temperatures are caused by the necessary 
    truncation of the phonon Hilbert space in ED.}
\label{f:sigmaT}       
\end{figure}

Figure~\ref{f:sigmaT} gives results for the 
finite-temperature optical conductivity of a Holstein polaron. 
Coherent transport related to diagonal transitions within 
the lowest polaron band is negligible at high temperatures. For
instance, the amplitude of the current matrix elements between the
degenerate states with momentum $K=\pm\pi/3$ ($K=0,\, \pm\pi/3,\,
\pm2\pi/3$, and $\pi$ are the allowed wave numbers of our 6-site system
with periodic boundary conditions) is of the order of $10^{-7}$ only.  
In the small polaron limit , where the polaronic sub-bands 
are roughly separated by the bare phonon frequency, non-diagonal
transitions become important for $T > \omega_0$.  
Let us consider the activated regime in more detail  
(see Fig.~\ref{f:sigmaT} (upper panel)). 
With increasing temperatures we observe a substantial
spectral weight transfer to lower frequencies, and 
an increase of the zero-energy transition probability 
in accordance with previous results~\cite{Na63}.  

In addition, we find a strong resonance
in the absorption spectra at about $\omega_0\sim 2t$, which can be easily
understood using a configurational coordinate picture~\cite{SWWAF05}.
In order to activate these transitions thermally, the electron has to
overcome the ``adiabatic'' barrier $\Delta=E_{1+}-E_0=\varepsilon_p/2-t$,
where we have assumed that the first relevant excitation is a state
with lattice distortion spread over two neighbouring sites and 
the particle mainly located at both these sites 
(in a symmetric ($+$) or antisymmetric ($-$) linear combination;
$E_{1,\pm} =\mp t-\varepsilon_p/2$). A
finite phonon frequency will relax this condition. From
Fig.~\ref{f:sigmaT}, we find the signature to occur above
$T > 0.5t$. Obviously this feature is absent in the 
standard small-polaron transport description 
which essentially treats the polaron as a quasiparticle 
without resolving its internal structure.

Now let us decrease the EP coupling strength $\lambda$ 
keeping $g^2=10$ fixed. Results for the optical response in the vicinity of 
the large to small polaron crossover 
are depicted in the lower panel of Fig.~\ref{f:sigmaT}.
Here the small polaron maximum has almost disappeared and the
$2t$-absorption feature can be activated at very low temperatures
($\Delta \to 0$ for the two-site model with $\lambda=1$).  
The gap observed at low frequencies and temperatures is clearly a
finite-size effect. The overall behaviour of $\sigma^{reg}(\omega;T)$ 
resembles that of polarons of intermediate size. 
At high temperatures these polarons will dissociate
readily and the transport properties are equivalent to those of
electrons scattered by thermal phonons. Let us emphasise that
many-polaron effects become increasingly important in the
large-to-small polaron transition region~\cite{Hoea05} 
(see also Sec.~\ref{sec:mp} below). As a result,
polaron transport might be changed entirely compared to the
one-particle picture discussed so far.

\paragraph{dc conductivity and thermopower}

We consider dc transport, 
or  $\sigma (\omega )$ in the limit $\omega \rightarrow 0$.
For simplicity, we consider only a single polaron, or a dilute system of polarons
where interactions can be neglected and bipolaron formation is prevented,
as by a large repulsive $U$.  We also neglect impurities, which can
localise or scatter a polaron.

At zero temperature, the conductivity or mobility of a polaron is infinite.
The polaron can be placed in a state of nonzero momentum by a weak electric field
acting for a short time.  This is an eigenstate, which carries current forever
and never decays.  At small temperatures $T \ll \omega _0 $, an exponentially small
number of phonons are thermally excited.  The conductivity becomes finite
due to scattering of a polaron off thermally excited phonons of density
$n_{ph} \sim e^{- \omega _0 / T}$.  The details depend on the EP scattering
process.

In 1D, when a polaron of momentum $k$ encounters a thermally excited phonon, in general
part of it is transmitted and part is backscattered.  Certain anomalies occur.  For example,
in the limit of small hopping $t$, as $g$ approaches 1, the backscattering of the polaron vanishes  and
the phonon is simultaneously transferred one site in the direction opposite the polaron momentum.  
The phonon thus recoils opposite to the direction expected, 
cf.  the collision of two balls.  This leads to a heat current in the opposite
direction as the polaron particle current, which should be observable in the thermopower.
A polaron-thermal phonon bound state also exists for sufficiently large $g$.  For this bound state, heat (a phonon excitation)
can be transported by an electric field, which again should be observable as a large contribution to the thermopower
of the opposite sign as the above.  For large $g$, this bound state or internal polaron excited state can have
a much smaller effective mass than the polaron ground state.  Perhaps surprisingly, as the temperature
increases, the polaron effective mass as measured by the 
low-frequency ac conductivity can decrease.

We next consider very high temperatures.  As $T$ increases, the typical phonon displacement increases
as $ \tilde K \langle \hat{x}^2 \rangle = k_B T $, 
where $\tilde K$ is the phonon spring constant.  For quasi-static phonons
(large phonon mass), this leads to a disorder potential for the electron 
that increases without bound
as $T$ increases.  The disorder Anderson localises the electron, leading to zero dc conductivity.
The disorder, however, is not quite static, and rearranges itself on a timescale $\tau \sim 1/ \omega _0$.
Once every time of order $\tau$, the diagonal energies of the electron site and a neighbouring site
become equal, and the electron can hop to a neighbouring site.  It is then diffusing with a 
diffusion constant  $\sim a^2 \omega _0$, where $a$ is the lattice constant.  Using the Einstein relation
relating diffusion and mobility, the high temperature resistivity becomes
\begin{equation}
\rho = {\pi k_B T \over n e^2 a^2 \omega_0} .
\label{eq:rho}
\end{equation}
The high temperature resistivity is metallic, i.e. $d \rho / dT > 0$, and can greatly
exceed the Ioffe-Regel limit.
Numerical studies to confirm or refute this scenario are incomplete.

\section{From few to many polarons} 
\label{sec:mp}

Let us now address the important issue of how the character of the 
(polaronic) quasiparticles may change if we increase the carrier density
$n=N_e/N$.   Consider first the case of zero electron-electron interaction.
Beginning with a noninteracting Fermi gas at $T=0$,
as the Holstein EP interaction $g$ is increased from zero,
a singlet superconductor is expected to form.  
As $g$ increases, the diameter of the Cooper pair decreases.
Eventually, the Cooper pair diameter becomes smaller than
the distance between Cooper pairs, and the behaviour crosses over
from BCS superconductivity to that of Bose condensation,
like that of $^4$He, where the hard core bosons are bipolarons
(bound states of two polarons).  In this limit, $T_c$ is
given approximately by the Bose condensation
temperature for ideal bosons of mass $m^*$, where $m^*$ is the bipolaron mass.
The limit of Bose condensation of bipolarons is not given correctly
by Eliashberg theory, which describes strong coupling, but not that strong.

\subsection{Bipolaron formation}
We investigate how two electrons coupled to phonons
may bind together to form a bipolaron, including the bipolaron 
effective mass, the crossover between two different types of 
bound states, and the dissociation into two polarons (see also~\cite{Ale,Aub}).
For problems with more than one electron, the Holstein 
Hamiltonian is generalised by adding a 
Hubbard electron-electron interaction term,
$U\sum_{j}n_{j\uparrow}n_{j\downarrow}$.
Basis states  for the many-body Hilbert  space can be written 
$\vert b \rangle = |j_1,j_2;  \dots ,n_{m}, n_{m+1}, \dots, \rangle$, 
where the up and down electrons are on sites 
$j_1$ and $j_2$, and there are $n_m$ phonons on site $m$.  
In a generalisation of the one electron VED method described above, 
a bipolaron variational
space is constructed beginning with an initial state where both
electrons   are  on the same   site   with no phonons,  and  operating
repeatedly  ($L$-times)   with   the off-diagonal   pieces ($t$ and $\bar g$)  
of  the Hamiltonian.  All translations of these
states are included on an infinite lattice.
The method is very
efficient in the intermediate coupling  regime, where  it provides
results that are variational in  the thermodynamic limit 
and bipolaron energies
that are accurate to  7 digits for the  case $L=18$ and size
of the Hilbert space $N_{st}=2.2 \times 10^6$ phonon and down
electron configurations 
for a given up electron position.  In 1D the size of the variational space
approximately doubles as $L$ is increased by one,
which is the same as for the one electron problem, although the
prefactor for two electrons is larger \cite{BKT00}.

For large phonon frequency $\omega _0$, the EP interaction leads to a
non-retarded attractive on-site interaction of strength $U_0 \equiv 2 \omega _0 g^2$.
One would expect that as the Hubbard repulsion $U$ becomes larger than this value,
the bipolaron would dissociate into two polarons.  As can be shown both analytically
and numerically, this is not what happens.  In the limit of small hopping $t$,
as $U$ exceeds $U_0$, the bipolaron crosses over from a state
S0 with both electrons primarily on the same site, to another bound state
S1 with the electrons primarily on nearest neighbour sites.  
Only for $U > 2 U_0$ does the bipolaron dissociate into two polarons.
The crossover from $S0$ to $S1$ bipolarons is important  in theories
of bipolaronic superconductivity applied to real materials,
since S1 bipolarons are generally orders of magnitude lighter than S0
bipolarons.   Since the superconducting $T_c$
in the dilute limit is inversely proportional to the effective mass,
the S1 regime usually provides a more compelling theory.

\begin{figure}
\centering
\includegraphics[width=0.7\textwidth, angle=0]{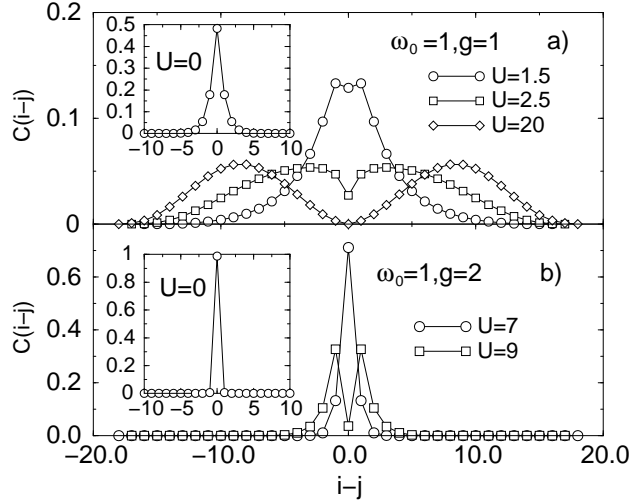}
\caption{Electron-electron correlation function $C(i-j)$ calculated at
$\omega _0=1$, a) $g=1$    and b)  $g =  2$   for different values of
$U$, with $L=18$. The two ordinate axes have a different range. Insets show
results for $U=0$. All curves are normalised, $\sum_i  C(i)=1$. }
\label{fig:corel}
\end{figure}

We now discuss numerical variational results for the singlet bipolaron
on an infinite 1D lattice.
We have been unable to demonstrate the existence of a bound triplet bipolaron for
the Holstein-Hubbard model.
Fig.~\ref{fig:corel} 
shows the ground state electron-electron density correlation
function $C(i-j)=\langle\psi_0\vert n_i n_j\vert\psi_0\rangle$, where
$n_i=n_{i\uparrow}+n_{i\downarrow}$
and $\vert \psi_0\rangle$ denotes the ground state wave function. 
At $g=1$, the bipolaron widens with
increasing $U$ and transforms into two unbound polarons
(which can only move a finite distance apart
in the variational space). The value
$U=1.5$ is below the transition to the unbound state at
$U_c=2.17$, calculated by comparing the polaron and bipolaron energies.
We see that the probability of electrons occupying the
same or neighbouring sites is almost equal.  In the unbound regime, the
nature of the correlation function changes significantly. 
At $U=1.5$, $C(i)$ falls off exponentially, while for $U > U_c$ the typical
distance between electrons is the order of the maximum allowed
separation $L$.  The electrons can be no farther apart than $L$ in the variational
space, although their centre of mass can be anywhere on an infinite lattice.
A state of separated polarons is clearly seen for $U=20$.

Two distinct regimes are seen at $g=2$ within the bipolaronic
region.  At $U=7 < U_0 \equiv 2 \omega _0 g^2 = 8$, 
the correlation function represents the S0
bipolaron, while at $U=9>U_0$ we find the largest probability for two
electrons to be on neighbouring sites, which is characteristic of
the S1 bipolaron.  In contrast to previous calculations
where phonons were treated classically \cite{PA98}, we find 
a crossover rather than a 
phase transition between the two regimes.  
Moving from strong towards intermediate
coupling, the S0 and S1 bipolarons consist of longer exponential
tails extending over many lattice sites, and the two regimes can
no longer be distinguished.
The precision  of  presented correlation  functions  in the bipolaron
regime   is  within the  size  of  the  plot symbols  in the  thermodynamic limit. 

\begin{figure}
\centering
\includegraphics[width=0.7\textwidth, angle=0]{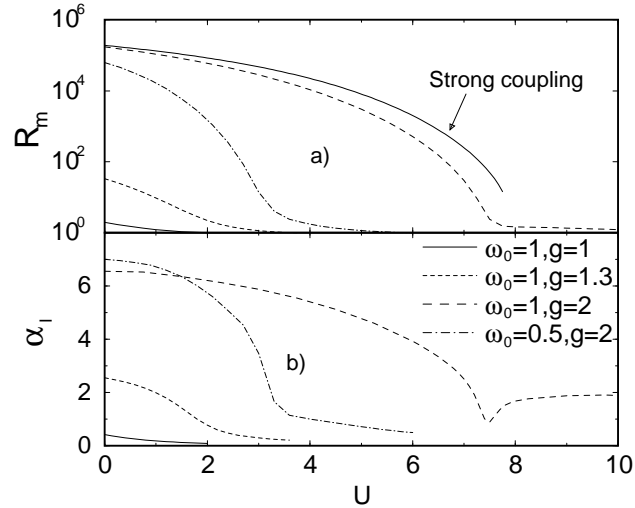}
\caption{a) The mass ratio  
$R_m=m_{bi}/2m_{po}$ vs.  $U$ and b) the bipolaron isotope effect
$\alpha_I$ vs. $U$.  Numerical results are for $L=18$.
Results for $R_m$ at $\omega _0 =0.5$ are obtained 
by extrapolating $L \to \infty$.  Precision in all curves is within
the line-width in the
thermodynamic limit, except for
$\alpha_I$ with $\omega _0=0.5$, where the error is estimated to be
$\pm 5 \%$. The thin line in (a) is 
the strong coupling expansion result for $\omega _0=1$, $g=2$.  
Polaron masses in units of
the noninteracting electron mass are $m_{po}=1.35, 1.76, 10.4, 3.06$ 
from top to bottom respectively. }
\label{fig:mass_bip}
\end{figure}

Figure~\ref{fig:mass_bip}a plots the  bipolaron  mass
ratio $R_m=m_{bi}/2m_{po}$  vs.  $U$   for different  values   of
$\omega _0$  and  $g$.  In all cases presented in Fig.~\ref{fig:mass_bip},
$R_m$ approaches 1 as $U$ approaches $U=U_c$ in agreement with a
state of  two free polarons.  At  fixed  $\omega _0=1$ the bipolaron mass ratio
increases by several orders of magnitude with increasing $g$ at $U=0$.
Increasing $U$ sharply decreases $R_m$ in the S0 regime.
Note that the mass scale is logarithmic. In the S1
regime with $U>U_0$, $R_m$ is small, as predicted by the strong
coupling result.

In the dilute bipolaron regime, the bipolaron isotope effect is the same as the
classic superconductivity isotope effect for $T_c$.
The bipolaron isotope effect, shown in Fig.~\ref{fig:mass_bip}b, 
is large in the
strong coupling ($\omega _0=1,g=2$) and small $U$ regime, where its value
is somewhat below the large $g$ strong coupling prediction 
$\alpha_{I,S0}\sim 2 g^2 - {1 \over 4} = 7.75$.  
With increasing $U$, $\alpha _I$ decreases and in the S1 regime
approaches $\alpha_{I,S1}=g^2/2= 2$. A kink is observed in the crossover
regime. 
With decreasing $g$ or $\omega _0$, $\alpha _I$ also decreases.

\begin{figure}
\centering
\includegraphics[width=0.7\textwidth, angle=0]{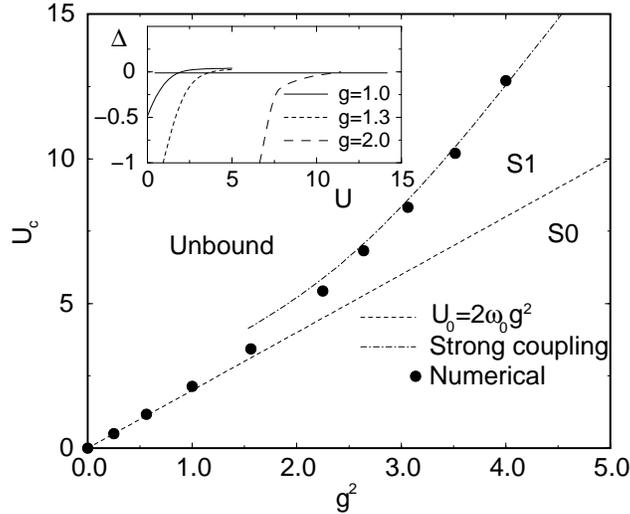}
\caption{Phase diagram and binding energy $\Delta$ in units of $t$ 
(inset) calculated  at
$\omega _0=1$.  Numerical results
are circles.  For  greater accuracy, results near
the weak and strong  coupling regime were obtained 
by extrapolating $L \rightarrow \infty$.}
\label{fig:diag_bip}
\end{figure}

The  phase diagram $U_c(g)$  is shown in
Fig.~\ref{fig:diag_bip} at fixed $\omega _0 = 1$. 
Numerical results, shown as circles, indicate the phase boundary
between two dissociated polarons each having energy $E_{po}$ and a
bipolaron bound state with energy $E_{bi}$. In the inset of
Fig.~\ref{fig:diag_bip} we show the bipolaron binding energy  defined
as $\Delta = E_{bi}-2 E_{po}$. The phase diagram is obtained from $\Delta=0$.
The dashed line,
given by  $U_0=2\omega _0 g^2$, is a  reasonable estimate for the
phase  boundary at small $g$.   At large  $g$  the dashed line roughly
represents  the crossover between a
massive S0 and lighter S1 bipolaron.  
The  S1 region grows with increasing  $g$.  
The   dot-dashed  line is the phase   boundary
between  S1 and the unbound polaron phase,   as  obtained by
degenerate  strong  coupling  perturbation theory.  Numerical  results
approach  this line at  larger  $g$.  The  dot-dashed  line asymptotically
approaches $U_c=4\omega _0 g^2$.  

\subsection{Many-polaron problem}
We consider how polarons evolve from the dilute to the concentrated limit,
in the regime where spinless or fully spin-polarised fermions prevent
bipolaron formation.  In 1D with open boundary conditions, the spinless fermion
problem is equivalent to infinite Hubbard $U$.
While for very strong EP coupling no significant changes 
are expected due to the existence of rather independent
small (self-trapped) polarons with negligible residual interaction
(assuming spinless fermions or strong enough electron-electron 
repulsion to prevent bipolaron formation), a
density-driven crossover from a state with large polarons to a metal with
weakly dressed electrons should occur in the intermediate-coupling 
regime. This issue has recently been investigated 
theoretically by ED~\cite{CGS99}, QMC~\cite{Hoea05}, and 
variational canonical transformation~\cite{LHF06} methods,
and is known to be of experimental relevance, e.g., in 
${\rm La_{2/3}(Sr/Ca)_{1/3}MnO_3}$ films~\cite{HMDLK04}.

In the spinless fermion Holstein model, the above-mentioned density-driven
transition from large polarons to weakly EP-dressed
electrons is expected to be possible only in 1D, where large polarons exist
at intermediate coupling. The situation is different for
Fr\"{o}hlich-type models~\cite{Fr54,AK99,DT01} with long-range EP
interaction, in which large-polaron states exist even for strong coupling and
in $D>1$.

To set the stage, we first comment on the evolution of 
the one-electron spectral function $A(k,\omega)$ with increasing 
electron density $n$ in the weak- and strong-coupling 
limiting cases~\cite{Hoea05}.   
In the former the spectra bear a close resemblance to 
the free-electron case for all $n$, i.e., there is a strongly
dispersive band running from $-2t$ to $2t$, which can be attributed to weakly
dressed electrons with an effective mass close to the non-interacting value.
As expected, the height (width) of the peaks increases (decreases)
significantly in the vicinity of the Fermi momentum.  In the opposite
strong-coupling limit the spectrum exhibits an almost dispersionless 
coherent polaron band $\forall\; n< 0.5$. Besides, there are two incoherent 
features located above and below the Fermi energy, 
broadened $\propto \varepsilon_p$, which are due to phonon-mediated 
transitions to high-energy electron states.  
The most important point, however, is the clear separation  
of the coherent band from the incoherent parts even at large $n$,
indicating that small polarons are well-defined quasiparticles in the 
strong-coupling regime, even at high carrier density.

\begin{figure}\centering
\includegraphics[width=1.0\linewidth]
{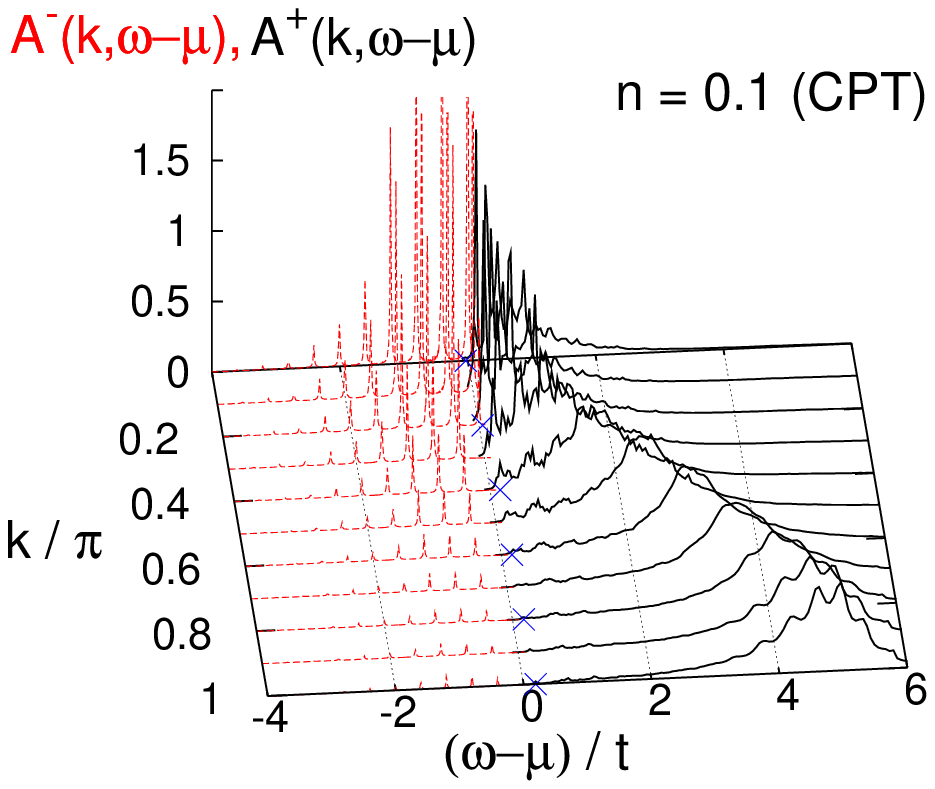}\\[0.5cm]
\includegraphics[width=1.0\linewidth]
{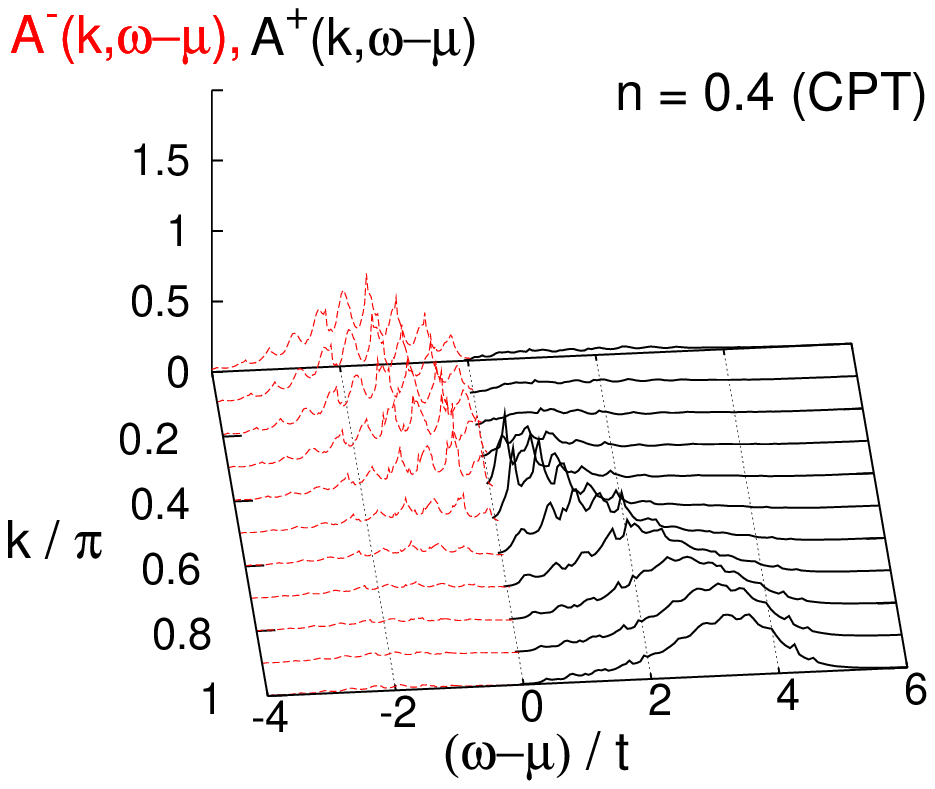}\\
\caption{Single particle spectral functions $A^-(k,\omega)$ (red dashed
  lines) and $A^+(k,\omega)$ (black solid lines) for two characteristic 
band fillings, $n=0.1$ and $n=0.4$, at $\omega_0/t=0.4$ 
and $\lambda=1$ ($T=0$). Results are obtained by CPT using $N_c=10$).
Blue crosses track the small-polaron band determined by ED.}
\label{f:manypol_akw_density}
\end{figure} 

Figure~\ref{f:manypol_akw_density} displays the inverse photoemission
[$A^+(k,\omega)$] and photoemission spectra [$A^-(k,\omega)$] at
intermediate EP coupling strength, determined by CPT.  
At low densities, $n=0.1$ (upper panel), we can easily identify a (coherent) 
polaron band crossing the Fermi energy level $E_F=\mu(T\to 0)$,
the latter being situated at the point where $A^-$ and $A^+$ intersect.
This large-polaron band has rather small electronic spectral weight especially
away from  $E_F$ and flattens at large $k$, as known from single-polaron 
studies (see Sec.~\ref{sec:sp}, Fig.~\ref{f:cpt_elecspe}). 
Below this band, there exist equally spaced phonon satellites, 
reflecting the Poisson distribution of phonons in the ground state.  
Above $E_F$ there is a broad dispersive incoherent feature 
whose maximum closely follows the dispersion relation of free particles.

As the density $n$ increases, a well-separated coherent polaron band 
can no longer be identified. At about  $n\simeq 0.3$ the deformation 
clouds of the (large) polarons start to overlap leading to a mutual
(dynamical) interaction between the particles. Increasing the carrier
density further, the polaronic quasiparticles dissociate, 
stripping their phonon cloud. This is the case shown in the lower
panel of Fig.~\ref{f:manypol_akw_density}.   
Now diffusive scattering of electrons and phonons 
seems to be the dominant interaction mechanism. 
As a result both the phonon peaks in $A^-(k,\omega)$ and 
the incoherent part of $A^+(k,\omega)$ are washed out, the spectra
broaden and ultimately merge into a single wide band. 
Most notably, the incoherent excitations now lie arbitrarily 
close to the Fermi level. Obviously the low-energy physics of the 
system can no longer be described by single-particle small-polaron theory.
\section{Polaronic effects in strongly correlated systems}
\label{sec:sc}
The interplay of electron-electron and electron-phonon interactions
in the formation of dressed quasiparticles is becoming the focus of 
attention in many contexts, including conducting polymers, 
ferroelectrics, halide-bridged transition-metal chain complexes,
and several important classes of perovskites. Especially research 
on high-${\rm T_c}$ superconductivity (HTSC) and colossal 
magnetoresistance (CMR) 
has spurred intense investigations of the competition or, if possible, 
of the cooperation of these two fundamental interactions
(for a recent review see~\cite{Eg06}, and references therein). 

Many experiments have indicated substantial EP interaction in the 
high-${\rm T_c}$ cuprates. The relevance of EP coupling can be seen 
from the experimental observation of phonon 
renormalisation~\cite{Ri94}.  
Ion channelling~\cite{SRBL89,Hagea90}, neutron scattering~\cite{Egea91} 
and photo-induced absorption measurements~\cite{Kiea88} 
proved the existence of large anharmonic lattice fluctuations, 
which may be responsible for 
local phonon-driven charge instabilities in the planar $\rm CuO_2$
electron system~\cite{Jo91,ZS92}. 
Photo-induced absorption experiments~\cite{MFVH90},
infrared spectroscopy~\cite{Caea94} and reflectivity 
measurements~\cite{FLKB93} indicate the formation of small polarons 
in the insulating parent compounds 
$\rm La_2CuO_{4+y}$ and $\rm Nd_2CuO_{4-y}$ of the
hole- and electron-doped superconductors $\rm La_{2-x}Sr_xCuO_{4+y}$
and $\rm Nd_{2-x}Ce_xCuO_{4-y}$, respectively. 
Recently angle-resolved photoemission spectroscopy
data were interpreted in terms of strong EP coupling
giving rise to self-localisation of holes (hole polarons)~\cite{Roea05}. 
Based on these experimental findings, several theoretical
groups~\cite{VPD91,Emi92b,Emi94,AK92,AR92b,ABMS93,Hi93,Mot93,AM94,Ra94,ICNC95}
promote a (bi)polaronic scenario for HTSC.

Even stronger evidence for polaron formation in doped 
charge transfer oxides is provided by experiments on 
the nickelates  $\rm La_{2-x}Sr_xNiO_4$~\cite{BE93,CCC93}. 
The isostructural compounds $\rm La_2CuO_4$ and $\rm La_2NiO_4$
show a remarkable difference upon the substitution of
La by Sr. Both materials become metallic
upon doping, but in the nickelates a nearly total substitution
of La for Sr is necessary. Also in 
$\rm La_{2-x}Sr_xNiO_{4+y}$ no superconductivity is found for
any $x$. A resolution of this problem might be given by
extended LDA calculations~\cite{AKZA92}, which show
that the nickelates are much more susceptible to a breathing
polaron instability than the cuprates. 
The reason is the much stronger magnetic confinement
effect of additional holes and nickel spins. These low--spin composite
holes are nearly entirely prelocalised and the EP coupling becomes
much more effective in forming polarons.
For the composition $\rm La_{1.5}Sr_{0.5}NiO_4$ 
(quarter filling, $x=0.5$), electron diffraction measurements 
show a commensurate superstructure spot at the $(\pi,\pi)$-point, 
which has been interpreted as a sign of truly 2D ordering 
of breathing-type polarons, i.e., as a polaronic 
superlattice. 

Localised lattice distortions are also suggested to play
an important role in determining the electronic and magnetic
properties of hole-doped manganese oxides of the form 
$\rm La_{1-x}[Sr,Ca]_xMnO_3$~\cite{ZCKM96,Mi98}. 
In the region $x_{MI}\sim 0.2<x<0.5$ these compounds 
show a transition from a metallic ferromagnetic 
low-temperature phase to an insulating 
paramagnetic high-temperature phase associated 
with a spectacularly large negative magnetoresistive 
response to an applied magnetic field~\cite{JTMFRC94}. 
Both breathing-mode collapsed ($\rm Mn^{4+}$) and (anti) 
Jahn-Teller distorted ($\rm Mn^{3+}$) sites are created 
simultaneously when the holes are localised in passing 
the metal-insulator transition~\cite{AP99,BPPSK00}.  
The relevance of small polaron transport above $T_c$ is obvious from
the activated behaviour of the conductivity~\cite{WMG98}. 
Consequently many theoretical studies focused on polaronic
approaches~\cite{AB99,Caea97,LM97,Quea98,KJN98,Mu99,Deea99,WLF03}. 
Polaronic features have been established by a variety of experiments. 
For example, high-temperature thermopower~\cite{JSRTHC96,PRCZSZ97} 
and Hall mobility measurements~\cite{JHSRDE97} confirmed the polaronic 
nature of charge carriers in the paramagnetic phase. More directly the 
existence of polarons has been demonstrated by atomic pair 
distribution~\cite{BDKNT96}, x-ray and neutron scattering 
studies~\cite{SWKT99,VROSLMSPFM99,Daea00}.
Interestingly it seems that the charge carriers partly retain their 
polaronic character well below $T_c$, as proved, e.g.,  
by neutron pair-distribution-function analysis~\cite{LEBRB97} and
resistivity measurements~\cite{ZSPK00}.     

Regardless of whether the EP coupling acts as a secondary 
pairing interaction for HTSC in the cuprates, is responsible 
for the charge ordering in the nickelates, or triggers
the CMR phenomenon in the manganites, EP and particularly 
polaronic effects need to be reconsidered for the case of
strong electronic correlations realised in these materials.    
For instance, Coulomb or spin exchange interactions may lead  
to a ``prelocalisation'' of the charge carriers.  
Then a rather weak EP coupling can cause polaronic band narrowing 
and that way might drive the system further into the strongly 
correlated regime. In the remaining part of this section we will 
keep track of this problem and present some exact 
results for composite spin/orbital-lattice polarons.
\subsection{Hole polarons in the Holstein $t-J$ model} 
\label{sec:htj}
Electronic motion in weakly doped Mott insulators like the HTSC 
cuprates is determined by the constraint of no double occupancy
of sites and antiferromagnetic exchange  
between nearest-neighbour spins. The generic model studied in this
context is the 2D $t-J$ Hamiltonian, 
\begin{equation}
H_{tJ}=-t \sum_{\langle i j \rangle \sigma} 
\Big(\tilde{c}_{i\sigma}^\dagger 
\tilde{c}_{j\sigma}^{} + {\rm H.c.}\Big)+ J \sum_{\langle i j\rangle}
\Big({\vec S}_i^{}{\vec S}_j^{} - \frac{1}{4}\tilde{n}_i^{}\tilde{n}_j^{}\Big)\,,
\label{tjm}
\end{equation}
acting in a projected Hilbert space, i.e. $\tilde{c}_{i\sigma}^{(\dagger)}=
c_{i\sigma}^{(\dagger)}(1-\tilde{n}_{i,-\sigma}^{})$,  
$\tilde{n}_{i}=\sum_{\sigma}
\tilde{c}_{i\sigma}^{\dagger}\tilde{c}_{i\sigma}^{}$, and 
${\vec S}_i=\sum_{\sigma,\sigma^{\prime}}\tilde{c}_{i\sigma}^{\dagger}
{\vec \tau}^{}_{\sigma\sigma^{\prime}}\tilde{c}^{}_{i\sigma^{\prime}}$.
Within the $t-J$ model the bare transfer amplitude of electrons ($t$) sets 
the energy scale for incoherent transport, while the Heisenberg 
interaction ($J$) allows for spin flips leading to coherent 
hole motion at the bottom of a band with an effective 
bandwidth determined by $J$. $J<t$ corresponds to the 
situation in the cuprates, e.g. $J/t\simeq 0.4$ 
with $t\simeq 0.3$~eV is commonly used to model
the quasi-2D $\rm La_{2-x}Sr_xCuO_4$ system.   

In order to study polaronic effects in systems 
exhibiting besides strong antiferromagnetic exchange a 
substantial EP coupling the Hamiltionian~(\ref{tjm}) 
is often supplemented by a Holstein-type interaction 
term
\begin{equation}
H=H_{tJ}- \sqrt{\varepsilon_p\omega_0}  
\sum_i \big(b_i^{\dagger} + b_i^{} \big)\,
\tilde{h}_i^{}
+\omega_0 \sum_i \big(b_i^{\dagger} b_i^{} + \frac{1}{2}\big)
\label{htjm}
\end{equation}
($\tilde{h}_i=1-\tilde{n}_i$ 
denotes the local density operator of the spinless hole).
The resulting Holstein $t-J$ model (HtJM)~(\ref{htjm}) 
takes the coupling to the hole as dominant source 
of the particle-lattice interaction. In the cuprate context an  
unoccupied site, i.e. a hole, 
corresponds to a Zhang-Rice singlet (formed 
by $\rm Cu \,3d_{x^2-y^2}$ and $\rm O\, 2p_{x,y}$ hole orbitals)
for which the coupling should be much stronger than for the occupied
($\rm Cu^{2+}$) site~\cite{FRMB93,Fer94,SS95}.  
The hole-phonon coupling constant is denoted by $\varepsilon_p=g^2\omega_0$,
and $\omega_0$ is the bare phonon frequency of an internal vibrational 
degree of freedom of lattice site $i$.

The changes of the quasiparticle properties due to the 
combined effects of hole-phonon/magnon correlations are expected to be 
very complex and as yet there exist no well-controlled 
analytical techniques to address this problem.  Naturally such a dressed 
hole quasiparticle will show the characteristics of both ``lattice'' 
and ``magnetic'' (spin) polarons.

We solved the Holstein $t-J$ Hamiltonian on finite lattices using ED, KPM
and a phonon Hilbert space truncation method.
To control our truncation  procedure we carefully checked 
the ground-state energy and the weight of the $m$-phonon states 
($|c^m|^2$) in the ground state as a function of the number of phonons 
retained ($M$). Convergence is assumed to be achieved 
if the relative error of both $E_0(M)$ and $|c^m|^2$ is less 
than $10^{-5}$. 
\begin{figure}
  \centering
\includegraphics[width=0.6\linewidth]{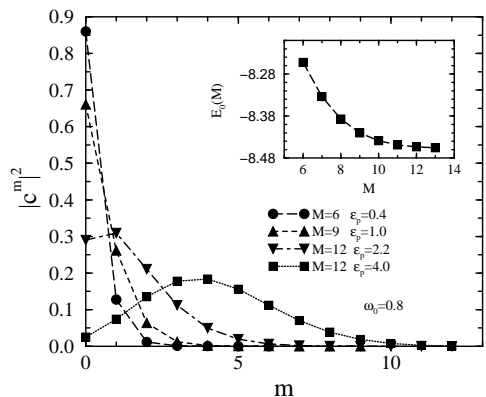}
\caption{Phonon-weight function  $|c^m|^2$ and 
ground-state energy $E_0(M)$ for the 2D HtJM with 
$J=0.4$ (throughout this section all energies 
will be measured in units of $t$).}
\label{f:pdis}
\end{figure}

Figure~\ref{f:pdis} shows $|c^m|^2$  for the single-hole case 
at weak, intermediate and strong EP couplings.    
The curves $|c^m|^2$
are bell-shaped and their maxima 
correspond to the most probable number of phonon quanta. 
The importance of multi-phonon states becomes apparent
especially in the adiabatic strong-coupling regime 
$\varepsilon_p\gg t,\; \omega_0$.  

\begin{figure}
  \centering
\includegraphics[width=0.8\linewidth]{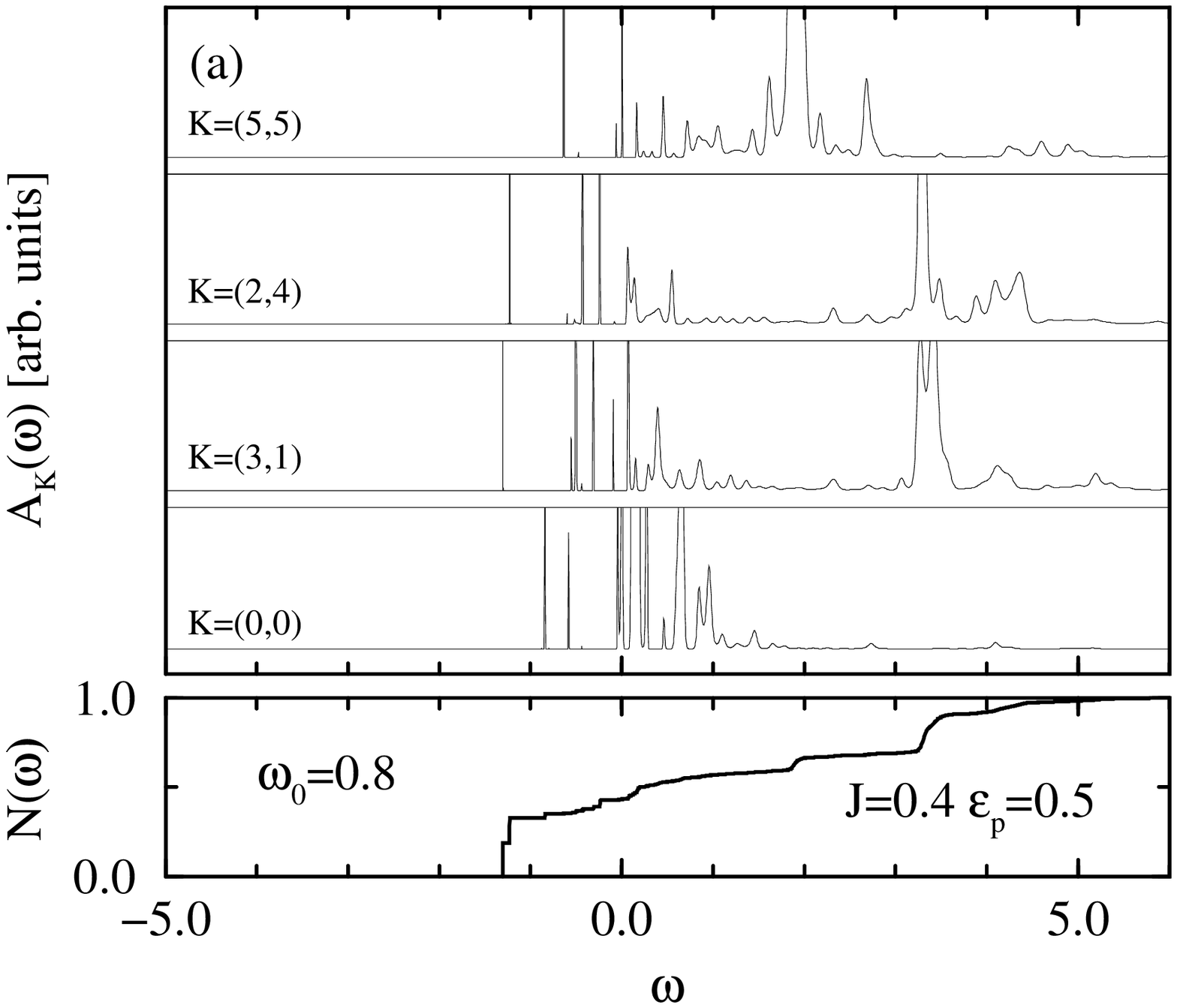}\\
\includegraphics[width=0.8\linewidth]{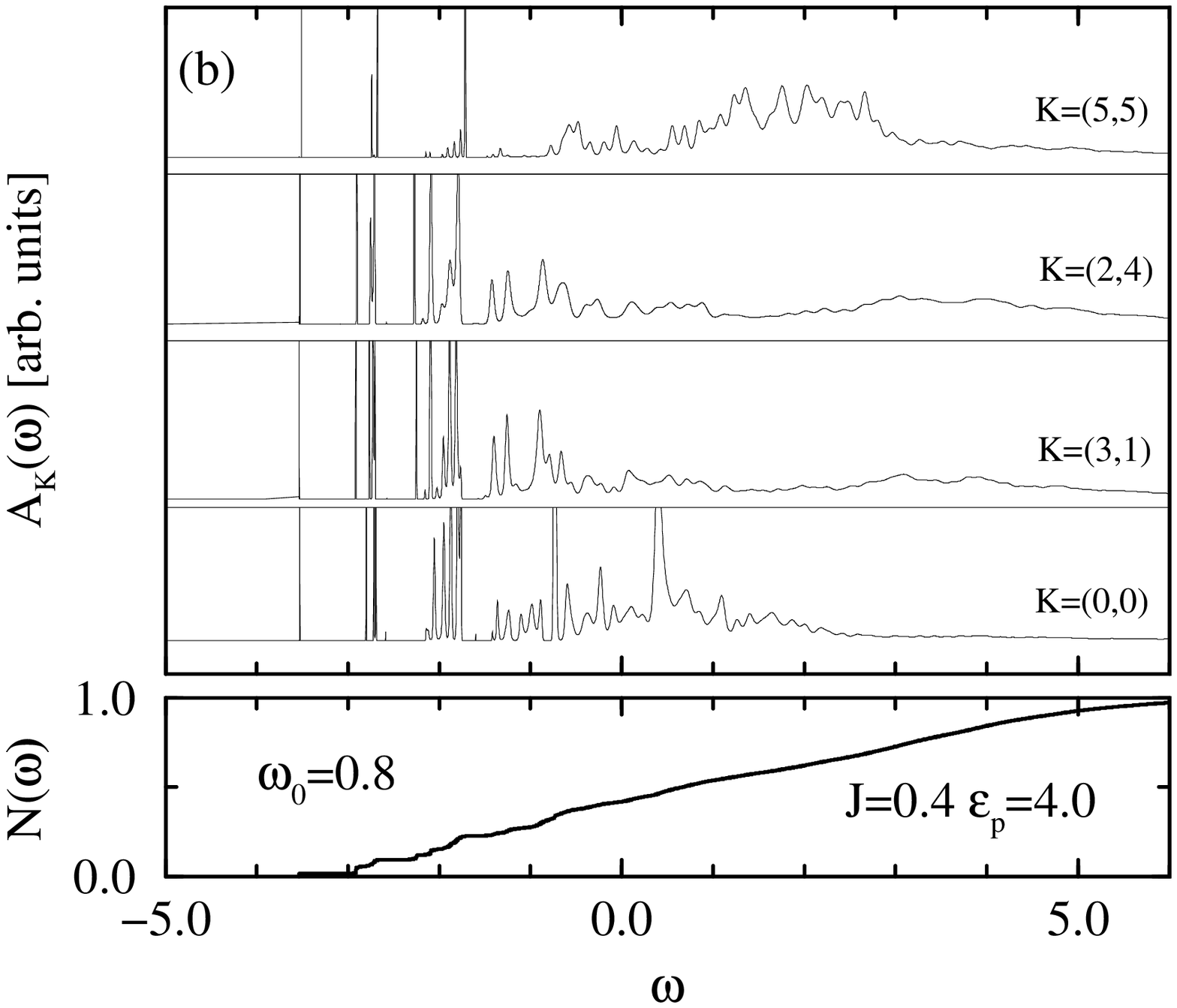}
\caption{Wavevector resolved single-hole spectral functions 
$A_{\vec K}(\omega)$ and integrated spectral weight $N(\omega)$ for 
the 2D Holstein $t-J$  model at weak (a) and strong (b) EP coupling. 
The results, taken from~Ref.~\cite{BWF98}, were obtained for 
a tilted $\sqrt{10}\times\sqrt{10}$ 
cluster with periodic boundary conditions  
(${\vec K}$-vectors are given in units of $(\pi/5,\pi/5)$). 
}
\label{f:akhtj}
\end{figure}

Let us start the analysis of the 2D HtJM with a discussion of the 
single-hole spectral function
\begin{equation}
A_{{\vec K}}(\omega)=\sum_{n,\sigma} 
\left|\langle {\mit \psi}_{n,{\vec K}}^{(N_e-1)}
|\tilde{c}^{}_{{\vec Q}-{\vec K},\sigma}| {\mit \psi}_{0,{\vec Q}}^{(N_e)}\rangle
\right|^2\;
\delta\left[\omega-(E_{n,{\vec K}}^{(N_e-1)}
-E_{0,{\vec Q}}^{(N_e)})\right]\,.
\label{akw}
\end{equation}
Figure~\ref{f:akhtj} displays $A_{{\vec K}}(\omega)$ for 
the allowed (nonequivalent) momenta ${\vec K}$ of a ten-site square lattice. 
To visualise the intensities (spectral weights) connected with the various 
peaks in each ${\vec K}$-sector we have also shown 
the integrated density of states
\begin{equation}
N(\omega)=\int_{-\infty}^{\omega}d\omega^{\prime}\frac{1}{N}
\sum_{\vec{K}}A_{{\vec K}}(\omega^{\prime})\,.
\label{nw}
\end{equation}  

In the absence of EP coupling, of course, we reproduced 
the single-particle spectrum of the pure $t-J$ model~\cite{Da94}.
Here one observes a quasiparticle pole corresponding directly to the 
coherent single-hole ground state (having momentum $(3\pi/5,\pi/5)$
on a ten-site lattice) separated by a pseudogap of 
about $J$ from the lower edge of a broad incoherent continuum being
$\simeq 6t$ wide. 
In the weak coupling regime the mass renormalisation of 
the coherent quasiparticle band due to the
hole-phonon Holstein coupling is small compared with 
that arising from hole-spin interactions (magnetic polaron regime). 
In particular,  the integrated density of states 
is barely changed from that of the pure $t-J$ model. 
The new structures, nevertheless observed in the $A_{{\vec K}}$ spectra
shown Fig.~\ref{f:akhtj}~(a), correspond to predominantly ``phononic'' side 
bands separated from the particle-spin excitations by multiples of the  
bare phonon frequency $\omega_0$. These phonon resonances 
have less and less ``electronic'' spectral weight the more
phonons are involved. This is because $A_{{\vec K}}(\omega)$ measures 
the overlap of these excited states with the state obtained 
by creating a hole in the zero-phonon Heisenberg ground state.  

\begin{figure}[b]
  \centering
      \includegraphics[width=0.49\linewidth]{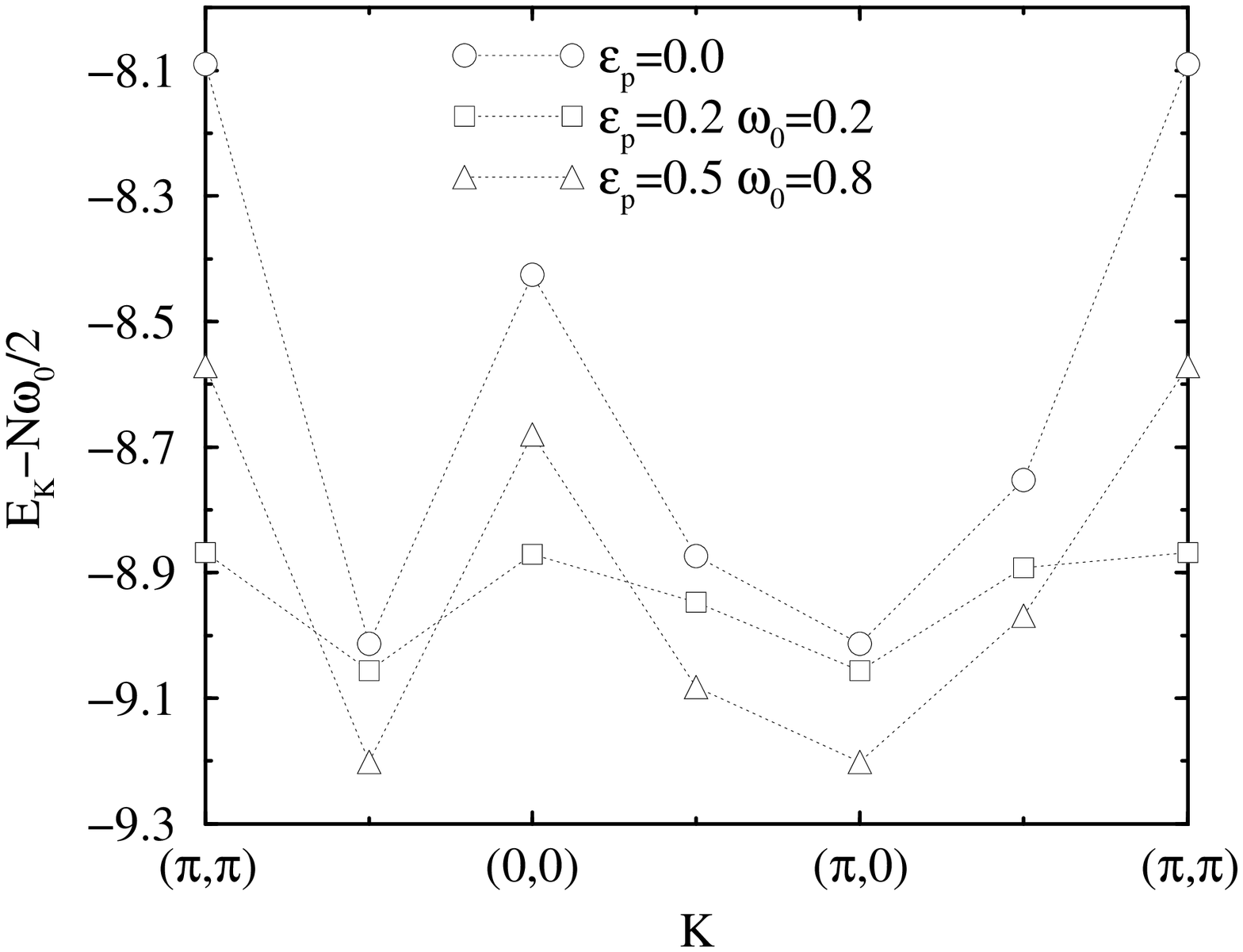}\hspace*{0.3cm}
\includegraphics[width=0.49\linewidth]{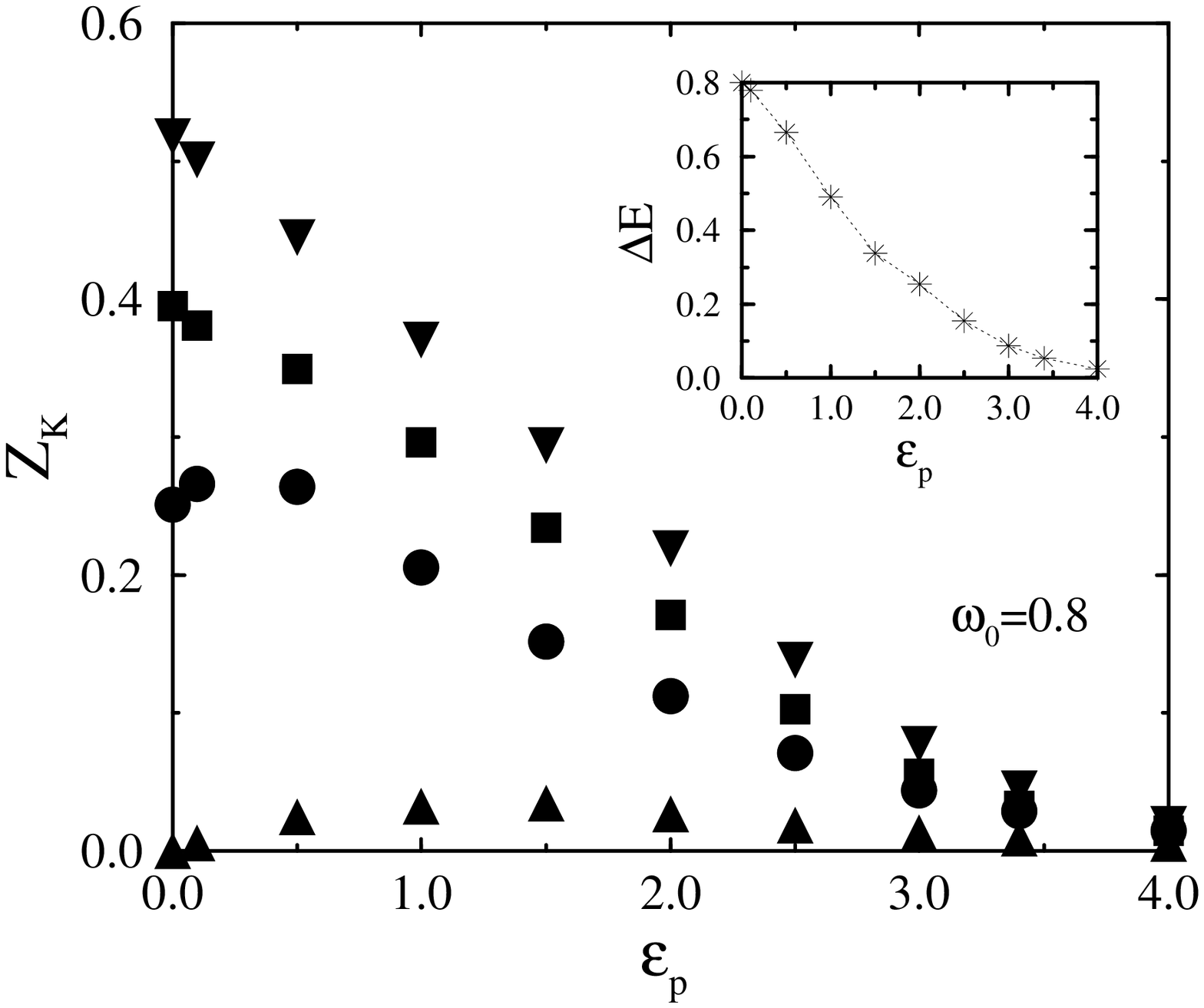}
\caption{Left panel: Band dispersion for the 2D HtJM on a 
16-site lattice ($J=0.4$). Right panel: Spectral weight 
factor $Z_{\vec K}$ as a function of the EP coupling strength $\varepsilon_p$ 
for the different ${\vec K}$ vectors of the ten-site lattice:  
(3,1) (triangles down), (2,4) (squares), (0,0) (circles), and
(5,5) (triangles up) [in units of ($\pi/5,\pi/5$)]. 
The inset shows the narrowing of the coherent 
bandwidth $\Delta E$ with increasing $\varepsilon_p$.}
\label{f:ekzk}
\end{figure}

With increasing $\varepsilon_p$ the lowest peaks in each ${\vec K}$-sector 
start to separate from the rest of the spectrum. These states become
very close in energy and finally a narrow well-separated lattice 
hole-polaron band evolves in the strong-coupling case (see
Fig.~\ref{f:akhtj}~(b)). Now the hole is  heavily dressed 
by phonons and the quasiparticle pole strength is 
strongly suppressed (cf. Fig.~\ref{f:ekzk}). 
At the same time spectral weight is transferred to the 
high-energy part and the whole spectrum
becomes incoherently broadened. Therefore we observe an 
overall smoothing of $N(\omega)$. 
The gap to the next higher energy band is of the order of $\omega_0$. 
This excitation will be triggered by
an one-phonon absorption process. 
The crossover to the lattice hole-polaron state 
is accompanied by a strong increase in 
the on-site hole-phonon correlations~\cite{WRF96},
indicating that the lattice polaron quasiparticle comprising 
a self-trapped hole and the phonon cloud is mainly
confined to a single lattice site (small hole polaron).
Most notably, compared to the non-interacting single-electron   
Holstein model~\cite{FLW97} or the spinless fermion 
Holstein-$t$ model~\cite{FRWM95}, 
the critical EP coupling strength for lattice polaron formation is 
considerably reduced due to magnetic prelocalisation effects.

To illustrate the formation of the lattice hole-polaron 
band in some more detail, we first determined 
the ``coherent'' band dispersion $E_{{\vec K}}$ of the 2D HtJM for 
a 16-site lattice. The band structure is 
shown in the left panel of Fig.~\ref{f:ekzk} 
along the principal directions in the Brillouin zone. 
The minima of the quasiparticle dispersion
are found to be located at the momenta ${\vec K}=(\pm\pi/2,\pm\pi/2)$ 
(the hidden symmetry of the 4$\times$4 cluster leads to an accidental
degeneracy with the  ${\vec K}=(\pm\pi,0)$, $(0,\pm \pi)$ states).  
At weak EP couplings the energy dispersion is not significantly 
changed from that of the standard $t-J$ model, 
provided that the phonon frequency exceeds 
the effective bandwidth of the magnetic lattice polaron 
($\omega_0=0.8\geq \,\Delta E_{tJ}$).

Next we evaluated the wave-function renormalisation factor for  
different band states, 
\begin{equation}
Z_{{\vec K}}=\frac{\left|
\langle \psi_{0,{\vec K}}^{(N_e-1)}
|\tilde{c}^{}_{{\vec Q}-{\vec K},\sigma}| \psi_{0,{\vec Q}}^{(N_e)}\rangle
\right|^2}{
\left|\langle \psi_{0,{\vec Q}}^{(N_e)}
|\tilde{c}^{\dagger}_{{\vec Q}-{\vec K},\sigma}\tilde{c}^{}_{{\vec Q}-{\vec K},
\sigma}
| \psi_{0,{\vec Q}}^{(N_e)}\rangle\right|^2}\,,
\label{zk}
\end{equation}
where $ |\psi_{0,{\vec Q}}^{(N_e-1)}\rangle$ denotes
the one-hole state being lowest in energy in the 
${\vec Q}$-sector. $Z_{\vec{K}}$
can be taken as a measure of the ``contribution'' of the hole 
(dressed at $\varepsilon_p=0$ by spin-wave excitations
only) to the composite spin/lattice polaron (having
total momentum ${\vec K}$). 
The data obtained at weak EP coupling unambiguously confirm 
the different nature of band states 
in this regime (see Fig.~\ref{f:ekzk} right panel). 
We found practically zero-phonon ``hole'' states 
at the band minima (${\vec K}=(3\pi/5,\pi/5)$, triangles down) 
and ``phonon'' states, which are only weakly affected by the hole, 
around the (flat) band maxima (${\vec K}=(\pi,\pi)$, triangles up).   
With increasing $\varepsilon_p$, a strong ``mixing'' of holes and phonons 
takes place, whereby both quantum objects completely lose their 
own identity. Concomitantly $Z_{{\vec K}}$ decreases for the ``hole--like''
states but increases (first of all) for the ``phonon-like'' states.   
At large $\varepsilon_p$, a small lattice hole polaron is formed, which, 
according to the numerical data, 
has an extremely small spectral weight. 
Then the question arises whether the lattice hole polaron is 
a ``good'' quasiparticle in the sense that
one can construct a quasiparticle  operator,
$\tilde{c}_{\vec{K}\sigma}\to\tilde{d}_{\vec{K}\sigma}$, 
having large spectral weight at the lowest pole in the spectrum.  
It was demonstrated that such a composite electron/hole-phonon 
(polaron) operator could be constructed for the 
Holstein model~\cite{FLW97} as well as for the $t-J$ model 
coupled to buckling/breathing modes~\cite{SPS97}.

\begin{figure}
  \centering
\includegraphics[width=0.49\linewidth]{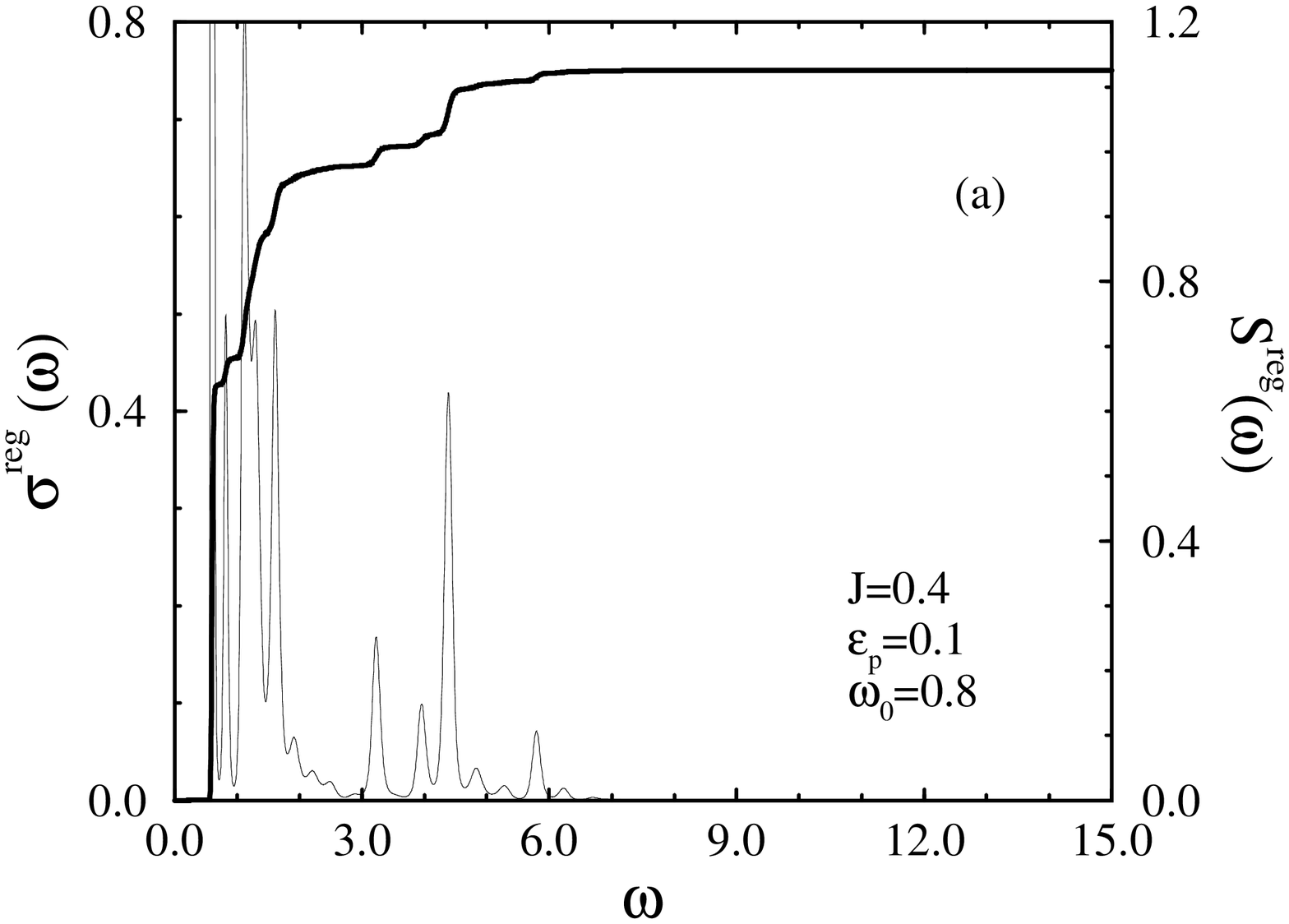}\hspace*{0.3cm}
\includegraphics[width=0.49\linewidth]{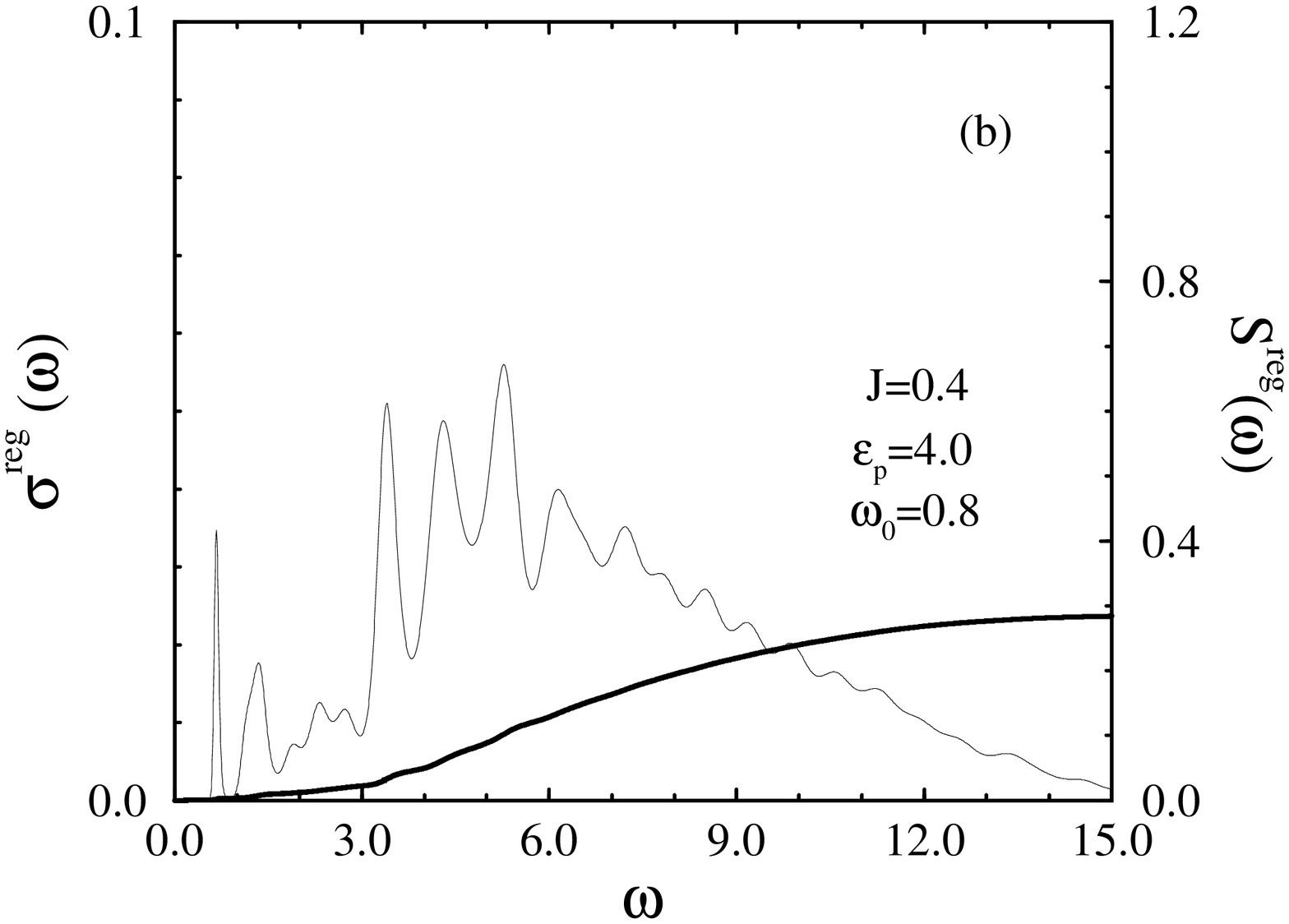}
\caption{Optical conductivity in the 2D HtJM with $J=0.4$. 
$\sigma^{reg}(\omega)$ and $S^{reg}(\omega)$ were calculated 
for a ten-site lattice with 15 phonons.}
\label{f:sigma}
\end{figure}

In Fig.~\ref{f:sigma} we show the optical response in the framework of 
the single-hole 2D HtJM. 
In the weak EP coupling regime and for phonon frequencies
$\omega_0 > \Delta E_{tJ}$, we recover the main
features of the optical absorption spectrum of the 2D $t-J$ model~\cite{Po91}, 
i.e., an ``anomalous'' broad mid-infrared 
band ($J<\omega<2t$), separated from 
the Drude peak (${\cal D}\delta (\omega)$; not shown) by a pseudo-gap 
$\simeq J$, and an ``incoherent'' tail up to $\omega\simeq 7t$
(cf. Fig.~\ref{f:sigma}~(a)). 
At larger EP couplings, the overlap with excited multi-phonon 
states is enlarged and the optical response is enhanced at higher energies. 
This redistribution of spectral weight from low to high energies
can be seen in Fig.~\ref{f:sigma}~(b). As expected the transition
to the lattice hole-polaron state, at about 
$\varepsilon_{p}^c(J=0.4, \omega_0=0.8)\simeq 2.0$, 
is accompanied by the development of a broad
maximum in $\sigma^{reg}(\omega)$, whereas the Drude weight as well as 
the low-frequency optical response become strongly suppressed.

Let us now make contact with the experimentally observed characteristics
of the mid-infrared (MIR) spectra in the doped 
perovskites~\cite{SAL95,Da94,TT92,AKR94b}).  
The simple 2D HtJM seems to contain the key ingredients 
to describe, at least qualitatively, the principal features of the optical 
absorption spectra of these compounds. This can be seen by comparing
Figs.~\ref{f:sigma}~(a) and (b), corresponding to the weak and
strong EP coupling situations realized in the 
cuprate ($\rm La_{2-x}Sr_xCuO_4$) and nickelate 
($\rm La_{2-x}Sr_xNiO_4$) systems, respectively.
The EP interaction ratio $\varepsilon_p/t$ in the nickelates 
is estimated to be about one order of magnitude larger than in 
the cuprates because of the much smaller transfer
amplitude ($t\simeq 0.08$ eV~\cite{BE93}). 
According to the internal structure of the low-spin state,
the hopping transport of spin-1/2 composite holes in a 
spin-1 background is rather complex; implying,
within an effective single-band description, a strong reduction
of the transition matrix elements~\cite{LF97}.
A striking feature of the absorption spectra in the cuprate
superconductors is the presence of a MIR  band,
centred at about 0.5~eV in lightly doped $\rm La_{2-x}Sr_xCuO_4$
(which, using $t\sim 0.3$~eV, means that $\omega\sim 1.5$).   
Such a strong MIR absorption is clearly visible in Fig.~\ref{f:sigma}~(a).
Obviously it is quite difficult to distinguish the spectral weight, produced 
by the dressing of the hole due to the ``bag'' of reduced 
antiferromagnetism in its neighbourhood~\cite{SWZ88}, 
from other (e.g. hole-phonon coupling) processes 
that may contribute to the MIR band observed experimentally.
The results presented for the HtJM in Fig.~\ref{f:sigma}~(b) support 
the claims, however,  that the MIR band in the cuprates has a
mainly ``electronic'' origin, i.e., the lattice polaron effects are 
rather weak~\cite{Da94}. The opposite is true for their isostructural
counterpart, the nickelate system, where the MIR absorption band 
has been ascribed by many investigators to ``polaronic''
origin~\cite{BE93,Caea97}. Within the HtJM such a situation
can be modelled by the parameter set used in Fig.~\ref{f:sigma}~(b).
If we fix the energy scale by  $t= 0.08$~eV,   
the maximum in the optical absorption is again located at about 0.5~eV.
The whole spectrum clearly shows lattice polaron characteristics, where
it seems that the lattice hole polarons are of small-to-intermediate 
size~\cite{Emi95}.
Most notably, we are able to reproduce the experimentally observed asymmetry
in the shape of the spectrum, in particular the very gradual decay 
of $\sigma^{reg}(\omega)$ at high energies. 
It is worth mentioning that this behaviour cannot 
be obtained from a simple fit to the analytical expressions
derived for the small polaron hopping conductivity~\cite{BE93,Re67,SPH91}.  
Exploiting the f-sum rule we found that there are almost no
contributions from band-like carriers in agreement with the
experimental findings~\cite{BE93,Caea97}.

\begin{figure}
  \centering
\includegraphics[width=0.7\linewidth]{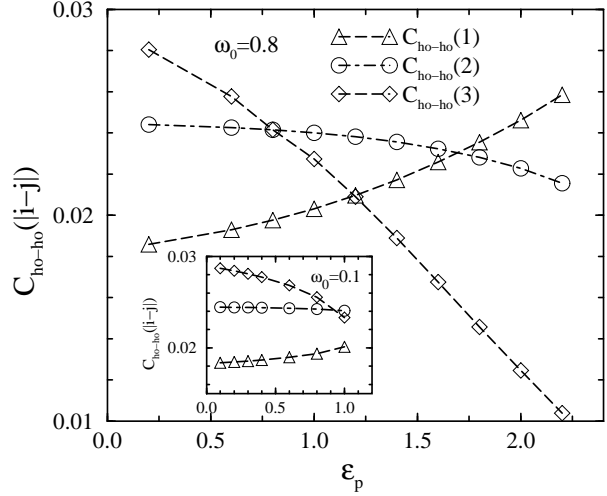}
\caption{Non-equivalent hole-hole pair correlation functions 
$C_{ho-ho}(|i-j|)$ in the two-hole ground state 
of the HtJM  at various $\varepsilon_p$. Here    
1, 2, and 3 label NN, next NN, and third NN distances,
respectively.}
\label{f:hhpaircor}
\end{figure}

Next let us briefly discuss  the two-hole problem. In order to study 
hole-binding effects, we have calculated
the hole-hole correlation function  
\begin{equation}
C_{ho-ho}^{}(|i-j|)=\langle \psi_0(\varepsilon_p,J)
|\tilde{h}_i^{} \tilde{h}_j^{}|\psi_0(\varepsilon_p,J)\rangle\,.
\label{choho}
\end{equation}  
Results for $C_{ho-ho}^{}(|i-j|)$ are presented in Fig.~\ref{f:hhpaircor}. 
At weak EP coupling, $C_{ho-ho}(|i-j|)$
becomes maximum at the largest distance of the ten-site lattice,
while in the intermediate EP coupling regime 
the preference is on next NN pairs.
As expected, further increasing $\varepsilon_p$, 
the maximum in $C_{ho-ho}(|i-j|)$ is shifted to the 
shortest possible distance, indicating hole-hole attraction. 
At $\varepsilon_p\gg 1 $, the two holes become ``self--trapped'' 
sharing a sizeable common lattice distortion, i.e., 
a nearly immobile hole-bipolaron is formed.  
The behaviour of $C_{ho-ho}$ is found to be qualitatively
similar for higher (lower) phonon frequencies (see inset), 
except that the crossings of different hole-hole correlation 
functions occur at larger (smaller) values of~$\varepsilon_p$,
which again shows the importance of both parameter ratios 
$\lambda=\varepsilon_p/2Dt$ and $g=\sqrt{\varepsilon_p/\omega_0}$. 

\begin{figure}
  \centering
      \includegraphics[width=0.4\linewidth]{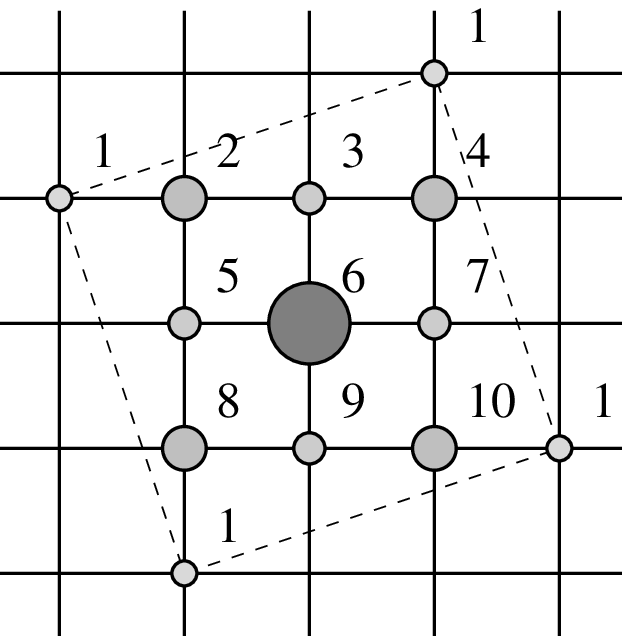}\hspace*{.5cm}
      \includegraphics[width=0.4\linewidth]{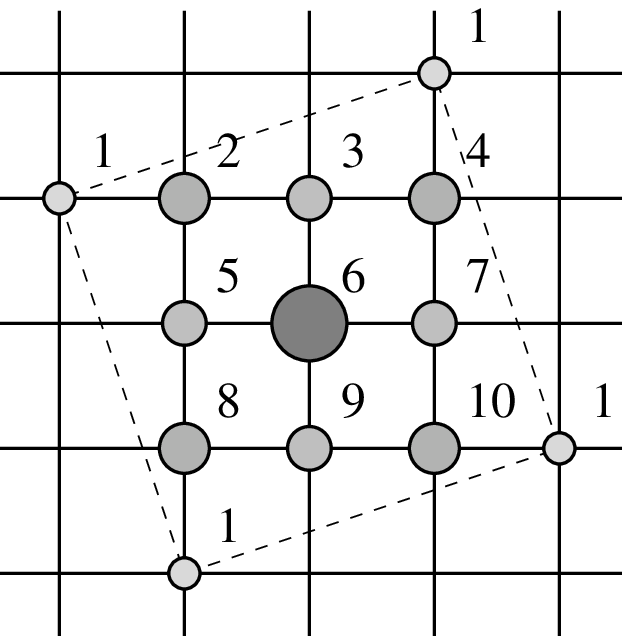}
\caption{$C_{ho-ho}(|6-j|)$ (left) 
and $C_{ho-ph}(|6-j|)$ (right) are displayed at $\varepsilon_p=3$ 
and $\omega_0=0.8$, where both diameter and gray level of the 
circles are proportional to the correlation strength.}
\label{f:hhpcor}
\end{figure}

Finally let us consider the quarter-filled band case.
Here, we have investigated the simpler spinless fermion model 
(total $S^z=S^z_{max}$). In accordance with previous approximate treatments 
based on an inhomogeneous variational Lang-Firsov
approach~\cite{FRWM95}, we found, as the EP coupling increases,
evidence for a transition from a free polaron state 
to a 2D polaronic superlattice, where the holes 
are self-trapped on every other site. 
This crossover is signalled by a pronounced peak in the charge
structure factor $S_c(\pi,\pi)$. 
To visualise the correlations in this state in more detail, 
in Fig.~\ref{f:hhpcor} we have depicted $C_{ho-ho}(|i-j|)$ and  
the corresponding hole-phonon density correlation function 
$C_{ho-ph}(|i-j|)=\langle \psi_0^{}|\tilde{h}_i^{} b^\dagger_j
b^{}_j|\psi_0^{}\rangle$ as a function of $|i-j|$.
Our exact results clearly show the phonon-dressing of the holes 
and the tendency towards CDW formation as observed, e.g., in $\rm
La_{1.5}Sr_{0.5}NiO_4$.  

\subsection{Lattice polarons in a generalised double-exchange model} 
\label{sec:dem}
The key elements of the electronic structure of the 
CMR ($\rm La_{1-x}[Sr,Ca]_xMnO_3$) compounds are
the partially filled 3d states. The cubic environment of the 
Mn sites within the perovskite lattice results in a crystal 
field splitting of Mn d-orbitals into $e_g$ and $t_{2g}$. 
In the case of zero doping ($x=0$) there are four electrons per 
Mn site which fill up the three $t_{2g}$ levels and one $e_g$ level, 
and by Hund's rule coupling, form a $S=2$ spin state. 
Doping will remove the electron from the $e_g$ level, and
by hopping via bridging oxygen sites the resulting holes 
acquire mobility.

Due to the specific symmetry of the manganese $d$ and oxygen $p$ 
orbitals, the transfer of the $e_g$-electrons shows a pronounced  
orbital anisotropy. In the limit of large on-site
Coulomb ($U$) and Hund  ($J_H$) interactions the electron
transfer is strongly affected by the spin of the core electrons
as well. Concentrating on the link between magnetic correlations 
and transport, early studies on lanthanum manganites 
attributed the low-$T$ metallic behaviour to 
Zener's  double-exchange (DE) mechanism~\cite{Ze51b,KO72a}, 
which maximises the hopping of a strongly Hund's rule coupled 
electron in a polarised spin background. 

As discussed above, it has been realized that physics 
beyond DE is important not only to explain the phase 
diagram of the manganites but also the CMR transition itself.
The orbital degeneracy in the ground state of Mn$^{3+}$ ions 
connects the system to the lattice, making it susceptible to Jahn-Teller 
polaron formation. 
There are two types of lattice distortions which are important 
in manganites. First the partially filled $e_g$ states of 
the $\rm Mn^{3+}$ ion are Jahn-Teller active, i.e., the
system can gain energy from a quadrupolar symmetric elongation 
of the oxygen octahedra which lifts the $e_g$ degeneracy. 
A second possible deformation is an isotropic shrinking of a  
$\rm MnO_6$ octahedron. This ``breathing''-type distortion
couples to changes in the $e_g$ charge density, i.e., 
is always associated with the presence of an $\rm Mn^{4+}$ ion.

Restricting the electronic Hilbert space to the large 
Hund's rule states given by the spin-$2$ orbital doublet
state $^{5}E$ [$t_2^3(^4A_2)e$] for Mn$^{3+}$ (d$^4$) 
and the spin-$\frac{3}{2}$  orbital singlet state $^{4}A_2$ [$t_2^3$] 
for Mn$^{4+}$ (d$^3$), within $2^{\rm nd}$ order perturbation theory 
the following Hamiltonian results (for details of the derivation
and notation see Ref.~\cite{WF02b,WF04b}): 
\begin{eqnarray}
\label{eho}
 H & = &H_{\rm DE} 
   +H_{\rm spin-orbital}^{\rm 2^{nd}\; order}
   +H_{\rm electron-JT} +H_{\rm electron-breathing}
   +H_{\rm phonon}\nonumber{}\\[0.2cm]
   & = & \sum_{i,\delta,\alpha,\beta}
  (a^{}_{i,\uparrow} a^{\dagger}_{i+\delta,\uparrow} 
  + a^{}_{i,\downarrow} a^{\dagger}_{i+\delta,\downarrow}) 
  \ t^{\delta}_{\alpha\beta}
  \ c^{\dagger}_{i,\alpha} n^{}_{i,\bar\alpha} 
  n^{}_{i+\delta,\bar\beta} c^{}_{i+\delta,\beta}\nonumber{}\\
& &+ \sum_{i,\delta,\kappa,\lambda} 
  (J^{\delta}_{\kappa\lambda}\ {\vec S}_{i} {\vec S}_{i+\delta}
  + \Delta^{\delta}_{\kappa\lambda})
  \ P^{\kappa}_{i} P^{\lambda}_{i+\delta}\nonumber{}\\
& &\,+\bar{g}\sum_i\left[
        (n^{}_{i,\varepsilon}-n^{}_{i,\theta})
        (b^{\dagger}_{i,\theta}+b^{}_{i,\theta})
        + (d^{\dagger}_{i,\theta} d^{}_{i,\varepsilon} +
        d^{\dagger}_{i,\varepsilon} d^{}_{i,\theta})
        (b^{\dagger}_{i,\varepsilon}+b^{}_{i,\varepsilon})\right]\nonumber{}\\
& &\,+ \tilde g\sum_i 
      (n^{}_{i,\theta} + n^{}_{i,\varepsilon} -
        2n^{}_{i,\theta}n^{}_{i,\varepsilon}) 
      (b^{\dagger}_{i,a_1}+b^{}_{i,a_1})\nonumber{}\\
&&\,+\bar{\omega}_0\sum_i\left[b^{\dagger}_{i,\theta}b^{}_{i,\theta}
        +b^{\dagger}_{i,\varepsilon} b^{}_{i,\varepsilon}\right] +
      \tilde\omega_0\sum_i b^{\dagger}_{i,a_1}b^{}_{i,a_1}\,.
\end{eqnarray}

The  effective low-energy Hamiltonian $H$ contains 
Schwinger bosons $a^{(\dagger)}_{i,\mu}$, i.e.
$2{\vec S}_{i} = 
a^{\dagger}_{i,\mu}\boldmath{\tau}^{}_{\mu\nu}a^{}_{i,\nu}$ 
($\mu,\nu\in\{\uparrow,\downarrow\}$),
fermionic holes $c^{(\dagger)}_{i,\alpha}$,
phonons $b^{(\dagger)}_{i,\alpha}$ ($\alpha\in\{\theta,\varepsilon\}$), and
orbital projectors $P^{\kappa (\lambda)}_{i}$
($\kappa,\lambda \in\{\xi,\eta,\zeta\}$). In Eq.~(\ref{eho}), 
the first term, being proportional to $t$, corresponds to the DE  
interaction~\cite{Ze51b,KO72a}. 
The second term appears to be a bit more involved, since 
a rather large number of accessible virtual 
excitations (proportional to $t^2$ and $t_{\pi}^2$) 
contribute (see Refs.~\cite{WF02b,WF04b}). 
However, in all cases it is basically the product of a
Heisenberg-type spin interaction and two orbital projectors. 

The coupling between the orbital degree of freedom of the $e_g$ 
electrons and the optical phonon modes to lowest order can be    
modelled  by the $E\otimes e$ Jahn-Teller Hamiltonian 
(third term) and a Holstein-type interaction (fourth term). 
The energy of the dispersionless optical phonons 
are given within the harmonic approximation (fifth term). 

\begin{figure}
  \centering
      \includegraphics[width=0.9\linewidth]{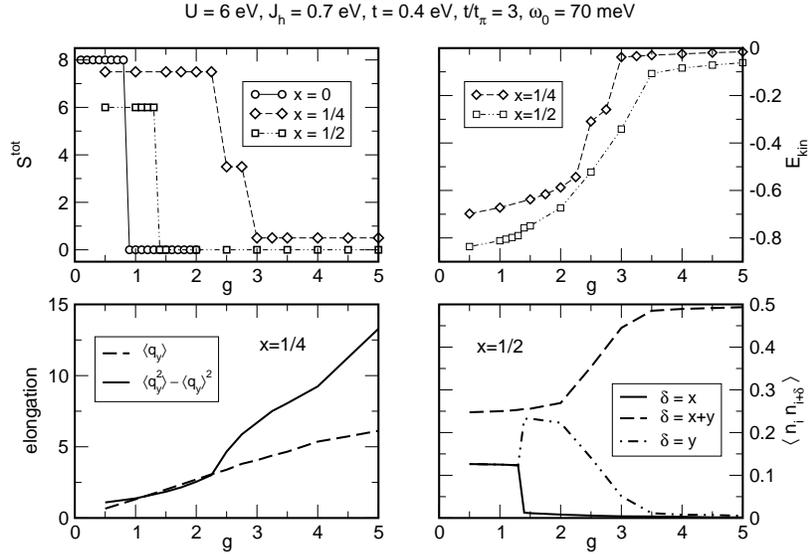}
\caption{Upper panels: Total spin $S^{tot}$ and kinetic energy
  $E_{kin}$ as a function of electron-phonon coupling 
  strength $g$ at various doping levels $x$. 
  Lower panels: Expectation values $\langle q_y\rangle$ and 
  $\langle q_y^2\rangle-\langle q_y\rangle^2$ 
  of the bond length in $y$ direction at $x=1/4$ (left) 
  and density-density correlations at $x=1/2$ (right).
  Results obtained by ED for the microscopic model~(\ref{eho})
  on a four site plaquette, where $g=\bar{g}/\omega_0$,
and $\omega_0=\bar{\omega}=\tilde{\omega}$ is assumed~\cite{WF04b}.}
\label{f:corr}
\end{figure}

\begin{figure}
  \centering
      \includegraphics[width=0.27\linewidth]{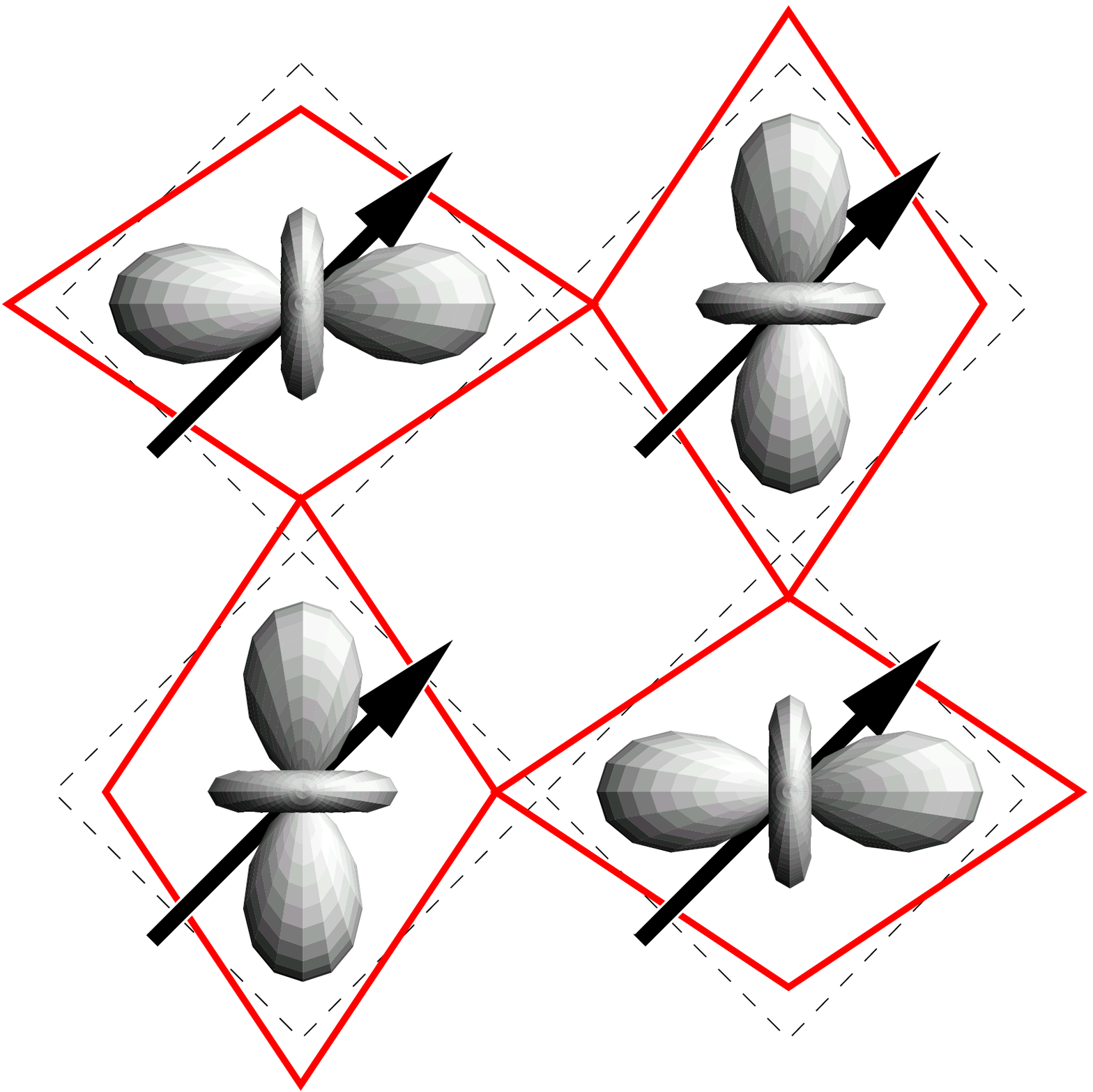} \hspace*{0.2cm}
      \includegraphics[width=0.27\linewidth]{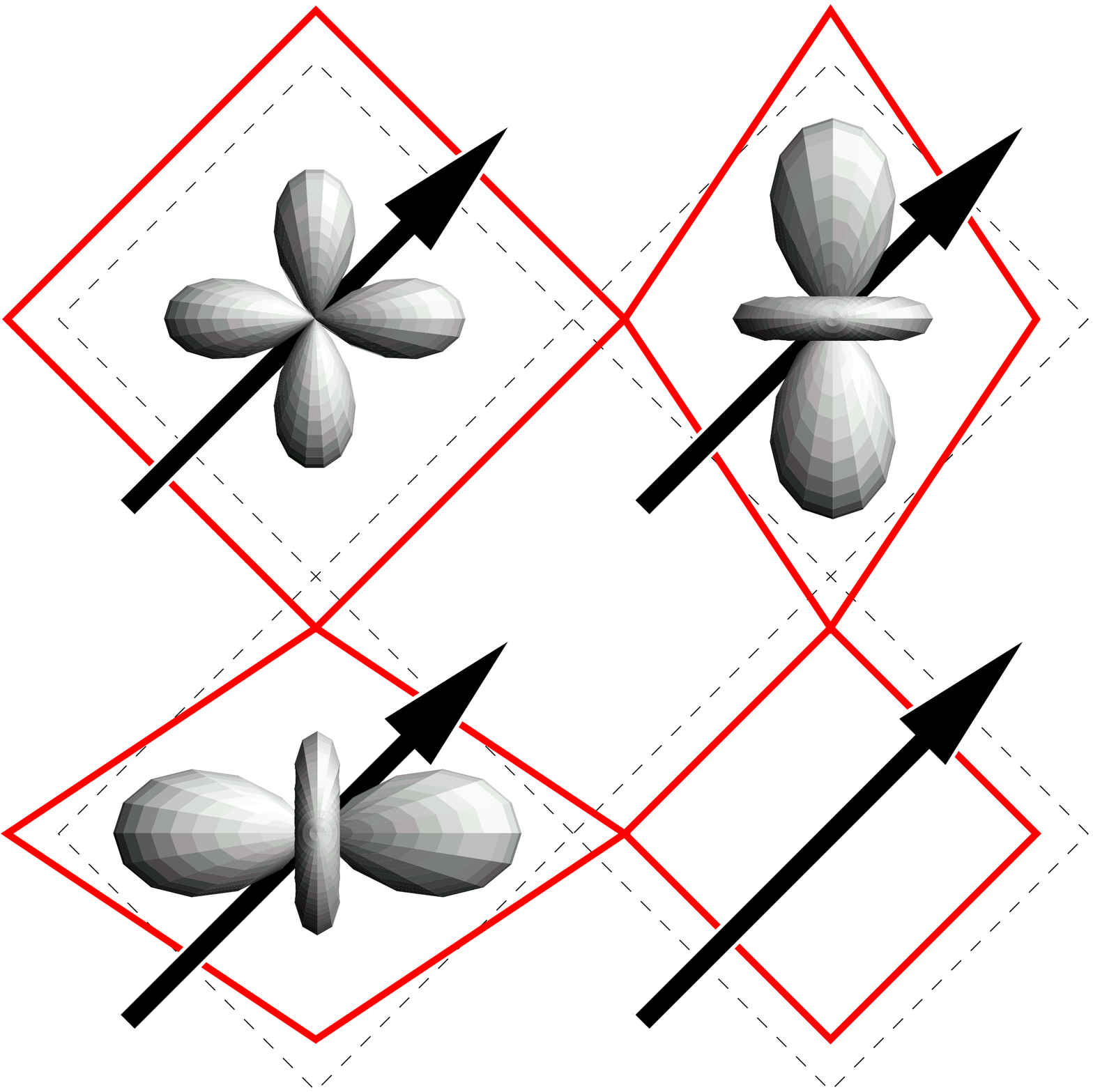} \hspace*{0.2cm}
      \includegraphics[width=0.27\linewidth]{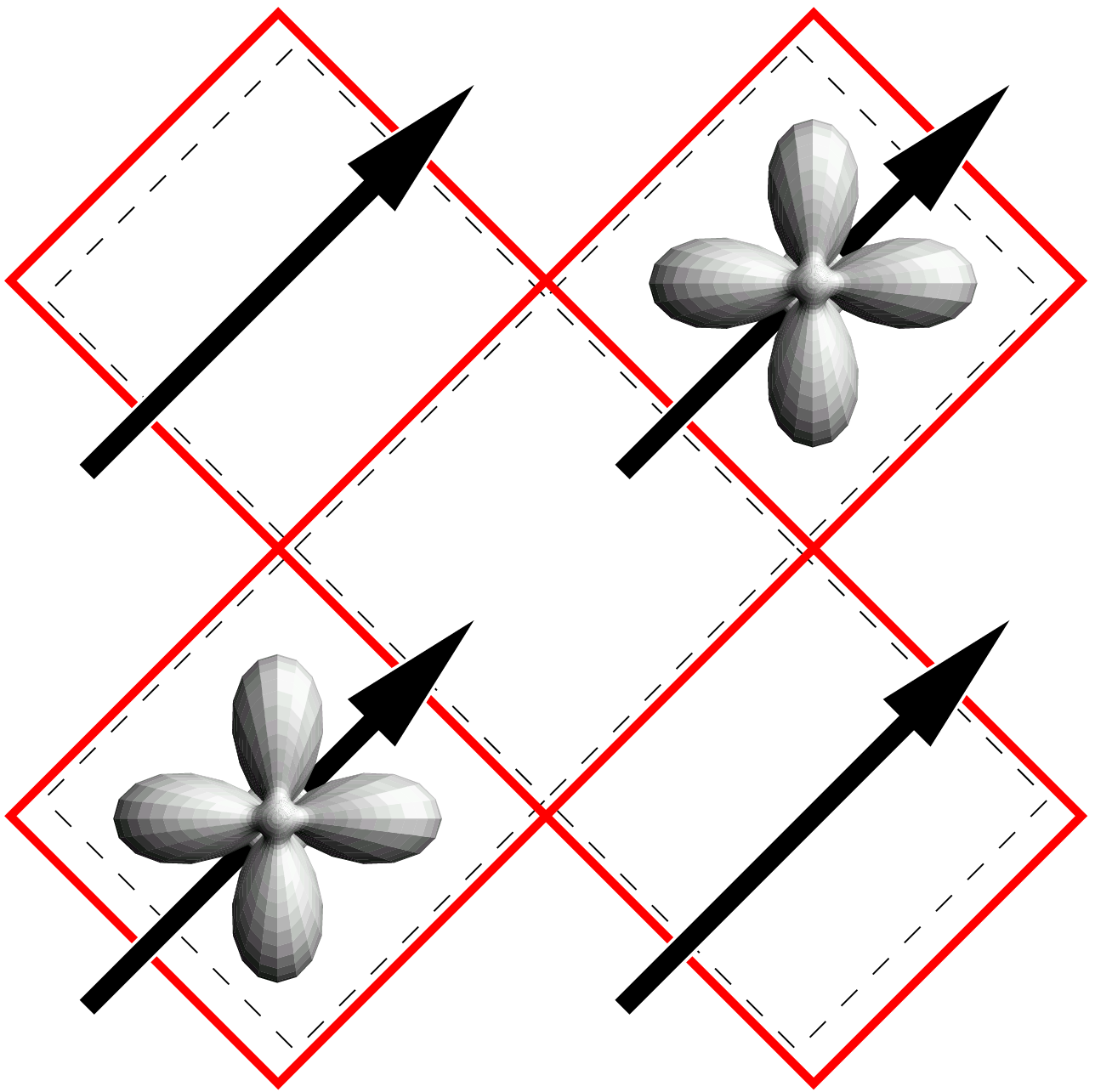}\\[0.2cm]
      \includegraphics[width=0.27\linewidth]{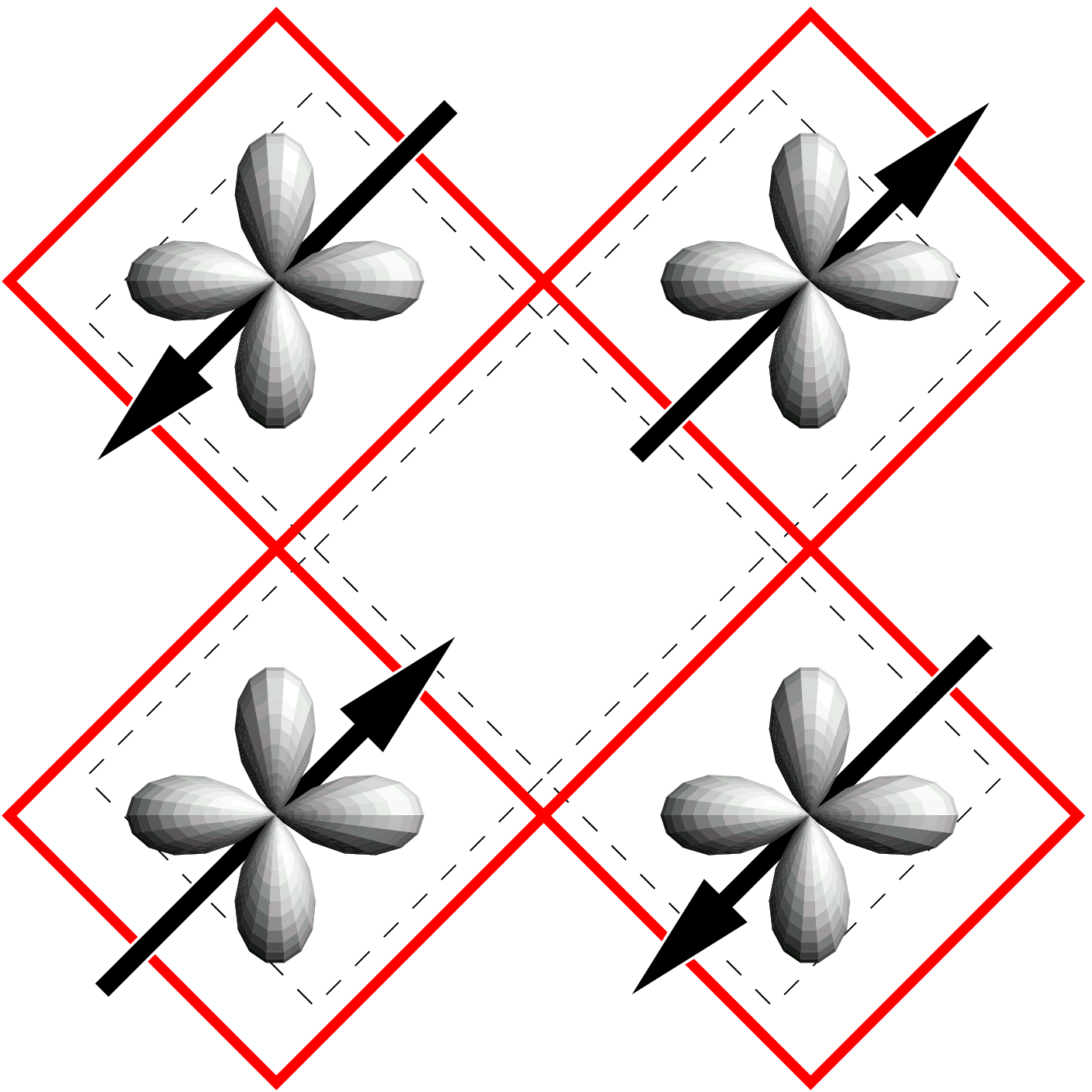}\hspace*{0.2cm}
      \includegraphics[width=0.27\linewidth]{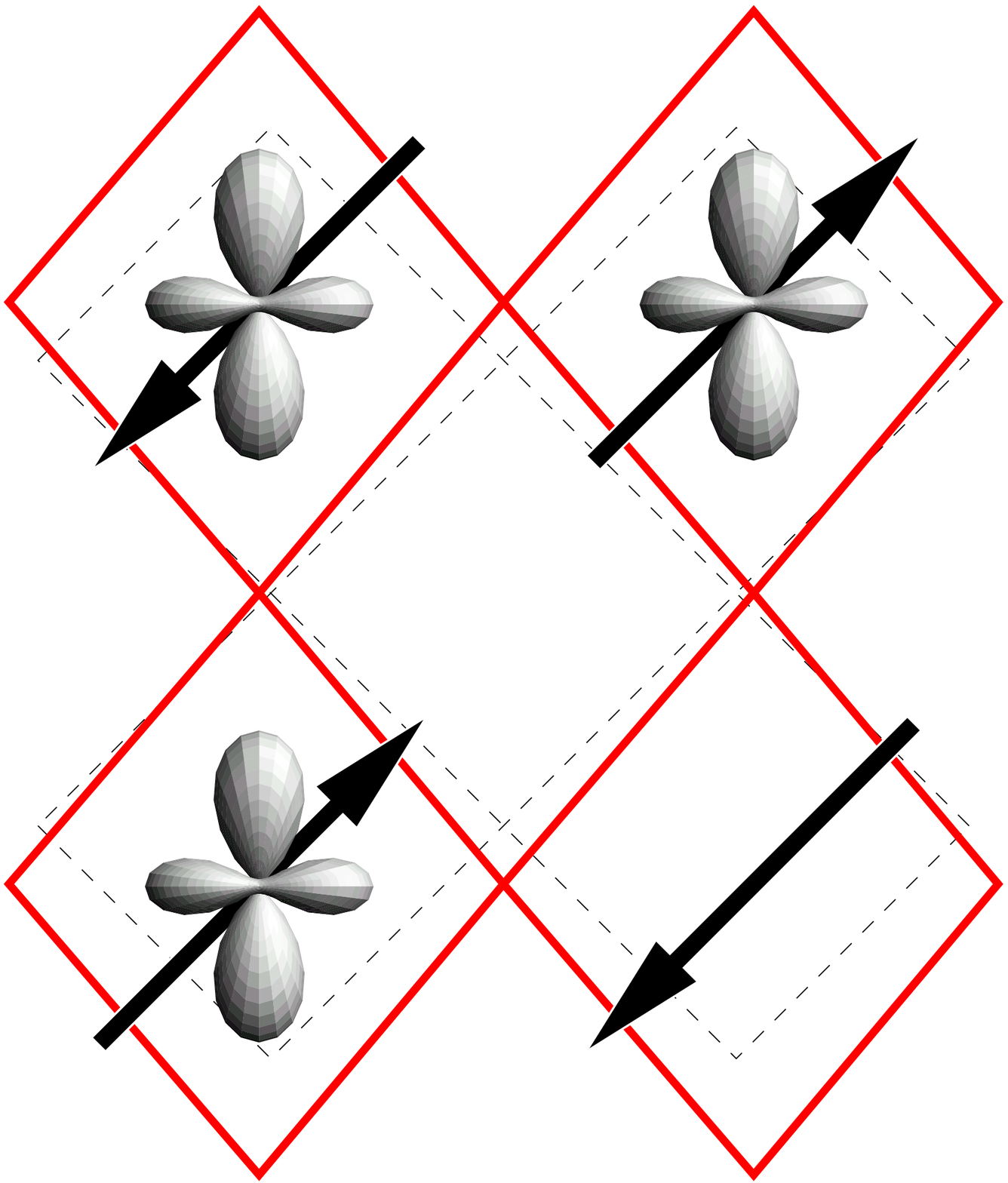}\hspace*{0.2cm}
      \includegraphics[width=0.27\linewidth]{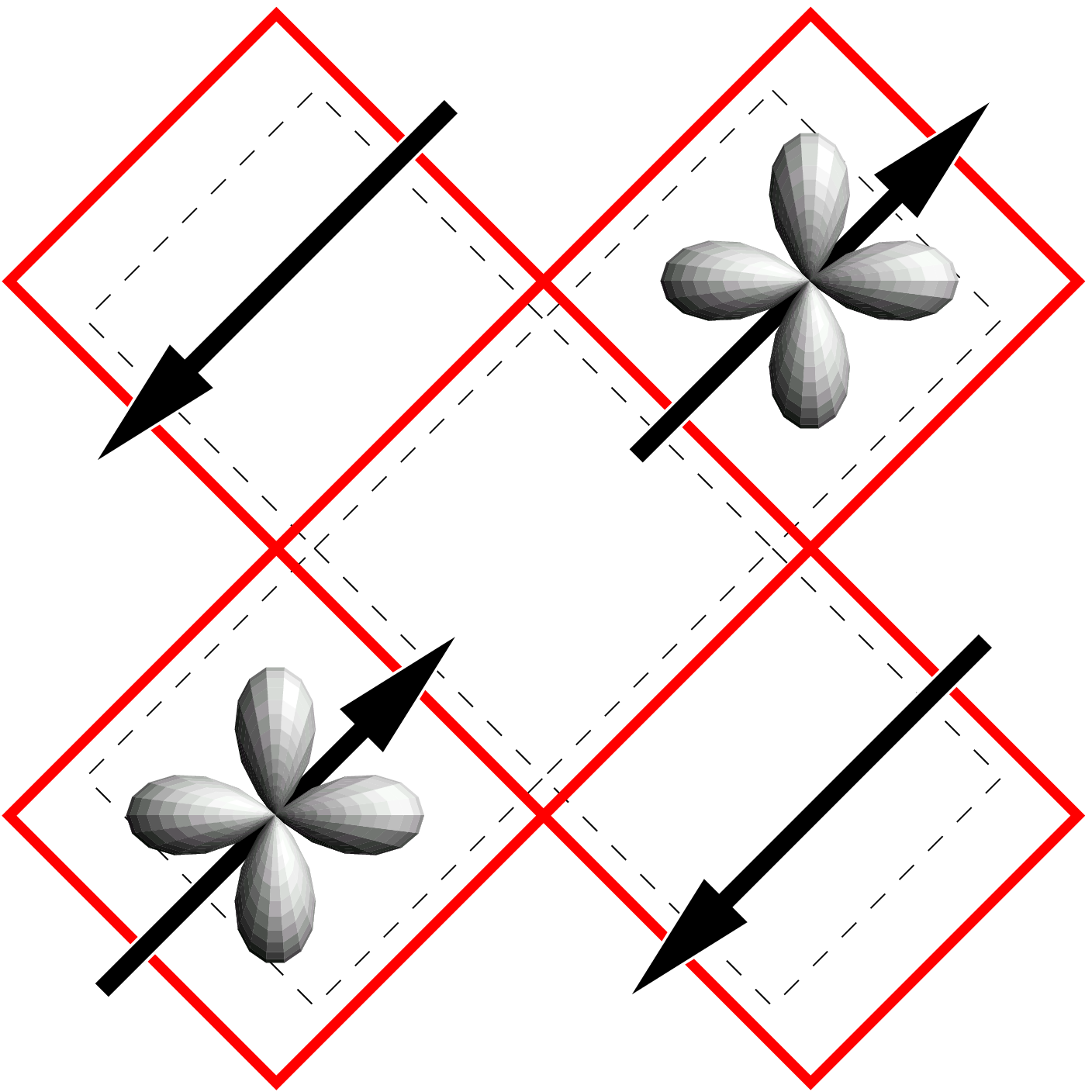}\\[0.2cm]
  \caption{Schematic evolution of charge, spin, orbital and 
    lattice correlation with increasing doping 
    [$x=0$, $x=0.25$ and $x=0.5$ (from left to right)]
    at rather weak (upper panels) and strong (lower panels)
    EP couplings.}
\label{f:orbis}
\end{figure}

For analytical methods the above Hamiltonian~(\ref{eho}) is far 
too complex to be understood in full detail, and even its numerical 
solution on finite lattices is hard. Using high performance computers 
and the phonon basis optimisation outlined in Sec.~\ref{sec:na}, 
we were able to calculate the ground-state properties of a 
small four-site cluster and to address, in particular, short-range 
correlations between the charge, spin, orbital and lattice degrees of 
freedom~\cite{WF02b,WF04b}. (See also [62].) Figures~\ref{f:corr} 
and~\ref{f:orbis} give a glimpse of these results. 
We assumed $t=0.4$~eV and $t/t_{\pi}=3$ for the hopping integrals 
and characterised the magnetic ``order'' according to the total 
spin of the ground state.

Undoped manganites ($\rm LaMnO_3$, $\rm PrMnO_3$) usually exhibit A-type 
antiferromagnetic order and strong Jahn-Teller distortion of
the ideal perovskite structure. The origin of the observed magnetic order has
been subject to discussions. While different band structure 
calculations~\cite{Saea95} emphasise the importance
of lattice distortions for the stability of antiferromagnetism,
purely electronic mechanisms were also favoured~\cite{FO99}.
In our microscopic model~(\ref{eho}), at $x=0$, only the 
second and third term will be active, and without EP interaction the 
competition of the spin-orbital contributions depends 
sensitively on the values of Coulomb and Hund's rule 
coupling. Starting from the "ferromagnetic" phase increasing either U or g changes the magnetic order of the ground
state to "antiferromagnetism''~\cite{WF02b} (see also
Fig.~\ref{f:corr} (upper left panel)). At larger 
EP interaction the system tends to develop static
Jahn-Teller distortions, which also fixes the orbital pattern and
subsequently the spin order. The change of orbital, spin and phonon 
correlations is illustrated schematically in 
Fig.~\ref{f:orbis} (left panels).

Our numerical calculations  corroborate the enhancement of 
ferromagnetic correlations for the weakly doped case 
(see Fig.~\ref{f:corr}, upper left panel; $x=0.25$). 
However, if strong electron-phonon coupling causes 
self-trapping of the carriers the spin
order switches back to antiferromagnetism. 
This coincidence can be seen by comparing
the total spin of the cluster and the kinetic energy in the ground state,
both depicted in the upper panels of Fig.~\ref{f:corr}. Obviously
the change in the magnetic order is accompanied by the appearance of a lattice
distortion and a signature in the fluctuation of the bond length
($\propto\langle q_{x/y}^2\rangle-\langle q_{x/y}\rangle^2$),  
which reminds of the data measured close to the critical
temperature by Booth~et~al.~\cite{BBKLCN98} (lower left panel). 
The orbital orientation at the sites surrounding the hole-site 
is sketched in Fig.~\ref{f:orbis} (middle panels). Increasing~$g$ 
isolates the lattice sites, each optimising EP interaction 
individually and uncorrelated with the neighbours. 

At half-filling ($x=0.5$) the picture is more involved. 
Here strong Coulomb and EP interactions tend to
order the charges in the diagonal direction, i.e., in an
AB-type structure (compare Fig.~\ref{f:corr}, lower right panel). 
This allows for a rather large antiferromagnetic spin 
exchange $\propto t^2/J_H$.  
Consequently ferromagnetic order is unstable at much lower values of~$g$. 
The ferromagnetic  to antiferromagnetic transition is not connected 
to charge  localisation and causes only a tiny jump of the kinetic 
energy. Considering the most relevant eigenstate of the bond 
orbital density matrix, we observe a symmetric order of 
complex orbitals along the diagonal~\cite{WF02b}. 
After charge localisation is achieved at large $g$,  
neighbouring sites are again uncorrelated  with respect to
orbital ordering and are in some real mixed-orbital state.

Of course, the results of this section will not provide 
a quantitative analysis or description of the real 3D HTSC and CMR 
materials. Nevertheless, ED of even such small systems 
gives valuable insights into the correlations and 
driving interactions behind the rich phase diagram of 
the cuprates and manganites. Moreover the above 
ED data may serve as a benchmark for approximate theories.

\section{Conclusions}
\label{sec:co}
In summary, we have performed an extensive numerical analysis 
of the Holstein model. Combining variational Lanczos
diagonalisation, density matrix renormalisation group,
kernel polynomial expansion, and cluster perturbation theory 
techniques we solved for properties of the Holstein polaron 
and bipolaron problems. 
Numerical solution of the Holstein model means that we 
determined the ground-state and low-lying excited states
with arbitrary precision in the thermodynamic limit for any 
dimension. Moreover, we calculated the spectral properties (e.g. 
photoemission and phonon spectra), optical response and thermal 
transport, as well as the dynamics of polaron formation. 
Our approach takes into account the full quantum dynamics of 
the electrons and phonons and yields unbiased results for all 
electron-phonon  interaction strengths and phonon frequencies, 
but is of particular value in the intermediate-coupling 
regime, where perturbation theories and other analytical techniques fail. 

Most importantly, polaron formation represents a continuous crossover 
of the ground state. Nevertheless we found indications for a 
true phase transition in the first excited state, where a polaron 
plus phonon system changes from unbound at weak electron-phonon 
coupling to bound 
at strong coupling. Obviously electron and phonon excitations become 
intimately, dynamically connected in the process of
polaron formation. Concerning ground-state and spectral properties
the (quasi-free) electron (small) polaron self-trapping
crossover is related to (i) a significant increase of the particle  
effective mass, (ii) a substantial narrowing of the polaronic band dispersion,
and (iii) a strongly suppressed (electronic) quasiparticle residue. 
At the same time we observe (iv) an enhancement of 
the (on-site) electron-phonon correlations and the formation 
of a phonon drag, (v) a loss of kinetic energy, and (vi) a 
drop of the Drude weight, accompanied (vii) by a maximum 
in the spectral weight contained in the regular part of the optical 
response. All these features are found to be much more pronounced
in higher dimensions. The study of bipolarons showed that the two-site 
bipolaron has a significantly reduced mass and isotope effect compared
to the on-site bipolaron, and is bound in the strong-coupling regime up 
to twice the on-site Coulomb repulsion.    

Although we have now achieved a rather complete picture of the single
Holstein polaron and bipolaron problem (perhaps dispersive phonons, 
longer-ranged electron-phonon interaction, finite-temperature 
and disorder effects deserve closer attention), 
the situation is disconcerting in the case of a finite carrier
density. Here a density-driven crossover from large polarons to
quasi-free electrons scattered by unbound phonons might occur in the 
non-adiabatic intermediate coupling regime. Besides 
electron-phonon coupling competes with sometimes strong electronic 
correlations, e.g., in quasi-1D MX chains, 
quasi-2D high-$T_c$ superconductors, 3D charge-ordered nickelates, or
bulk colossal magneto-resistance manganites. The corresponding microscopic
models contain (extended) Hubbard, Heisenberg or double-exchange terms,
and maybe also a coupling to orbital degrees of freedom, so that they can
hardly be solved even numerically with the same precision as the Holstein
model. First exact results obtained for the case of a single carrier 
on finite lattices give strong evidence that the tendency towards
lattice (electron-, hole- or Jahn-Teller-) polaron formation 
is enhanced in strongly correlated electron systems. 
A more thorough investigation of these 
systems materials and models will definitely be a great challenge 
for solid-state theory in the near future. 
\section*{Acknowledgements}
We would like to thank A. Alvermann, I. Batisti\'{c}, B. B\"auml, 
A. R. Bishop, J.~Bon\v{c}a, F. X. Bronold, H. B\"uttner, 
F. G\"ohmann, G.~Hager, 
M. Hohenadler, D.~Ihle, E.~Jeckelmann, 
T.~Katra\v{s}nik, L.-C.~Ku, J.~Loos, G.~Schubert, H. R\"oder, R. N. Silver,
A.~Wei{\ss}e and G.~Wellein for valuable discussions. 
This work was supported by Deutsche Forschungsgemeinschaft through
SPP1073 FE398/1-3 and Grant No. 436 TSE 113/33/0-3, and the U.S. 
Department of Energy. Furthermore, we acknowledge generous computer
granting by Leibniz-Rechenzentrum M\"unchen and Norddeutscher Verbund
f\"ur Hoch- und H\"ochstleistungsrechnen (HLRN).
\bibliographystyle{spphys}
\bibliography{./ref}

\begin{thebibliography}{100}
\providecommand{\url}[1]{{#1}}
\providecommand{\urlprefix}{URL }
\expandafter\ifx\csname urlstyle\endcsname\relax
  \providecommand{\doi}[1]{DOI \discretionary{}{}{}#1}\else
  \providecommand{\doi}{DOI \discretionary{}{}{}\begingroup
  \urlstyle{rm}\Url}\fi

\bibitem{La33}
L.D. Landau, Phys. Z. Sowjetunion \textbf{3}, 664 (1933)

\bibitem{Pe46b}
S.I. Pekar, Zh. Eksp. Teor. Fiz. \textbf{16}, 341 (1946)

\bibitem{Fi75}
Y.A. Firsov, \emph{Polarons} (Izd. Nauka, Moscow, 1975)

\bibitem{Ra82}
E.I. Rashba, in \emph{Excitons}, ed. by E.I. Rashba, M.D. Sturge
  (North-Holland, Amsterdam, 1982), p. 543

\bibitem{PW84}
Y.E. Perlin, M.~Wagner (eds.), \emph{The Dynamical Jahn-Teller Effect in
  Localized Systems}.
\newblock No.~7 in Modern Problems in Condensed Matter Sciences (North-Holland,
  Amsterdam, 1984)

\bibitem{SS93}
A.L. Shluger, A.M. Stoneham, J. Phys. Condens. Matter \textbf{5}, 3049 (1993)

\bibitem{Fr54}
H.~Fr\"ohlich, Adv. Phys. \textbf{3}, 325 (1954)

\bibitem{Fr74}
H.~Fr\"ohlich, Fortschr. Phys. \textbf{22}, 159 (1974)

\bibitem{Dev}
J.T. Devreese, this volume, and references therein

\bibitem{Ho59a}
T.~Holstein, Ann. Phys. (N.Y.) \textbf{8}, 325 (1959)

\bibitem{Ho59b}
T.~Holstein, Ann. Phys. (N.Y.) \textbf{8}, 343 (1959)

\bibitem{PD82}
F.M. Peeters, J.T. Devreese, Phys. Status Solidi B \textbf{112}, 219 (1982)

\bibitem{Sp87a}
H.~Spohn, Ann. Phys. (N.Y.) \textbf{175}, 278 (1987)

\bibitem{Loe88}
H.~L\"owen, Phys. Rev. B \textbf{37}, 8661 (1988)

\bibitem{GL91}
B.~Gerlach, H.~L\"owen, Rev. Mod. Phys. \textbf{63}, 63 (1991)

\bibitem{Emi95}
D.~Emin, in \emph{Polarons and bipolarons in High--$T_c$ superconductors and
  related materials}, ed. by E.K.H. Salje, A.S. Alexandrov, W.Y. Liang
  (Cambridge University Press, Cambridge, 1995)

\bibitem{WF98a}
G.~Wellein, H.~Fehske, Phys. Rev. B \textbf{58}, 6208 (1998)

\bibitem{Ra06}
J.~Ranninger, in \emph{Polarons in Bulk Materials and Systems With Reduced
  Dimensionality}, \emph{International School of Physics Enrico Fermi}, vol.
  161, ed. by G.~Iadonisi, J.~Ranninger, G.~De~Filippis (IOS Press, Amsterdam,
  2006), \emph{International School of Physics Enrico Fermi}, vol. 161, pp.
  1--25

\bibitem{De63}
J.T. Devreese, Bull. Soc. Belge de Phys. (Ser. III) \textbf{4}, 259 (1963)

\bibitem{TNYS90}
N.~Tsuda, K.~Nasu, A.~Yanese, K.~Siratori, \emph{Electronic Conduction in
  Oxides} (Springer-Verlag, Berlin, 1990)

\bibitem{BEMB92}
Y.~Bar-Yam, T.~Egami, J.M. de~Leon, A.R. Bishop, \emph{Lattice Effects in
  High--$T_c$ Superconductors} (World Scientific, Singapore, 1992)

\bibitem{SAL95}
E.K.H. Salje, A.S. Alexandrov, W.Y. Liang, \emph{Polarons and Bipolarons in
  High Temperature Superconductors and Related Materials} (Cambridge University
  Press, Cambridge, 1995)

\bibitem{CCC93}
C.H. Chen, S.W. Cheong, A.S. Cooper, Phys. Rev. Lett. \textbf{71}, 2461 (1993)

\bibitem{Mi98}
A.J. Millis, Nature \textbf{392}, 147 (1998)

\bibitem{JHSRDE97}
M.~Jaime, H.T. Hardner, M.B. Salamon, M.~Rubinstein, P.~Dorsey, D.~Emin, Phys.
  Rev. Lett. \textbf{78}, 951 (1997)

\bibitem{Eg06}
T.~Egami, in \emph{Polarons in bulk materials and systems with reduced
  dimensionality}, \emph{International School of Physics Enrico Fermi}, vol.
  161, ed. by G.~Iadonisi, J.~Ranninger, G.D. Filipis (IOS Press, Amsterdam,
  2006), \emph{International School of Physics Enrico Fermi}, vol. 161, pp.
  101--117

\bibitem{Em73}
D.~Emin, Adv. Phys. \textbf{22}, 57 (1973)

\bibitem{To61}
Y.~Toyozawa, Prog. Theor. Phys. \textbf{26}, 29 (1961)

\bibitem{Mi58}
A.B. Migdal, Sov. Phys. JETP \textbf{7}, 999 (1958)

\bibitem{LF62}
I.G. Lang, Y.A. Firsov, Zh. Eksp. Teor. Fiz. \textbf{43}, 1843 (1962)

\bibitem{Ma95}
F.~Marsiglio, Physica C \textbf{244}, 21 (1995)

\bibitem{AM94b}
A.S. Alexandrov, N.F. Mott, \emph{High Temperature Superconductors and Other
  Superfluids} (Taylor \& Francis, London, 1994)

\bibitem{Su74}
H.~Sumi, J. Phys. Soc. Jpn. \textbf{36}, 770 (1974)

\bibitem{CPFF97}
S.~Ciuchi, F.~de~Pasquale, S.~Fratini, D.~Feinberg, Phys. Rev. B \textbf{56},
  4494 (1997)

\bibitem{Ma93}
F.~Marsiglio, Phys. Lett. A \textbf{180}, 280 (1993)

\bibitem{AKR94}
A.S. Alexandrov, V.V. Kabanov, D.K. Ray, Phys. Rev. B \textbf{49}, 9915 (1994)

\bibitem{MR97}
E.V.L. de~Mello, J.~Ranninger, Phys. Rev. B \textbf{55}, 14872 (1997)

\bibitem{WRF96}
G.~Wellein, H.~R\"oder, H.~Fehske, Phys. Rev. B \textbf{53}, 9666 (1996)

\bibitem{BWF98}
B.~B{\"a}uml, G.~Wellein, H.~Fehske, Phys. Rev. B \textbf{58}, 3663 (1998)

\bibitem{RL83}
H.~De~Raedt, A.~Lagendijk, Phys. Rev. B \textbf{27}, 6097 (1983)

\bibitem{BVL95}
E.~Berger, P.~Val\'{a}\v{s}ek, W.~v.~d. Linden, Phys. Rev. B \textbf{52}, 4806
  (1995)

\bibitem{KP97}
P.E. Kornilovitch, E.R. Pike, Phys. Rev. B \textbf{55}, R8634 (1997)

\bibitem{HEL04}
M.~Hohenadler, H.G. Evertz, W.~von~der Linden, Phys. Rev. B \textbf{69}, 024301
  (2004)

\bibitem{Kor}
P.~Kornilovitch, this volume

\bibitem{HL}
M.~Hohenadler, W.~von~der Linden, this volume

\bibitem{MN}
A.S. Mishchenko, N.~Nagaosa, this volume, and references therein

\bibitem{RBL98}
A.H. Romero, D.W. Brown, K.~Lindenberg, J. Chem. Phys \textbf{109}, 6540 (1998)

\bibitem{JW98b}
E.~Jeckelmann, S.R. White, Phys. Rev. B \textbf{57}, 6376 (1998)

\bibitem{ZJW98}
C.~Zhang, E.~Jeckelmann, S.R. White, Phys. Rev. Lett. \textbf{80}, 2661 (1998)

\bibitem{Je02b}
E.~Jeckelmann, Phys. Rev. B \textbf{66}, 045114 (2002)

\bibitem{CW85}
J.K. Cullum, R.A. Willoughby, \emph{Lanczos Algorithms for Large Symmetric
  Eigenvalue Computations}, vol. I \& II (Birkh\"auser, Boston, 1985)

\bibitem{SRVK96}
R.N. Silver, H.~R\"oder, A.F. Voter, D.J. Kress, J. of Comp. Phys.
  \textbf{124}, 115 (1996)

\bibitem{WWAF06}
A.~Wei{\ss}e, G.~Wellein, A.~Alvermann, H.~Fehske, Rev. Mod. Phys. \textbf{78},
  275 (2006)

\bibitem{SPP00}
D.~S\'en\'echal, D.~Perez, M.~Pioro-Ladri\`ere, Phys. Rev. Lett. \textbf{84},
  522 (2000)

\bibitem{HAL03}
M.~Hohenadler, M.~Aichhorn, W.~von~der Linden, Phys. Rev. B \textbf{68}, 184304
  (2003)

\bibitem{WFWB00}
A.~Wei{\ss}e, H.~Fehske, G.~Wellein, A.R. Bishop, Phys. Rev. B \textbf{62},
  R747 (2000)

\bibitem{CFIP06}
V.~Cautaudella, G.D. Filippis, G.~Iadonisi, C.A. Perroini, in \emph{Polarons in
  Bulk Materials and Systems With Reduced Dimensionality}, ed. by G.~Iadonisi,
  J.~Ranninger, G.~De~Filippis (IOS Press, Amsterdam, 2006), International
  School of Physics Enrico Fermi, pp. 119--130

\bibitem{CFP}
V.~Cataudella, G.D. Filippis, C.A. Perroini, this volume

\bibitem{TKB04}
S.A. Trugman, L.C. Ku, J.~Bon\v{c}a, J. Supercond. \textbf{17}, 193 (2004)

\bibitem{BTB99}
J.~Bon\v{c}a, S.A. Trugman, I.~Batisti\'{c}, Phys. Rev. B \textbf{60}, 1633
  (1999)

\bibitem{KTB02}
L.C. Ku, S.A. Trugman, J.~Bon\v{c}a, Phys. Rev. B \textbf{65}, 174306 (2002)

\bibitem{EBKT03}
S.~El~Shawish, J.~Bon\v{c}a, L.C. Ku, S.A. Trugman, Phys. Rev. B \textbf{67},
  014301 (2003)

\bibitem{RL82}
H.~De~Raedt, A.~Lagendijk, Phys. Rev. Lett. \textbf{49}, 1522 (1982)

\bibitem{CSG97}
M.~Capone, W.~Stephan, M.~Grilli, Phys. Rev. B \textbf{56}, 4484 (1997)

\bibitem{WF97}
G.~Wellein, H.~Fehske, Phys. Rev. B \textbf{56}, 4513 (1997)

\bibitem{Feea94}
H.~Fehske, D.~Ihle, J.~Loos, U.~Trapper, H.~B\"uttner, Z. Phys. B \textbf{94},
  91 (1994)

\bibitem{FWHWB04}
H.~Fehske, G.~Wellein, G.~Hager, A.~Wei{\ss}e, A.R. Bishop, Phys. Rev. B
  \textbf{69}, 165115 (2004)

\bibitem{BKT00}
J.~Bon\v{c}a, T.~Katra\v{s}nik, S.A. Trugman, Phys. Rev. Lett. \textbf{84},
  3153 (2000)

\bibitem{SHBWF05}
S.~Sykora, A.~H{\"u}bsch, K.W. Becker, G.~Wellein, H.~Fehske, Phys. Rev. B
  \textbf{71}, 045112 (2005)

\bibitem{AP98}
D.~Augier, D.~Poilblanc, Eur. Phys. J. B \textbf{1}, 19 (1998)

\bibitem{Da75}
E.R. Davidson, J. of Comp. Phys. \textbf{17}, 87 (1975)

\bibitem{SR97}
R.N. Silver, H.~R\"oder, Phys. Rev. E \textbf{56}, 4822 (1997)

\bibitem{St96}
W.~Stephan, Phys. Rev. B \textbf{54}, 8981 (1996)

\bibitem{FLW97}
H.~Fehske, J.~Loos, G.~Wellein, Z. Phys. B \textbf{104}, 619 (1997)

\bibitem{LHF06}
J.~Loos, M.~Hohenadler, H.~Fehske, J. Phys. Condens. Matter \textbf{18}, 2453
  (2006)

\bibitem{FAHW06}
H.~Fehske, A.~Alvermann, M.~Hohenadler, G.~Wellein, in \emph{Polarons in Bulk
  Materials and Systems With Reduced Dimensionality}, \emph{International
  School of Physics Enrico Fermi}, vol. 161, ed. by G.~Iadonisi, J.~Ranninger,
  G.~De~Filippis (IOS Press, Amsterdam, 2006), \emph{International School of
  Physics Enrico Fermi}, vol. 161, pp. 285--296

\bibitem{Go82}
A.A. Gogolin, Phys. Status Solidi B \textbf{109}, 95 (1982)

\bibitem{EH76}
D.~Emin, T.~Holstein, Phys. Rev. Lett. \textbf{36}, 323 (1976)

\bibitem{JE83}
D.R. Jennison, D.~Emin, Phys. Rev. Lett. \textbf{51}, 1390 (1983)

\bibitem{KM93}
V.V. Kabanov, O.Y. Mashtakov, Phys. Rev. B \textbf{47}, 6060 (1993)

\bibitem{KG96}
J.A. Kenrow, T.K. Gustafson, Phys. Rev. Lett. \textbf{77}, 3605 (1996)

\bibitem{Geea98}
N.H. Ge, et~al., Science \textbf{279}, 202 (1998)

\bibitem{SSKYK01}
A.~Sugita, T.~Saito, H.~Kano, M.~Yamashita, T.~Kobayashi, Phys. Rev. Lett.
  \textbf{86}, 2158 (2001)

\bibitem{DVBS00}
S.L. Dexheimer, A.D.V. Pelt, J.A. Brozik, B.I. Swanson, Phys. Rev. Lett.
  \textbf{84}, 4425 (2000)

\bibitem{TNSSTK98}
S.~Tomimoto, H.~Nansei, S.~Saito, T.~Suemoto, J.~Takeda, , S.~Kurita, Phys.
  Rev. Lett. \textbf{81}, 417 (1998)

\bibitem{AT02}
R.D. Averitt, A.J. Taylor, J. Phys. Condens. Matter \textbf{14}, R1357 (2002)

\bibitem{DRT00}
G.P. Donati, G.~Rodriguez, A.J. Taylor, J. Opt. Soc. Am. B \textbf{17}, 1077
  (2000)

\bibitem{KAK02}
B.~Krummheuer, V.M. Axt, T.~Kuhn, Phys. Rev. B \textbf{65}, 195313 (2002)

\bibitem{BLW86}
D.W. Brown, K.~Lindenberg, B.~J.West, J. Chem. Phys \textbf{84}, 1574 (1986)

\bibitem{Ku03}
L.C. Ku, Ph.D. thesis, UCLA (2003)



\bibitem{Emi93}
D.~Emin, Phys. Rev. B \textbf{48}, 13691 (1993)

\bibitem{AM95}
A.S. Alexandrov, N.F. Mott, \emph{Polarons and Bipolarons} (World Scientic,
  Singapore, 1995)

\bibitem{WMG98}
D.C. Worledge, L.~Mi\'{e}ville, T.H. Geballe, Phys. Rev. B \textbf{57}, 15267
  (1998)

\bibitem{Mah00}
G.D. Mahan, \emph{Many-Particle Physics} (Kluwer Academic/Plenum Publishers,
  New York, 2000)

\bibitem{RH67}
H.G. Reik, D.~Heese, J. Phys. Chem. Solids \textbf{28}, 581 (1967)

\bibitem{Lo88}
J.~Loos, Z. Phys. B \textbf{71}, 161 (1988)

\bibitem{Fi95}
Y.A. Firsov, Semiconductors \textbf{29}, 515 (1995)

\bibitem{Fir}
Y.A. Firsov, this volume

\bibitem{FC03}
S.~Fratini, S.~Ciuchi, Phys. Rev. Lett. \textbf{91}, 256403 (2003)

\bibitem{FC06}
S.~Fratini, S.~Ciuchi, Phys. Rev. B \textbf{74}, 075101 (2006)

\bibitem{SWWAF05}
G.~Schubert, G.~Wellein, A.~Wei{\ss}e, A.~Alvermann, H.~Fehske, Phys. Rev. B
  \textbf{72}, 104304 (2005)

\bibitem{KMF69}
E.K. Kudinov, D.N. Mirlin, Y.A. Firsov, Fiz. Tverd. Tela \textbf{11}, 2789
  (1969)

\bibitem{Na63}
E.L. Nagaev, Sov. Phys. Solid State \textbf{4}, 1611 (1963)

\bibitem{BB85}
H.~B\"ottger, V.V. Bryksin, \emph{Hopping conduction in solids} (Akademie
  Verlag, Berlin, 1985)

\bibitem{Hoea05}
M.~Hohenadler, D.~Neuber, W.~von~der Linden, G.~Wellein, J.~Loos, H.~Fehske,
  Phys. Rev. B \textbf{71}, 245111 (2005)

\bibitem{Ale}
A.S. Alexandrov, this volume

\bibitem{Aub}
S.~Aubry, this volume

\bibitem{PA98}
L.~Proville, S.~Aubry, Physica (Amsterdam) \textbf{133D}, 307 (1998)

\bibitem{CGS99}
M.~Capone, M.~Grilli, W.~Stephan, Eur. Phys. J. B \textbf{11}, 551 (1999)

\bibitem{HMDLK04}
C.~Hartinger, F.~Mayr, J.~Deisenhofer, A.~Loidl, T.~Kopp, Phys. Rev. B
  \textbf{69}, 100403(R) (2004)

\bibitem{AK99}
A.S. Alexandrov, P.E. Kornilovitch, Phys. Rev. Lett. \textbf{82}, 807 (1999)

\bibitem{DT01}
J.T. Devreese, J.~Tempere, Phys. Rev. B \textbf{64}, 104504 (2001)

\bibitem{Ri94}
H.~Rietschel, J. Low Temp. Phys. \textbf{95}, 293 (1994)

\bibitem{SRBL89}
R.P. Sharma, L.E. Rehn, P.M. Baldo, J.Z. Lin, Phys. Rev. Lett. \textbf{62},
  2869 (1989)

\bibitem{Hagea90}
T.~Haga, et~al., Phys. Rev. B \textbf{41}, 826 (1990)

\bibitem{Egea91}
T.~Egami, et~al., in \emph{Electronic Structure and Mechanisms of High
  Temperature Superconductivity}, ed. by J.~Ashkenazi, G.~Vezzoli (Plenum
  Press, New York, 1991)

\bibitem{Kiea88}
Y.H. Kim, et~al., Phys. Rev. B \textbf{38}, 6478 (1988)

\bibitem{Jo91}
J.D. Jorgenson, Physics Today \textbf{44}, 34 (1991)

\bibitem{ZS92}
J.~Zhong, H.B. Sch\"uttler, Phys. Rev. Lett. \textbf{69}, 1600 (1992)

\bibitem{MFVH90}
D.~Mihailovi\'{c}, C.M. Foster, K.~Voss, A.J. Heeger, Phys. Rev. B \textbf{42},
  7989 (1990)

\bibitem{Caea94}
P.~Calvani, Solid State Commun. \textbf{91}, 113 (1994)

\bibitem{FLKB93}
J.P. Falck, A.~Levy, M.A. Kastner, R.J. Birgenau, Phys. Rev. B \textbf{48},
  4043 (1993)

\bibitem{Roea05}
O.~R\"osch, O.~Gunnarson, X.~Zhou, T.~Yoshida, T.~Sasagawa, A.~Fujimori,
  Z.~Hussain, Z.X. Shen, S.~Uchida, Phys. Rev. Lett. \textbf{95}, 227002 (2005)

\bibitem{VPD91}
G.~Verbist, F.M. Peeters, J.T. Devreese, Physica Scripta \textbf{T 39}, 66
  (1991)

\bibitem{Emi92b}
D.~Emin, Phys. Rev. B \textbf{45}, 5525 (1992)

\bibitem{Emi94}
D.~Emin, Phys. Rev. Lett. \textbf{72}, 1052 (1994)

\bibitem{AK92}
A.S. Alexandrov, A.B. Krebs, Usp. Fiz. Nauk \textbf{162}, 1 (1992)

\bibitem{AR92b}
A.S. Alexandrov, J.~Ranninger, Physica C \textbf{198}, 360 (1992)

\bibitem{ABMS93}
A.S. Alexandrov, A.M. Bratkovsky, N.F. Mott, E.K.H. Salje, Physica C
  \textbf{215}, 359 (1993)

\bibitem{Hi93}
J.E. Hirsch, Phys. Rev. B \textbf{47}, 5351 (1993)

\bibitem{Mot93}
N.F. Mott, J. Phys. Condens. Matter \textbf{5}, 3487 (1993)

\bibitem{AM94}
A.S. Alexandrov, N.F. Mott, Int. J. Mod. Phys. B \textbf{8}, 2075 (1994)

\bibitem{Ra94}
J.~Ranninger, Acta Phys. Pol. A \textbf{85}, 89 (1994)

\bibitem{ICNC95}
G.~Iadonisi, V.~Cataudella, D.~Ninno, M.L. Chiofalo, Phys. Lett. A
  \textbf{196}, 359 (1995)

\bibitem{BE93}
X.X. Bi, P.C. Eklund, Phys. Rev. Lett. \textbf{70}, 2625 (1993)

\bibitem{AKZA92}
V.I. Anisimov, M.A. Korotin, J.~Zaanen, O.K. Andersen, Phys. Rev. Lett.
  \textbf{68}, 345 (1992)

\bibitem{ZCKM96}
G.~Zhao, K.~Conder, H.~Keller, K.A. M\"uller, Nature \textbf{381}, 676 (1996)

\bibitem{JTMFRC94}
S.~Jin, T.H. Tiefel, M.~McCormack, R.A. Fastnach, R.~Ramesh, L.H. Chen, Science
  \textbf{264}, 413 (1994)

\bibitem{AP99}
P.B. Allen, V.~Perebeinos, Phys. Rev. B \textbf{61}, 10747 (1999)

\bibitem{BPPSK00}
S.J.L. Billinge, T.~Proffen, V.~Petkov, J.L. Sarrao, S.~Kycia, Phys. Rev. B
  \textbf{62}, 1203 (2000)

\bibitem{AB99}
A.S. Alexandrov, A.M. Bratkovsky, Phys. Rev. Lett. \textbf{82}, 141 (1999)

\bibitem{Caea97}
P.~Calvani, P.~Dore, S.~Lupi, A.~Paolone, P.~Maselli, P.~Guira, B.~Ruzicka,
  S.W. Cheong, W.~Sadowski, J. Supercond. \textbf{10}, 293 (1997)

\bibitem{LM97}
J.D. Lee, B.I. Min, Phys. Rev. B \textbf{55}, 12454 (1997)

\bibitem{Quea98}
M.~Quijada, J.~\v{C}erne, J.R. Simpson, H.D. Drew, K.H. Ahn, A.J. Millis,
  R.~Shreekala, R.~Ramesh, M.~Rajeswari, T.~Venkatesan, Phys. Rev. B
  \textbf{58}, 16093 (1998)

\bibitem{KJN98}
K.H. Kim, J.H. Jung, T.W. Noh, Phys. Rev. Lett. \textbf{81}, 1517 (1998)

\bibitem{Mu99}
K.A. M\"uller, J. Supercond. \textbf{12}, 3 (1999)

\bibitem{Deea99}
D.S. Dessau, T.~Saitoh, C.H. Park, Z.X. Shen, P.~Villella, N.~Hamada,
  Y.~Morimoto, Y.~Tokura, J. Supercond. \textbf{12}, 273 (1999)

\bibitem{WLF03}
A.~Wei{\ss}e, J.~Loos, H.~Fehske, Phys. Rev. B \textbf{68}, 024402 (2003)

\bibitem{JSRTHC96}
M.~Jaime, M.B. Salamon, M.~Rubinstein, R.E. Treece, J.S. Horwitz, D.B. Chrisey,
  Phys. Rev. B \textbf{54}, 11914 (1996)

\bibitem{PRCZSZ97}
T.T.M. Palstra, A.P. Ramirez, S.W. Cheong, B.R. Zegarski, P.~Schiffer,
  J.~Zaanen, Phys. Rev. B \textbf{56}, 5104 (1997)

\bibitem{BDKNT96}
S.J.L. Billinge, R.G. DiFrancesco, G.H. Kwei, J.J. Neumeier, J.D. Thompson,
  Phys. Rev. Lett. \textbf{77}, 715 (1996)

\bibitem{SWKT99}
S.~Shimomura, N.~Wakabayashi, H.~Kuwahara, Y.~Tokura, Phys. Rev. Lett.
  \textbf{83}, 4389 (1999)

\bibitem{VROSLMSPFM99}
L.~Vasiliu-Doloc, S.~Rosenkranz, R.~Osborn, S.K. Sinha, J.W. Lynn, J.~Mesot,
  O.H. Seeck, G.~Preosti, A.J. Fedro, J.F. Mitchell, Phys. Rev. Lett.
  \textbf{83}, 4393 (1999)

\bibitem{Daea00}
P.~Dai, J.A. Fernandez-Baca, N.~Wakabayashi, E.W. Plummer, Y.~Tomioka,
  Y.~Tokura, Phys. Rev. Lett. \textbf{85}, 2553 (2000)

\bibitem{LEBRB97}
D.~Louca, T.~Egami, E.L. Brosha, H.~R\"oder, A.R. Bishop, Phys. Rev. B
  \textbf{56}, R8475 (1997)

\bibitem{ZSPK00}
G.M. Zhao, V.~Smolyaninova, W.~Prellier, H.~Keller, Phys. Rev. Lett.
  \textbf{84}, 6086 (2000)

\bibitem{FRMB93}
H.~Fehske, H.~R\"oder, A.~Mistriotis, H.~B\"uttner, J. Phys. Condens. Matter
  \textbf{5}, 3635 (1993)

\bibitem{Fer94}
R.~Fehrenbacher, Phys. Rev. B \textbf{49}, 12230 (1994)

\bibitem{SS95}
A.~Sherman, M.~Schreiber, Phys. Rev. B \textbf{52}, 10621 (1995)

\bibitem{Da94}
E.~Dagotto, Rev. Mod. Phys. \textbf{66}, 763 (1994)

\bibitem{FRWM95}
H.~Fehske, H.~R\"oder, G.~Wellein, A.~Mistriotis, Phys. Rev. B \textbf{51},
  16582 (1995)

\bibitem{SPS97}
T.~Sakai, D.~Poilblanc, D.J. Scalapino, Phys. Rev. B \textbf{55}, 8445 (1997)

\bibitem{Po91}
D.~Poilblanc, Phys. Rev. B \textbf{44}, 9562 (1991)

\bibitem{TT92}
D.B. Tanner, T.~Timusk, in \emph{Physical Properties of High--Temparature
  Superconductors III}, ed. by D.M. Ginsberg (World Scientific, Singapore,
  1992), p. 363

\bibitem{AKR94b}
A.S. Alexandrov, V.V. Kabanov, D.K. Ray, Physica C \textbf{224}, 247 (1994)

\bibitem{LF97}
J.~Loos, H.~Fehske, Phys. Rev. B \textbf{56}, 3544 (1997)

\bibitem{SWZ88}
J.R. Schrieffer, X.G. Wen, S.C. Zhang, Phys. Rev. Lett. \textbf{60}, 944 (1988)

\bibitem{Re67}
H.G. Reik, Z. Phys. \textbf{203}, 364 (1967)

\bibitem{SPH91}
Y.Y. Suzuki, P.~Pincus, A.J. Heeger, Phys. Rev. B \textbf{44}, 7127 (1991)

\bibitem{Ze51b}
C.~Zener, Phys. Rev. \textbf{82}, 403 (1951)

\bibitem{KO72a}
K.~Kubo, N.~Ohata, J. Phys. Soc. Jpn. \textbf{33}, 21 (1972)

\bibitem{WF02b}
A.~Wei{\ss}e, H.~Fehske, Eur. Phys. J. B \textbf{30}, 487 (2002)

\bibitem{WF04b}
A.~Wei{\ss}e, H.~Fehske, New J. Phys. \textbf{6}, 158 (2004)

\bibitem{Saea95}
D.D. Sarma, N.~Shanthi, S.R. Barman, N.~Hamada, H.~Sawada, K.~Terakura, Phys.
  Rev. Lett. \textbf{75}, 1126 (1995)

\bibitem{FO99}
L.F. Feiner, A.M. Ole\'s, Phys. Rev. B \textbf{59}, 3295 (1999)

\bibitem{BBKLCN98}
C.H. Booth, F.~Bridges, G.H. Kwei, J.M. Lawrence, A.L. Cornelius, J.J.
  Neumeier, Phys. Rev. Lett. \textbf{80}, 853 (1998)

\end{thebibliography}
%
\printindex
\end{document}